\documentclass[iop,apj,useAMS,usenatbib]{emulateapj-rtx4}
\usepackage{amsmath,epsfig,url,natbib}
\usepackage{graphicx,epstopdf,float,color, longtable, gensymb,
  multirow, booktabs, float}
\usepackage{multirow}
\usepackage{gensymb}
\usepackage{courier}
\usepackage[hang,flushmargin]{footmisc}
\usepackage{hyperref}
\hypersetup{colorlinks=true, citecolor=blue, filecolor=magenta, urlcolor=cyan}
\def\lsim{\mathrel{\rlap{\lower4pt\hbox{\hskip1pt$\sim$}}
    \raise1pt\hbox{$<$}}}                % less than or approx. symbol
\def\gsim{\mathrel{\rlap{\lower4pt\hbox{\hskip1pt$\sim$}}
    \raise1pt\hbox{$>$}}}                % greater than or approx. symbol

\submitted{}
\begin{document}
\title{CANDELS Multiwavelength Catalogs: Source Identification and
  Photometry in the CANDELS COSMOS Survey Field}

\author{H. Nayyeri\altaffilmark{1,2}, S. Hemmati\altaffilmark{2,3},
  B. Mobasher\altaffilmark{2}, H. C. Ferguson\altaffilmark{4},
  A. Cooray\altaffilmark{1}, G. Barro\altaffilmark{5,6},
  S. M. Faber\altaffilmark{5}, M. Dickinson\altaffilmark{7},
  A. M. Koekemoer\altaffilmark{4}, M. Peth\altaffilmark{8},
  M. Salvato\altaffilmark{9}, M. L. N. Ashby\altaffilmark{10},
  B. Darvish\altaffilmark{2,11},  J. Donley\altaffilmark{12},
  M. Durbin\altaffilmark{4}, S. Finkelstein\altaffilmark{13},
  A. Fontana\altaffilmark{14}, N. A. Grogin\altaffilmark{4},
  R. Gruetzbauch\altaffilmark{15}, 
  K. Huang\altaffilmark{16}, A. A. Khostovan\altaffilmark{2},
  D. Kocevski\altaffilmark{17}, D. Kodra \altaffilmark{18}, B. Lee\altaffilmark{19},
  J. Newman\altaffilmark{18}, C. Pacifici\altaffilmark{20,21},
  J. Pforr\altaffilmark{22,23}, M. Stefanon\altaffilmark{24},
  T. Wiklind\altaffilmark{25}, S. P. Willner\altaffilmark{10},
  S. Wuyts\altaffilmark{26}, M. Castellano\altaffilmark{14},
  C. Conselice\altaffilmark{27}, T. Dolch\altaffilmark{28},
  J. S. Dunlop\altaffilmark{29}, A. Galametz\altaffilmark{9},
  N. P. Hathi\altaffilmark{22}, R. A. Lucas\altaffilmark{4}, H. Yan\altaffilmark{30}}

\altaffiltext{1}{Department of Physics and Astronomy, University of California Irvine, Irvine, CA 92697}
\altaffiltext{2}{Department of Physics and Astronomy, University of California Riverside, Riverside, CA 92521}
\altaffiltext{3}{Infrared Processing and Analysis Center, California Institute of Technology, MS 100-22, Pasadena, CA 91125}
\altaffiltext{4}{Space Telescope Science Institute, 3700 San Martin Drive, Baltimore, MD 21218, USA}
\altaffiltext{5}{UCO/Lick Observatory, Department of Astronomy and Astrophysics, University of California, Santa Cruz, CA 95064, USA}
\altaffiltext{6}{Department of Astronomy, University of California at Berkeley, Berkeley, CA 94720-3411, USA}
\altaffiltext{7}{National Optical Astronomy Observatory, Tucson, AZ 85719, USA}
\altaffiltext{8}{Department of Physics and Astronomy, The Johns
  Hopkins University, 366 Bloomberg Center, Baltimore, MD 21218, USA}
\altaffiltext{9}{Max-Planck-Institut fur Extraterrestrische Physik,
  Giessenbachstrasse 1, D-85748 Garching bei Munchen, Germany}
\altaffiltext{10}{Harvard Smithsonian Center for Astrophysics, 60
  Garden Street, Cambridge, MA 02138, USA}
\altaffiltext{11}{Cahill Center for Astrophysics, California Institute of Technology,
1216 East California Boulevard, Pasadena, CA 91125, USA}
\altaffiltext{12}{Los Alamos National Laboratory, Los Alamos, NM
87544, USA}
\altaffiltext{13}{Department of Astronomy, The University of Texas at Austin, Austin, TX 78712, USA}
\altaffiltext{14}{INAF - Osservatorio Astronomico di Roma, via
  Frascati 33, I-00040 Monte Porzio Catone (RM), Italy}
\altaffiltext{15}{Center for Astronomy and Astrophysics, Observatorio Astronomico de Lisboa, Tapada da Ajuda, 1349-018 Lisboa, Portugal}
\altaffiltext{16}{Department of Physics, University of California, Davis, CA 95616, USA}
\altaffiltext{17}{Colby College, 4000 Mayflower Hill, Waterville, Maine 04901}
\altaffiltext{18}{Department of Physics and Astronomy and PITT PACC, University of Pittsburgh, Pittsburgh, PA 15260, USA}
\altaffiltext{19}{Department of Astronomy, University of Massachusetts,
  710 North Plesant Street, Amherst, MA 01003, USA}
\altaffiltext{20}{NASA Postdoctoral Program Fellow}
\altaffiltext{21}{Goddard Space Flight Center, Code 665, Greenbelt, MD, USA}
\altaffiltext{22}{Aix Marseille Universit´e, CNRS, LAM (Laboratoire
  d’Astrophysique de Marseille) UMR 7326, 13388, Marseille, France}
\altaffiltext{23}{ESA/ESTEC SCI-S, Keplerlaan 1, 2201 AZ, Noordwijk, The Netherlands}
\altaffiltext{24}{Huygens Laboratory Niels Bohrweg 2 2333 CA Leiden}
\altaffiltext{25}{Department of Physics, Catholic University of America, 20064 Washington DC}
\altaffiltext{26}{Department of Physics, University of Bath, Claverton
  Down, Bath, BA2 7AY, UK}
\altaffiltext{27}{School of Physics and Astronomy, University of
  Nottingham, Nottingham, UK}
\altaffiltext{28}{Department of Physics, Hillsdale College 33 E. College
  St. Hillsdale, MI 49242}
\altaffiltext{29}{Institute for Astronomy, University of Edinburgh,
  Royal Observatory, Edinburgh EH9 3HJ, UK}
\altaffiltext{30}{Department of Physics and Astronomy, University of
  Missouri, Columbia, MO 65211, USA}

\journalinfo{Accepted to the Astrophysical Journal Supplement Series}
\submitted{}

\begin{abstract}

We present a multi-wavelength photometric catalog in the COSMOS field
as part of the observations by the Cosmic Assembly Near-infrared Deep
Extragalactic Legacy Survey (CANDELS). The catalog is based on {\it Hubble
Space Telescope} Wide Field Camera 3 ({\it HST}/WFC3) and Advanced Camera for
Surveys (ACS) observations of the COSMOS field (centered at RA: $10^h00^m28^s$,
Dec:$+02^{\circ}12^{\prime}21^{\prime\prime}$). The final catalog has
38671 sources with photometric data in
forty two bands from UV to the infrared ($\rm \sim 0.3-8\,\mu
m$). This includes broad-band photometry from the {\it HST}, CFHT, Subaru, VISTA and {\it Spitzer Space
Telescope} in the visible, near infrared and infrared bands along with
intermediate and narrow-band photometry from Subaru and medium band data from Mayall
NEWFIRM. Source detection was conducted in the WFC3 F160W band (at
1.6\,$\mu$m) and photometry is generated using the Template FITting
algorithm. We further present a catalog of the physical properties of sources as identified
in the {\it HST} F160W band and measured from the multi-band photometry by
fitting the observed spectral energy distributions of sources against templates.

\end{abstract}


\maketitle

\section{Introduction}

The Cosmic Assembly Near-infrared Deep Extragalactic Legacy Survey
(CANDELS: PI. S. Faber and H. Ferguson; see \citealp {Grogin2011} and
\citealp {Koekemoer2011}) is the largest Multi-Cycle Treasury (MCT)
program ever approved on the
{\it HST}, with more than 900 orbits, and it was designed to use deep
observations by the Wide Field Camera 3 (WFC3) and Advanced Camera for Surveys (ACS)
instruments to study galaxy formation and evolution throughout
cosmic time in five fields in many different
bands. The observations were done by
the {\it HST}/WFC3 as the main mode with ACS observations in parallel. The
CANDELS images are publicly available, and multi
wavelength photometric catalogs are made available by the CANDELS team
following the release of the images. All CANDELS photometric catalogs
were selected based on the WFC3 F160W band
and this is the reference image for all the other {\it HST}
and non-{\it HST} data. This provides a data-set with consistent
photometry and physical properties across all the fields targeted as
part of the survey. The CANDELS catalogs for the first two observed
fields of UDS and GOODS-S \citep{Galametz2013, Guo2013} are already
publicly available\setcounter{footnote}{0}\footnote{\url{http://vizier.cfa.harvard.edu/viz-bin/VizieR?-source=J/ApJS/206/10}
  and \url{http://vizier.cfa.harvard.edu/viz-bin/VizieR?-source=J/ApJS/207/24}} and the
three fields of COSMOS (this work), EGS (Stefanon et al. 2016) and
GOODS-N (Barro et al., in prep.) are in progress. The CANDELS observations
are aimed at achieving several major science goals that could only be
attained with data at the depth and resolution of CANDELS. These
include studying the most distant objects in the Universe at the epoch
of reionization in the {\it cosmic dawn}
(e.g. \citealp{Finkelstein2012, Grazian2012, Yan2012, Lorenzoni2013,
  Oesch2013, Duncan2014, Bouwens2015, Giallongo2015, Finkelstein2015,
  Song2015, Roberts2015, Caputi2015, Mitchell2015}), understanding galaxy
formation and evolution during the peak epoch of star formation in the
{\it cosmic high noon} (e.g. \citealp{Bruce2012, Bell2012,
  Kocevski2012, Wuyts2013, Lee2013, Hemmati2014,
  Barro2014b, Whitaker2014, Williams2015, Hemmati2015}) and
studying star-formation from deep UV
observations and cosmological studies from supernova
observations (e.g. \citealp{Jones2013, Teplitz2013, Rodney2014,
  Strolger2015, Rodney2016}). These main science goals are described in more detail by
\citet{Grogin2011} and \citet{Koekemoer2011}.

One of the major goals of modern observational cosmology is to study
the formation and evolution of galaxies with cosmic time. Recent
advances in this frontier have been enabled by the availability of
observations in different wavelengths, targeting different
populations of galaxies (e.g. \citealp{York2000, Giavalisco2004, Skrutskie2006,
  Lawrence2007, Scoville2007, Wolf2003, Bell2004, Faber2007,
  Ilbert2013, Khostovan2015, Hemmati2016, Vasei2016, Shivaei2015}). The advent of the {\it
  HST} benefited many such studies by making it possible to have the deepest
observations of the sky in multiple bands. In particular the installation
of the Wide Field Camera 3 (WFC3) on board {\it HST} initiated a new
stage for studying galaxy evolution at new extremes \citep{McLeod2015, Oesch2016}. 

\begin{figure*}
\centering
\leavevmode
\includegraphics[trim=1cm 0cm 0cm 0cm,scale=0.43]{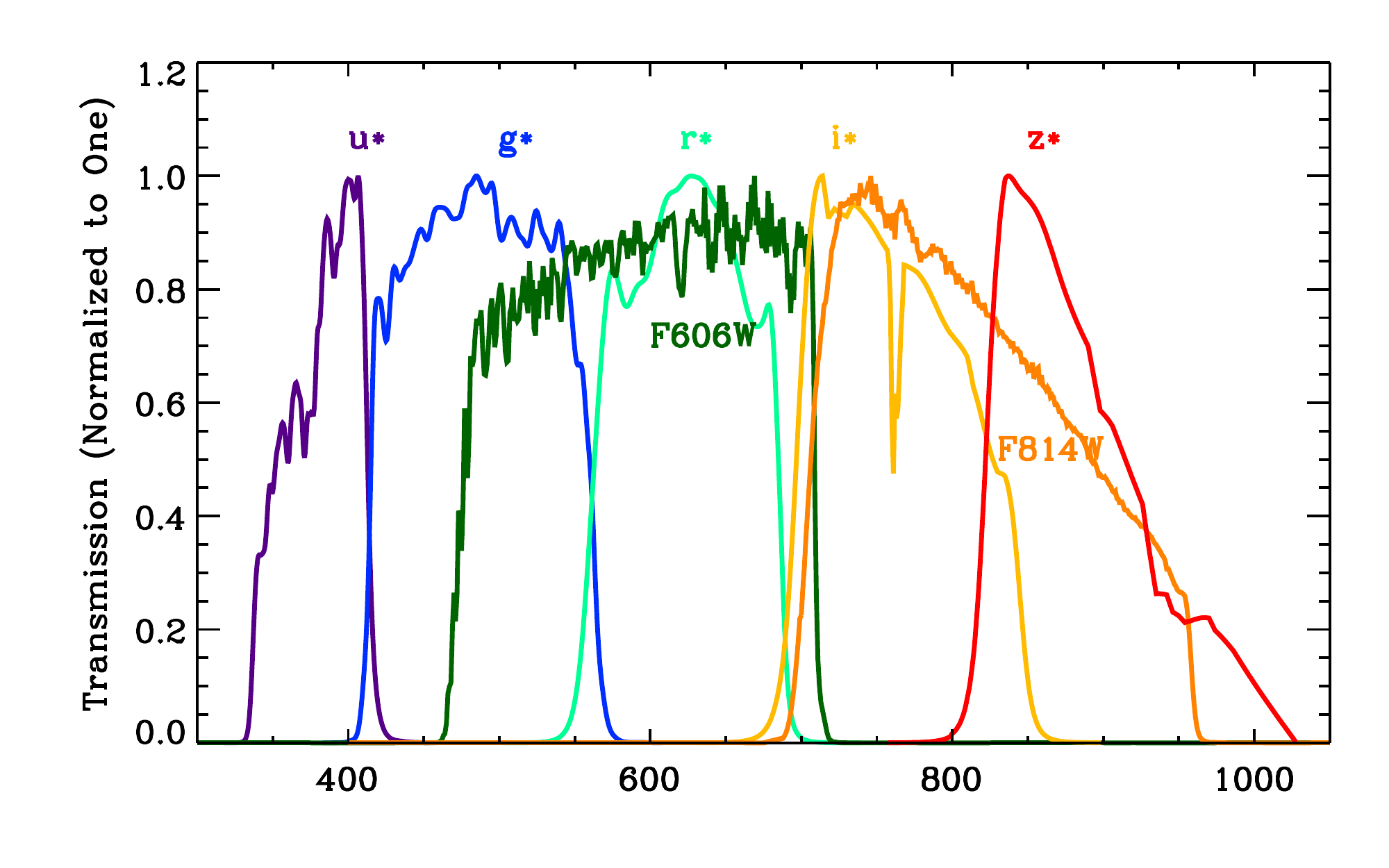}
\includegraphics[trim=1cm 0cm 0cm 0cm,scale=0.43]{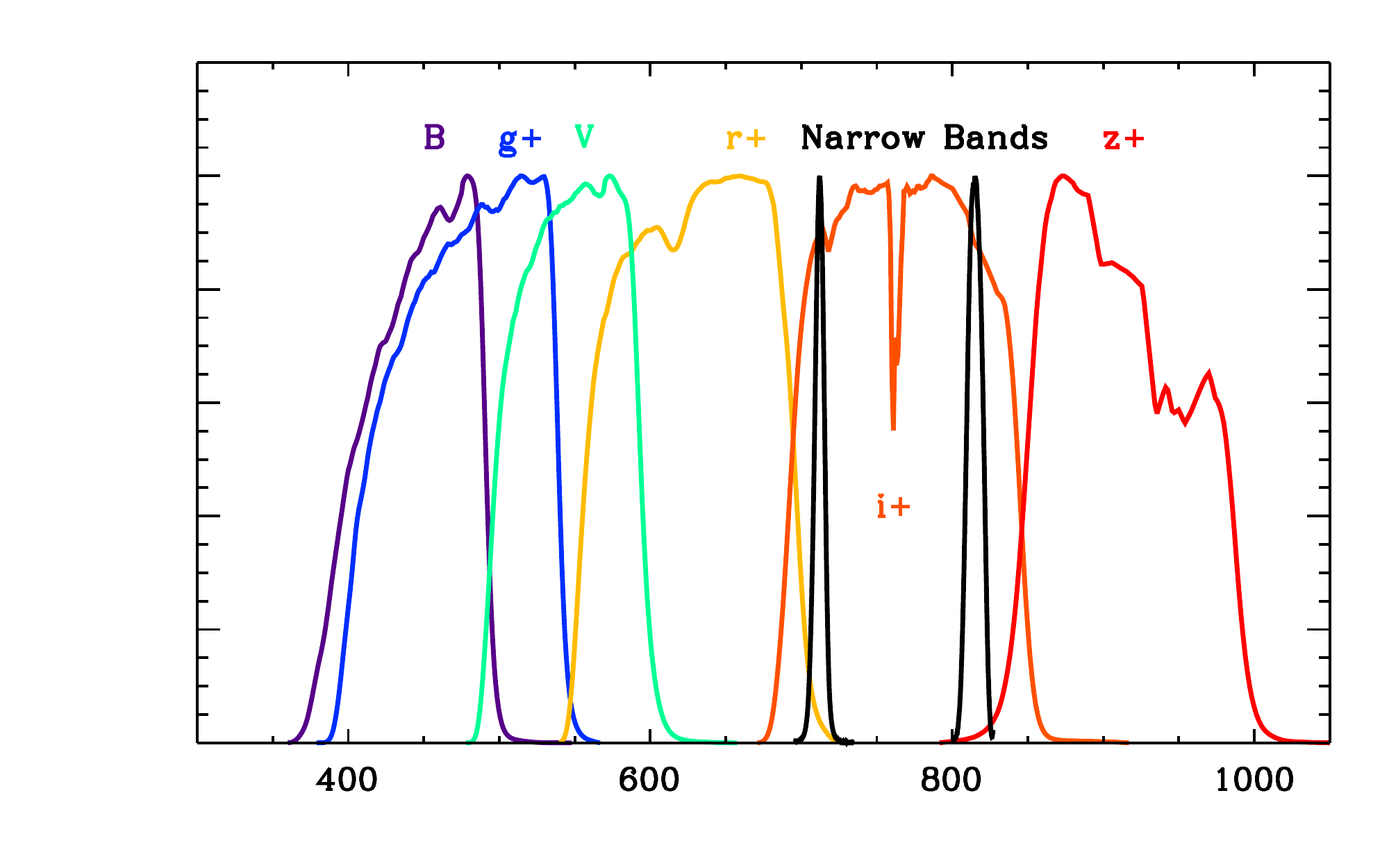} \\
\includegraphics[trim=1cm 0cm 0cm 0cm,scale=0.43]{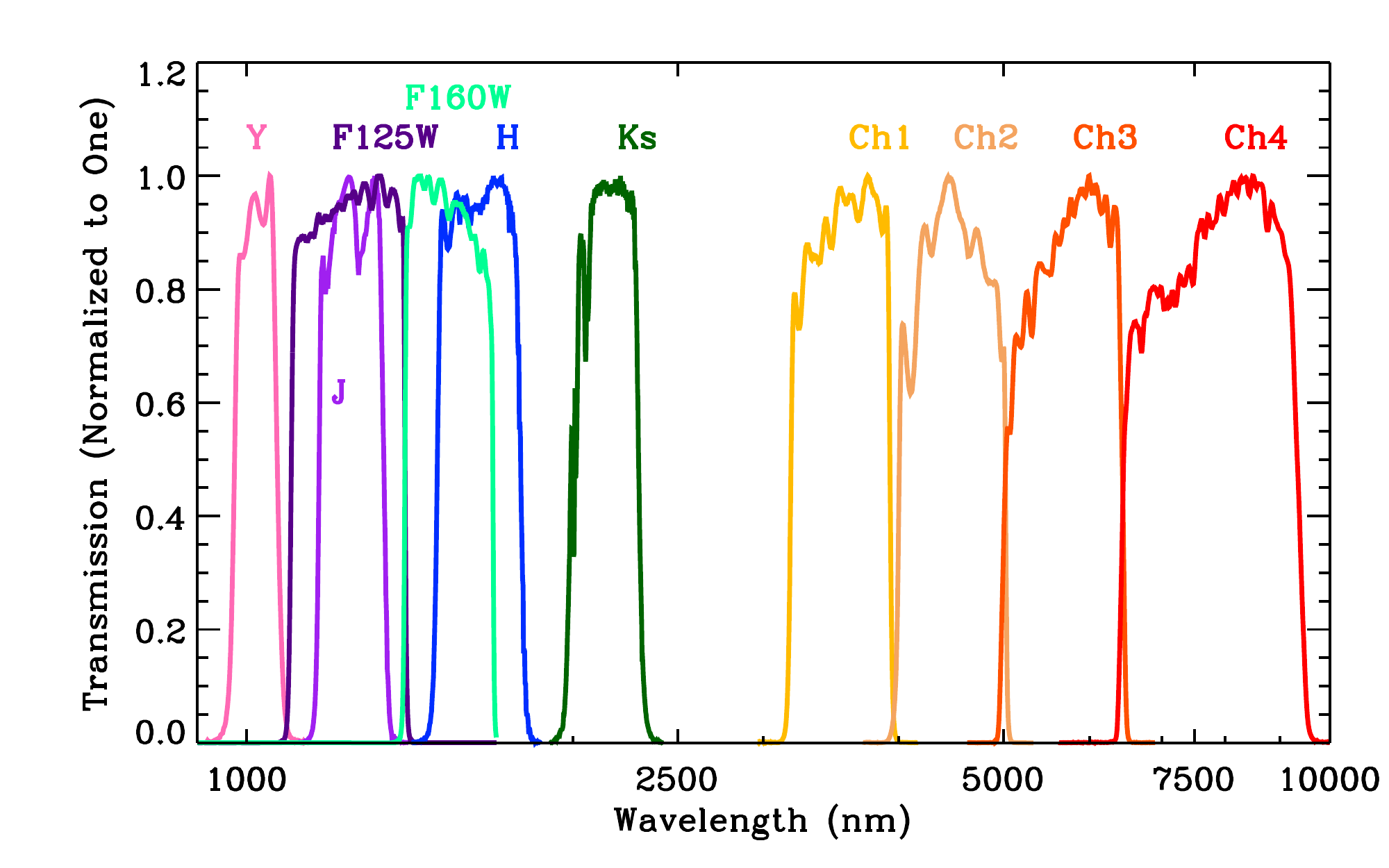}
\includegraphics[trim=1cm 0cm 0cm 0cm,scale=0.43]{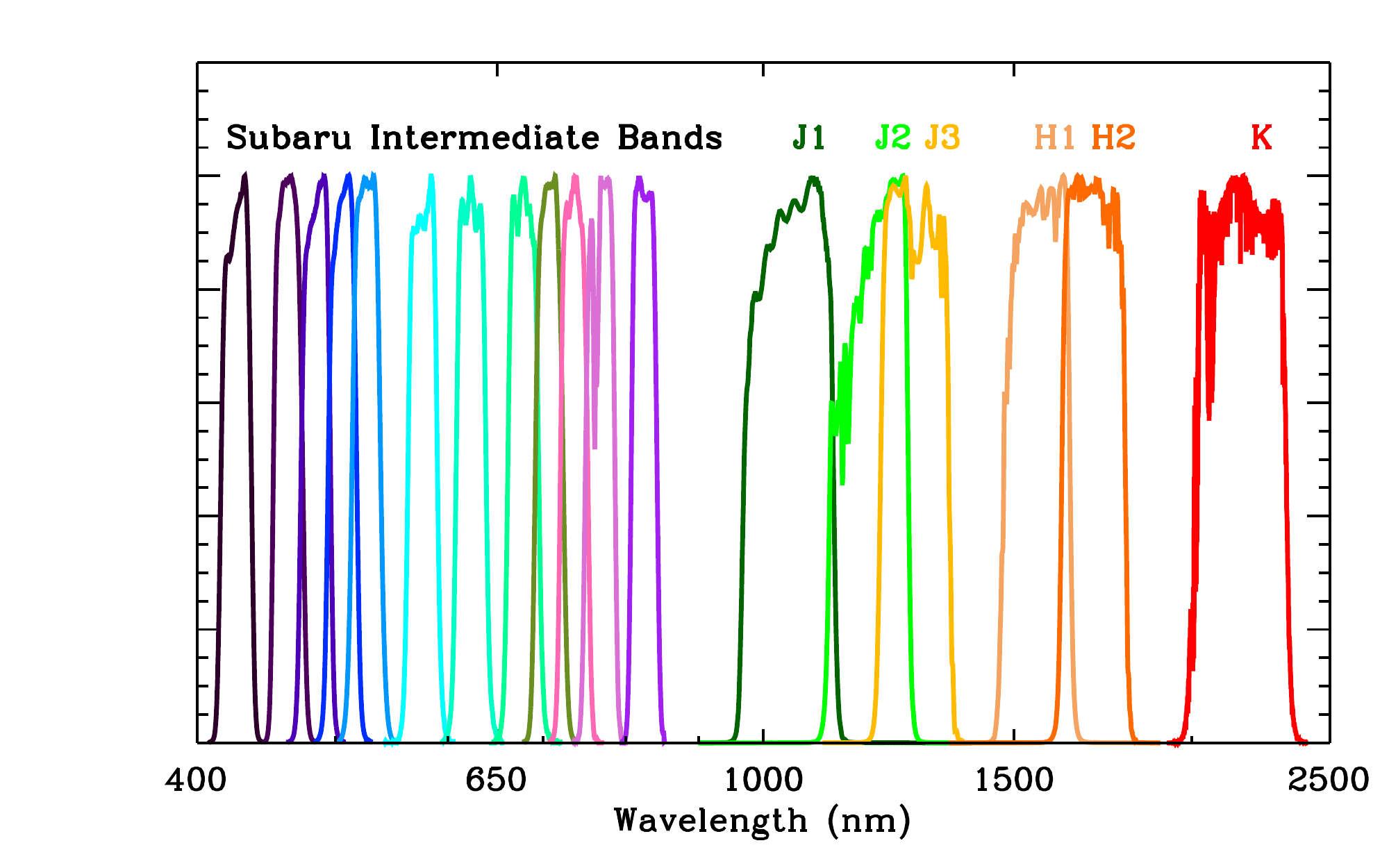}
\caption{The transmission curves for the CFHT and ACS visible (top
  left), Subaru optical broad-band and narrow-band (top right), WFC3 and UltraVISTA near infrared and
  {\it Spitzer} infrared (bottom left) and the Subaru intermediate and NEWFIRM
  medium band filters (bottom right) used in the COSMOS
  CANDELS TFIT catalog. These cover observations at $\rm \sim 0.3-8\,\mu
m$ in forty two filters. The filters are adopted from \url{http://cosmos.astro.caltech.edu/page/filterset}. The effective wavelength of the filters are reported in Tables 1 and 2.}
\end{figure*}

Galaxy populations at
different look-back times range from very blue star forming
galaxies to red dusty or very old systems. Understanding the
evolution of these populations relies on the availability of
multi-wavelength photometric data from the bluest to the reddest bands
possible. The CANDELS multi-wavelength catalogs combine the best and
deepest observations by the {\it HST} with the deepest ground-based
observations and {\it Spitzer Space Telescope}
data \citep{Galametz2013, Guo2013}. These catalogs of tens of thousands of extragalactic
sources, consistently measured across many bands
from $\rm \sim 0.3-8\,\mu m$, bring a unique opportunity to study galaxy
evolution.

The Cosmic Evolution Survey (COSMOS; \citealp{Scoville2007}) centered at RA:$10^h00^m28^s$,
Dec:$+02^{\circ}12^{\prime}21^{\prime\prime}$) is a 2\,deg$^2$
field located near the celestial equator. It was initially picked to maximize the visibility
from observatories from both hemispheres and was specifically chosen to
avoid any bright X-ray, UV or radio sources \citep{Scoville2007} and
to be large enough for studies of large scale structure (e.g. \citealp
{Scoville2007b, Kovac2010, Scoville2013, Darvish2014, Darvish2015a}). 

The COSMOS field was targeted by CANDELS in a north-south strip, lying
within the central ultra-deep strip of the UltraVISTA imaging
\citep{McCracken2012} and hence also the {\it Spitzer} SEDS imaging
\citep{Ashby2013} in order to ensure the best possible supporting
data at longer near-infrared wavelengths. The {\it HST} observations
cover an area of $\simeq 216$\,arcmin$^2$ in the WFC3/IR with parallel
ACS observations. The catalog was selected
in the {\it HST}/WFC3 F160W band and has multi-band data for 38671
objects from $\sim$0.3 to 8\,$\mu$m. These fluxes are measured
consistently across all these bands and in agreement with photometry
measurement techniques adopted by all CANDELS catalogs. We use the
unprecedented depth and resolution provided
by the {\it HST} for measuring the flux of the faintest targets across
all bands. By fitting this multi-waveband information with
template libraries, we also measured photometric redshift and
stellar mass for each object.

The paper is organized as follow. Section 2 presents the data used to
compile the catalog. Section 3 describes our
photometry of the {\it HST}, {\it Spitzer} and ground-based
bands. Section 4 is devoted to data quality checks for our TFIT
photometry. Physical parameter estimation using the measured
photometry is presented in Section
5. In Section 6 we investigate the applications of our
deep photometry on studies of high redshift star forming and quiescent
galaxies. We summarize our results in Section
7. Throughout this paper we assume a cosmological model
with $\text{H}_0=70\:\text{kms}^{-1}\text{Mpc}^{-1}$, $\Omega_m=0.3$ and
$\Omega_\Lambda=0.7$. All magnitudes are in the AB system where
$\text{m}_{\text{AB}}=23.9-2.5 \times \text{log}(f_{\nu}/1\mu
\text{Jy})$ \citep {Oke1983}.

The CANDELS COSMOS photometry catalog will be publicly available through the
CANDELS website\footnote{\url{http://candels.ucolick.org/}} along with the physical properties estimates and all
the documentations. These will also be available on the Mikulski
Archive for Space Telescopes (MAST)\footnote{\url{https://archive.stsci.edu/}}, via the online version of the
catalog and through Centre de Donnees astronomiques de Strasbourg
(CDS). We will also make these data available through the Rainbow
Database\footnote{\url{https://rainbowx.fis.ucm.es/Rainbow\_navigator\_public/}}
\citep{Perez2008, Barro2011}.

\begin{figure*}

\centering
\includegraphics[trim=1cm 13.5cm 0cm 2cm,scale=0.95]{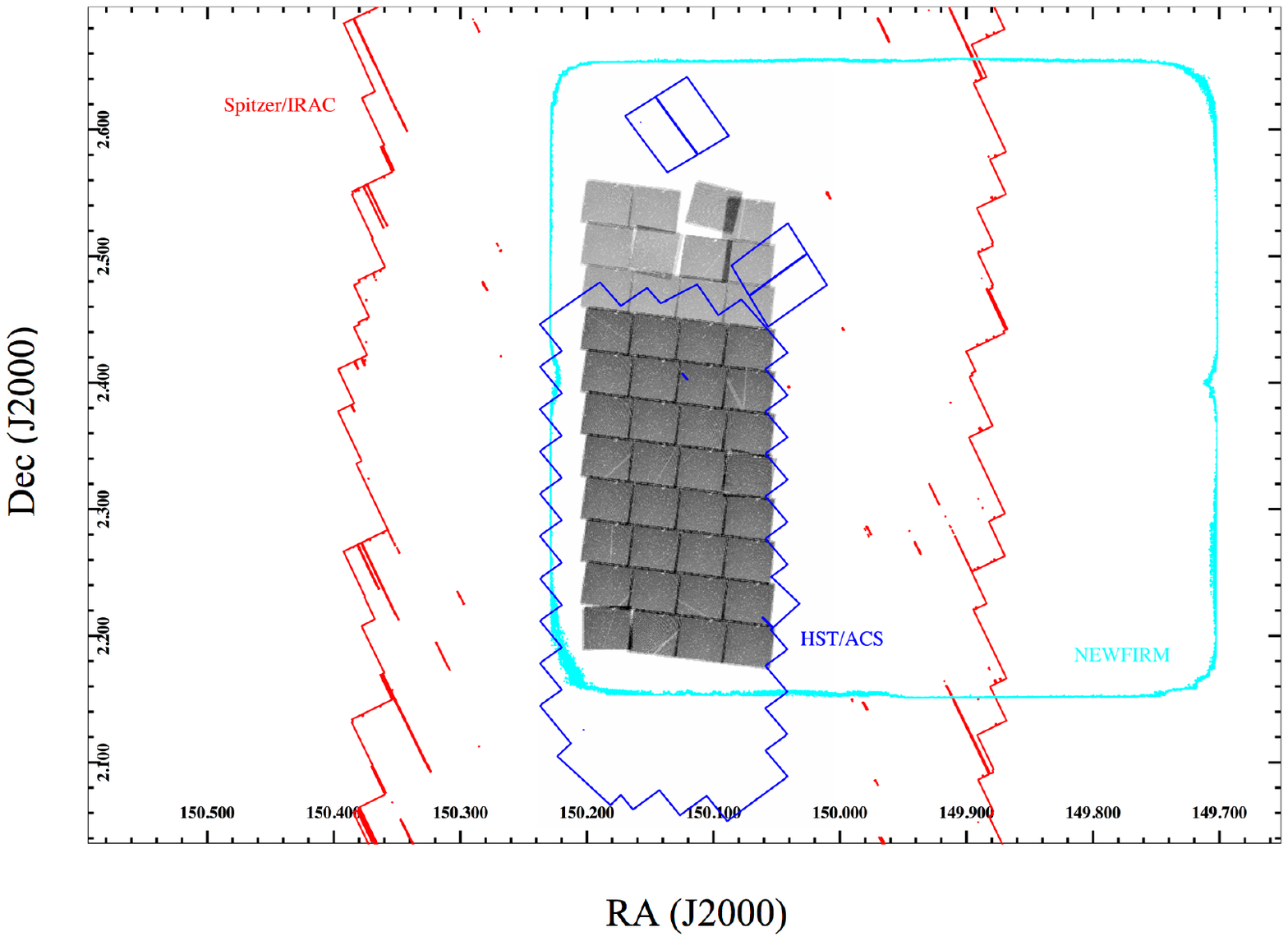}
\caption{Sky coverage of data in the COSMOS field. The WFC3
  F160W mosaics are shown as the grey shaded region. The entire WFC3
  footprint is covered by the CFHT, Subaru and UltraVISTA observations (which
  are much larger than the scales of this Figure).}

\end{figure*}

\section{Data}

The COSMOS field \citep{Scoville2007} has been observed in many
different wavelengths from the X-ray to the far infrared. There are
observations from the CFHT/MegaPrime
in the $u^*$, $g^*$, $r^*$, $i^*$ and $z^*$
\citep{Gwyn2012}, from the Subaru/Suprime-Cam in the $B$, $g^+$, $V$, $r^+$,
$i^+$ and $z^+$ \citep{Taniguchi2007}, from the
VLT/VISTA in the $Y$, $J$, $H$ and $K_s$
bands \citep{McCracken2012}, from {\it Spitzer} in the four IRAC
bands \citep{Sanders2007, Ashby2013}, VLA \citep{Schinnerer2007}, XMM
\citep{Hasinger2007, Cappelluti2007}, Chandra
\citep{Elvis2009, Civano2016} and GALEX \citep{Schiminovich2005}. There are also
numerous medium and narrow-band observations available in the
COSMOS from the Mayall NEWFIRM \citep{Whitaker2011} and Subaru
Suprime-Cam \citep{Taniguchi2015}. The full COSMOS
field has been observed with {\it HST} in F814W \citep{Koekemoer2007}
and contains more than 2 million galaxies with multi-band data from
the UV to the far-IR \citep{Mobasher2007, Capak2007,
  Ilbert2009, Ilbert2013}. Furthermore, COSMOS has been followed up
spectroscopically by the larger and most sensitive
telescopes/instruments like VLT/VIMOS (e.g. \citealp{Lilly2009,
  Lefevre2015}) and Keck/DEIMOS and MOSFIRE among others\footnote{For
  a complete list of ancillary data on the entire field visit:
  \url{http://astro.caltech.edu/~cosmos}}. This makes it possible to
study different populations of galaxies from blue
and young systems at shorter wavelengths to red dusty or old objects
at longer wavelengths. These ancillary data are accompanied by high
resolution observations from CANDELS
using the {\it Hubble Space Telescope} in both visible and near
infrared. Figure 1 shows the transmission curves for all bands
included in the CANDELS COSMOS catalog.

\subsection{CANDELS {\it HST} Observations}

The CANDELS COSMOS field was observed by the Wide Field Camera 3
(WFC3) in F125W and F160W ($J_{125}$ and $H_{160}$) and in parallel by
the  Advanced Camera for
Surveys (ACS) in the F606W and F814W filters ($V_{606}$ and
$i_{814}$). The WFC3 observations covered a rectangular grid of $4 \times 11$ tiles ($\sim
8^{\prime}.6 \times 23^{\prime}.8$) running north to south allowing for
maximum contiguous coverage in the near infrared. The field was
observed over two epochs with each
tile observed for one orbit in each epoch. The one orbit observations
were divided into two exposures in F125W ($\sim 1/3$ orbit depth) and
F160W  ($\sim 2/3$ orbit depth) along with parallel ACS observations
in the F606W and F814W \citep{Koekemoer2011, Grogin2011}. The exposures in each orbit
were dithered using a small scale pattern providing half-pixel
subsampling of the PSF and also ensuring that the hot pixels and
persistences were moved around. The output is a calibrated and
astrometry-corrected mosaics of all the exposures in the four
individual {\it HST} bands. The astrometry is based on the
  CFHT/MegaCam $i^*$ imaging supplemented by deep Subaru/Suprime-Cam
  $i^+$ imaging with absolute astrometry registered to the VLA 20\,cm
  survey of the COSMOS field (\citealp{Schinnerer2007}; see also \citealp{Koekemoer2007}). This is also the
  adopted reference grid by the COSMOS team \citep{Capak2007}. All the
  ground-based and {\it Spitzer} data (described in the next Section)
  were aligned to the {\it HST} data astrometry using \texttt{SWARP}. For the
  present work we used the V0.5 release of the {\it HST}/ACS and WFC3
  data available from the CANDELS
  website\footnote{\url{http://candels.ucolick.org/data_access/Latest_Release.html}}.
The observation depth, effective wavelength and the PSF information for
each of the {\it HST} filters are summarized in Table 1. 

\begin{figure*}[t]
\centering
\includegraphics[trim=1cm 0cm 1cm 0cm, scale=0.3]{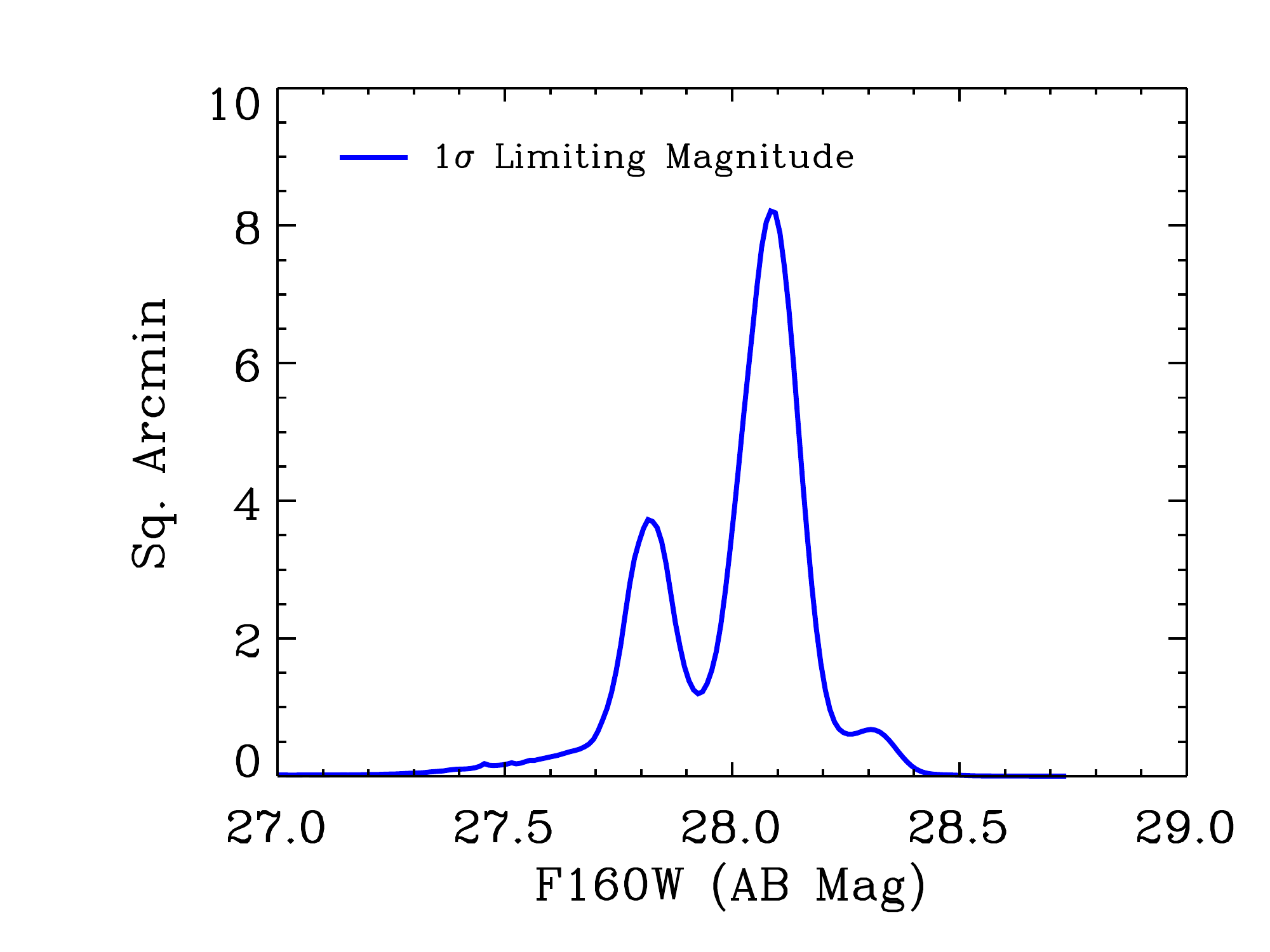}
\includegraphics[trim=1cm 0cm 1cm 0cm, scale=0.3]{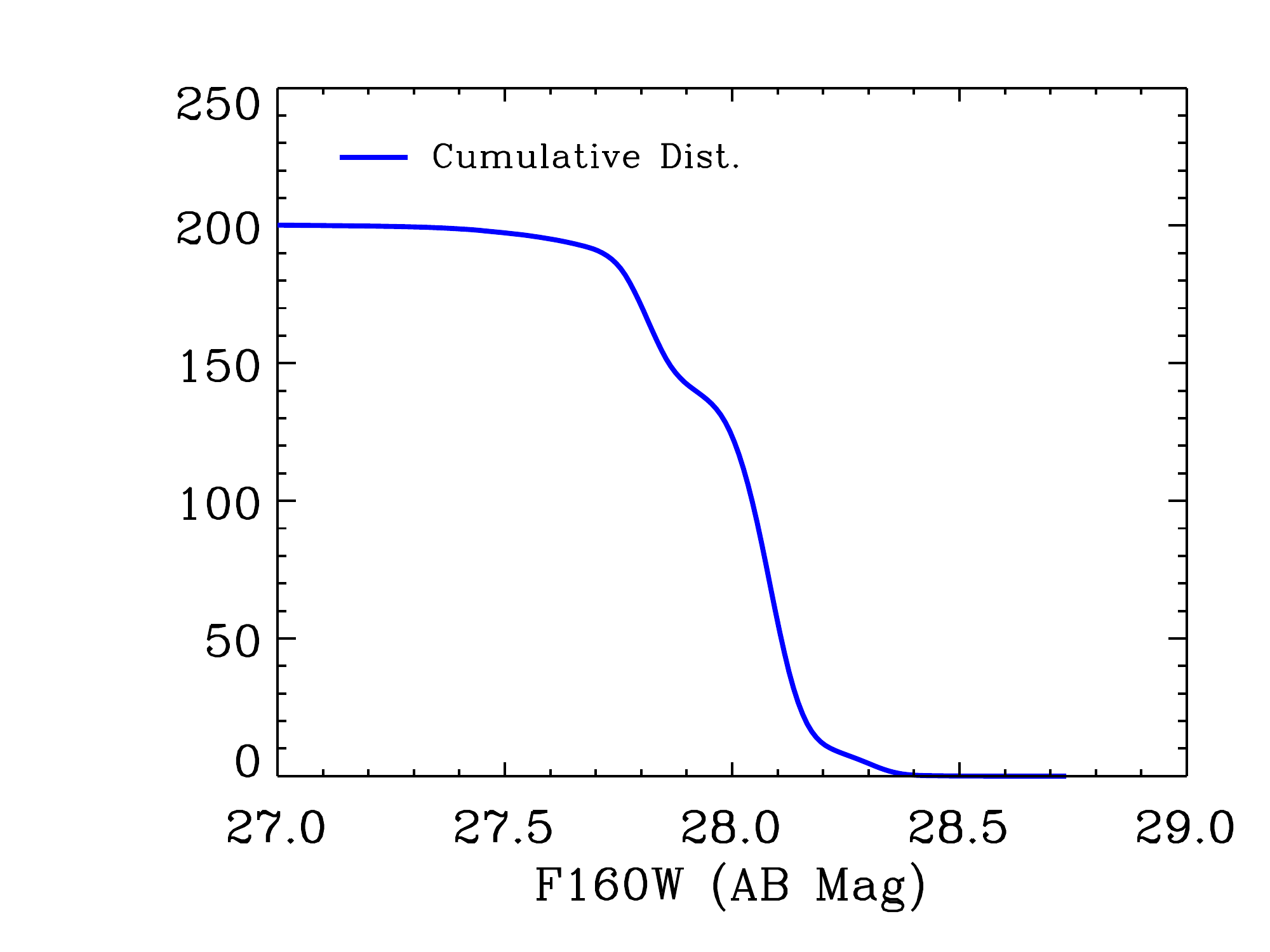}
\includegraphics[trim=1cm 0cm 1cm 0cm, scale=0.3]{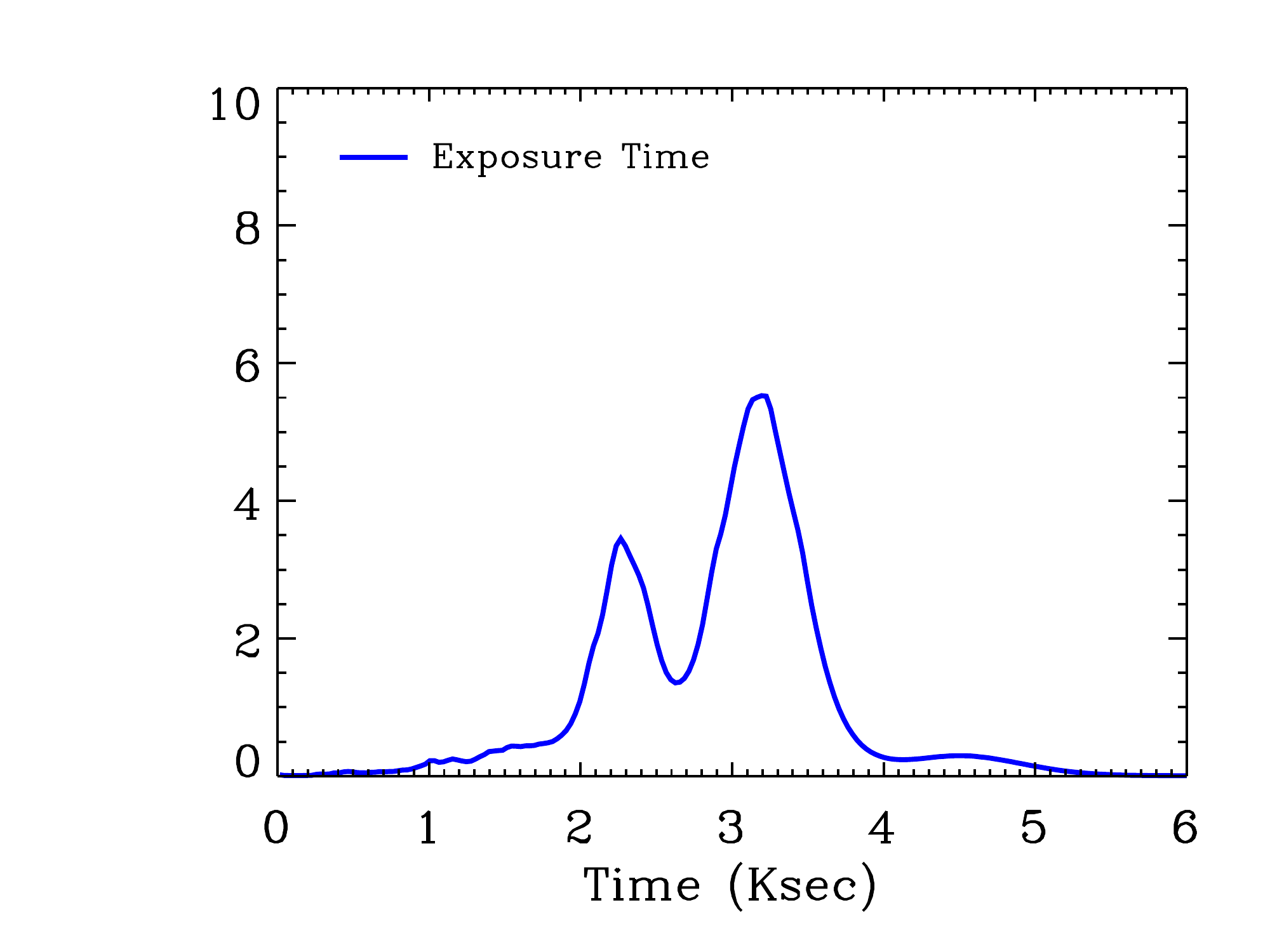}
\caption{Left: The 1$\sigma$ limiting magnitude distribution (per bin of 0.01) in the WFC3 F160W
  detection band in each pixel normalized to an area of
  1\,arcsec$^2$. Middle: Cumulative distribution of the area with a
  sensitivity greater than a given 1$\sigma$ limiting
  magnitude. Right: Distribution of the exposure time.}
\end{figure*}

\subsection{Ground-based Observations}

The full COSMOS field was targeted
by the Canada-France-Hawaii Telescope (CFHT) 3.6\,m telescope MegaPrime instrument in the
$u^*$, $g^*$, $r^*$, $i^*$ and $z^*$
optical bands as part of the CFHT Legacy
Survey\footnote{\url{http://www.cfht.hawaii.edu/Science/CFHTLS/}} with
COSMOS being in the second Deep field. The MegaCam/MegaPrime camera used for
the observations has a pixel scale of
0.187\,arcsec\,pixel$^{-1}$\citep{Boulade2003}. The final images are
from MegaPipe\footnote{\url{http://www.cadc-ccda.hia-iha.nrc-cnrc.gc.ca/en/megapipe/cfhtls/index.html}}
\citep{Gwyn2008} with zero point adjustments. The image processing and
stacking of the data is further described by \citet{Gwyn2012}. In
  this work we used the CFHTLS D2 mosaics 2009 release accessible from
  the CADC\footnote{\url{http://www.cadc-ccda.hia-iha.nrc-cnrc.gc.ca/en/search/?collection=CFHTMEGAPIPE&noexec=true}}.

\begin{figure*}
\centering
\includegraphics[scale=0.42]{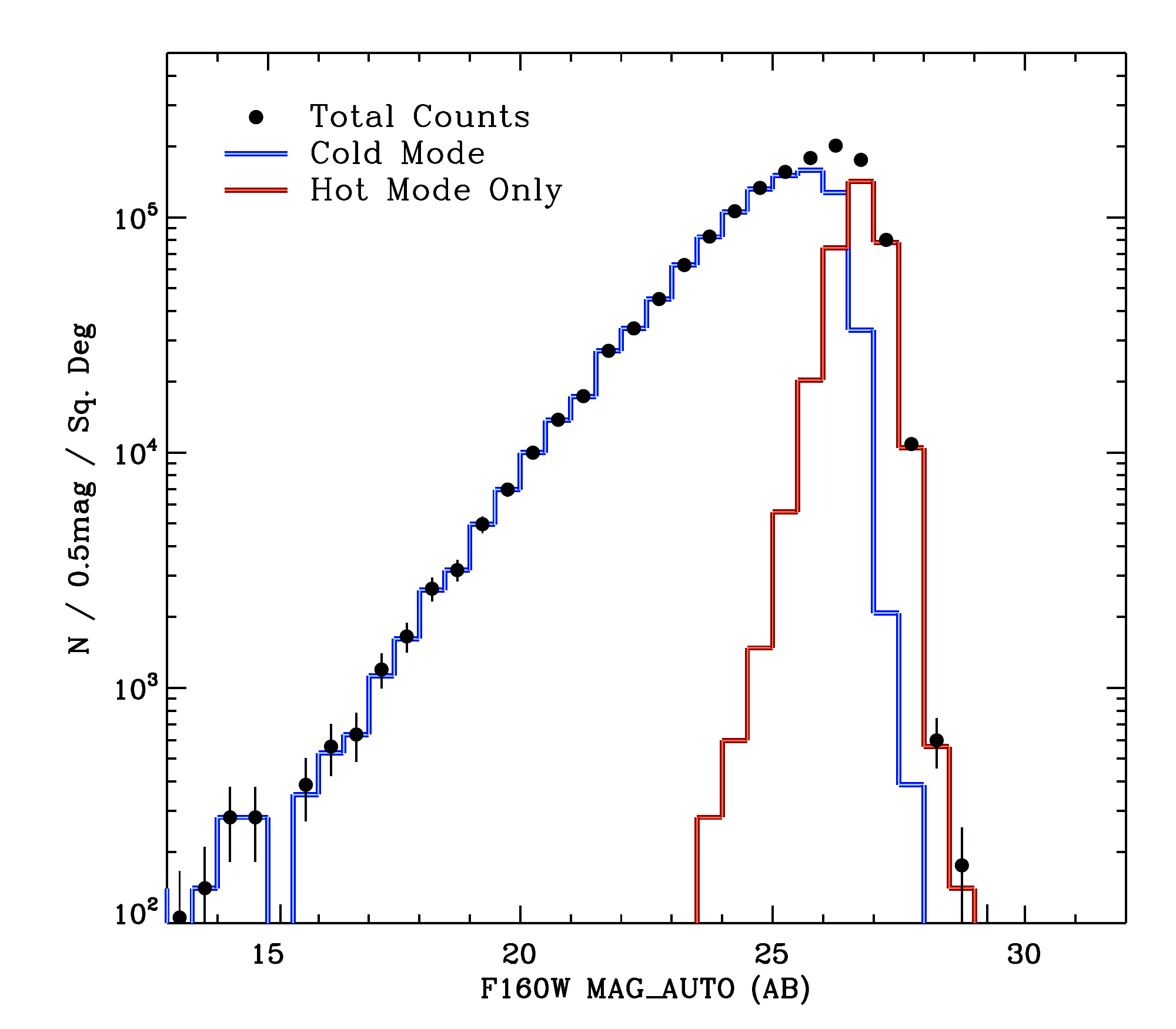}
\includegraphics[scale=0.42]{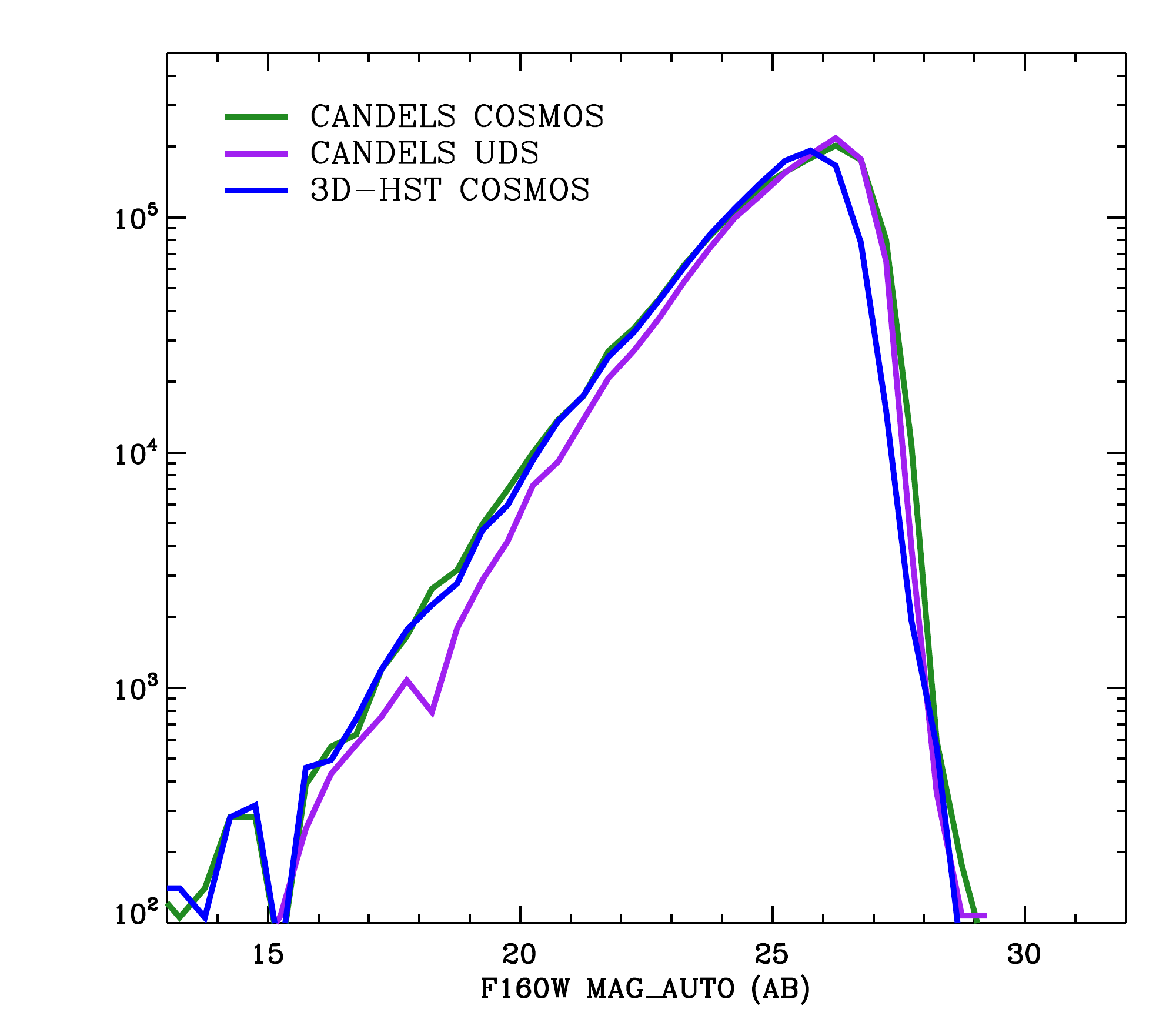}
\caption{Left: The number of galaxies in the CANDELS COSMOS catalog in bins
  of F160W magnitude (black filled circles). The
  number counts of the cold-mode selected galaxies (bright sample) and
hot-mode only selected galaxies (faint sample) are shown in blue and red
respectively. The uncertainties associated with the total counts are
Poisson errors. The counts and the associated uncertainties are
reported in Table 3. Right: Number counts of the combined CANDELS
COSMOS catalog compared to the CANDELS UDS \citep{Galametz2013} and
the 3D-HST COSMOS \citep{Skelton2014}.}
\end{figure*}

\begin{figure*}[t]
\centering
\includegraphics[trim=0cm 0cm 0cm 0cm,scale=0.2]{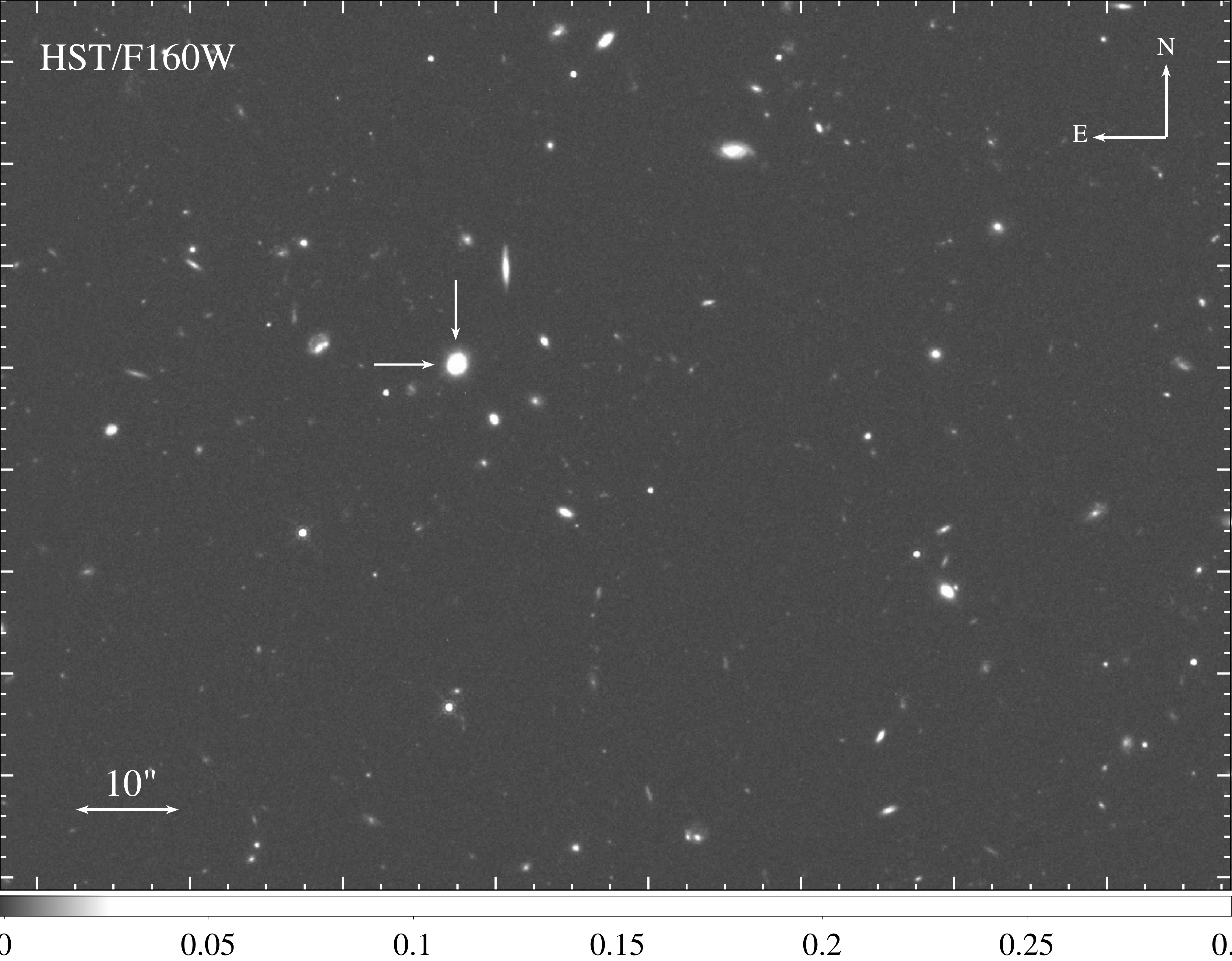}
\includegraphics[trim=0cm 0cm 0cm 0cm,scale=0.2]{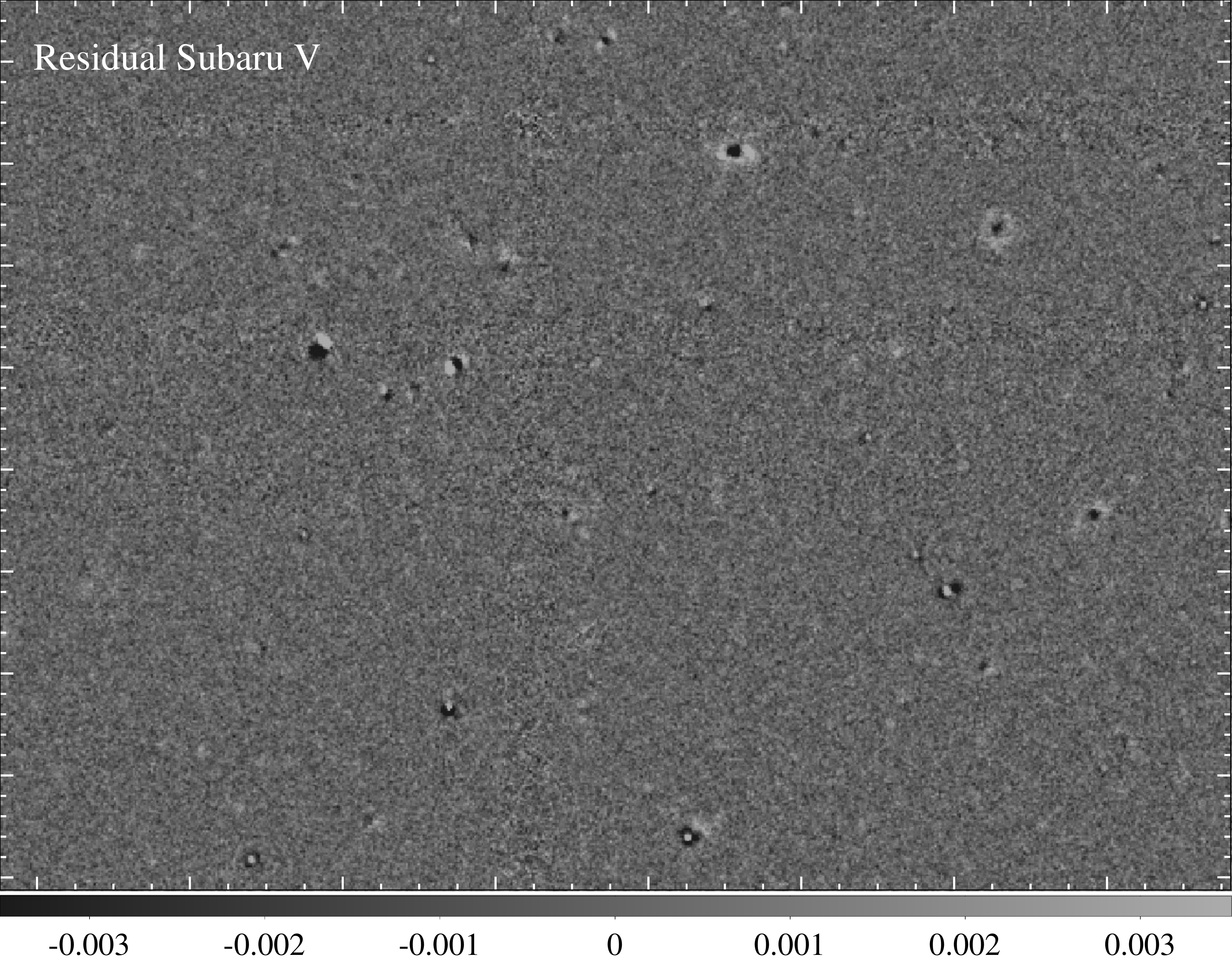}
\includegraphics[trim=0cm 0cm 0cm 0cm,scale=0.2]{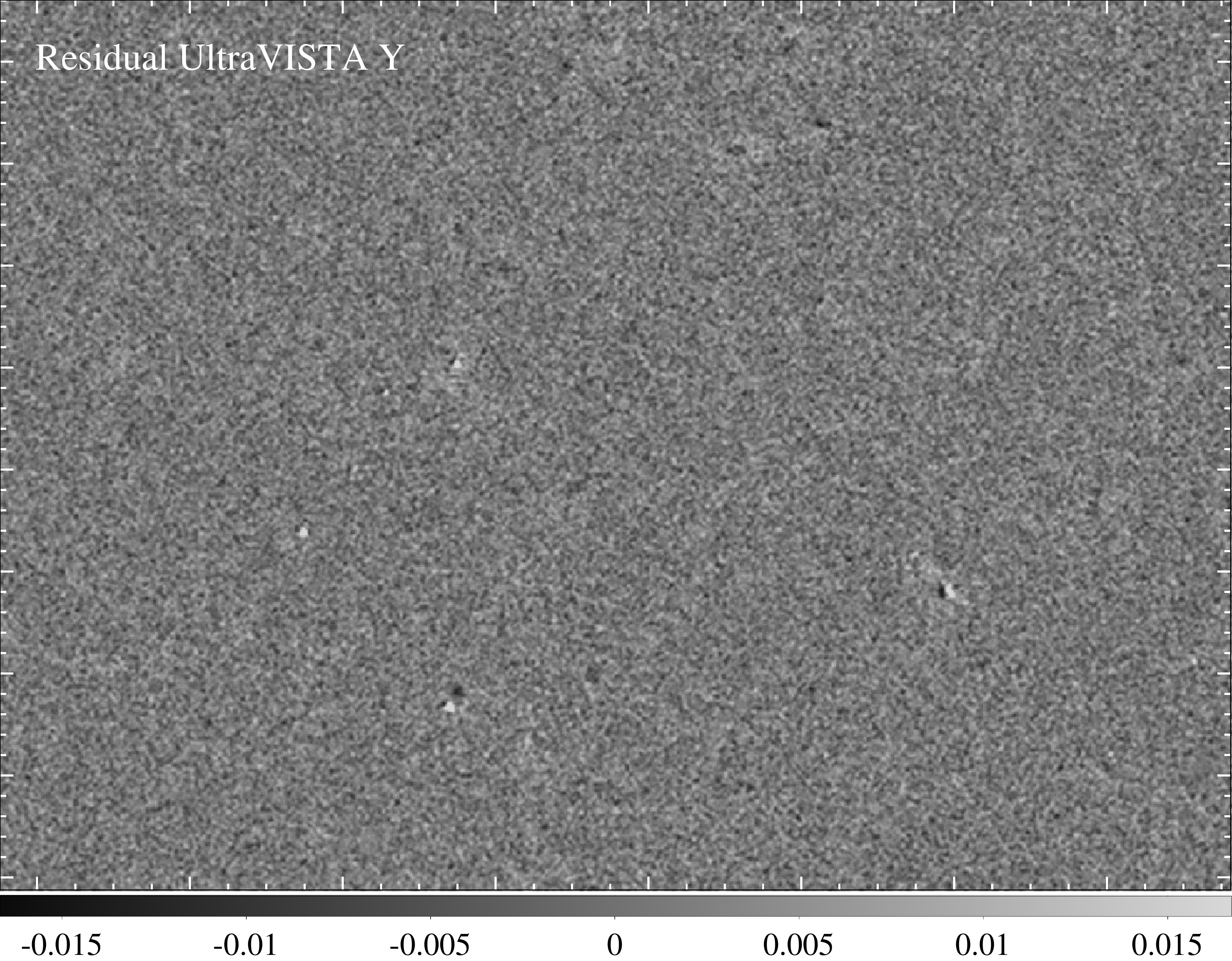}
\includegraphics[trim=0cm 0cm 0cm 0cm,scale=0.2]{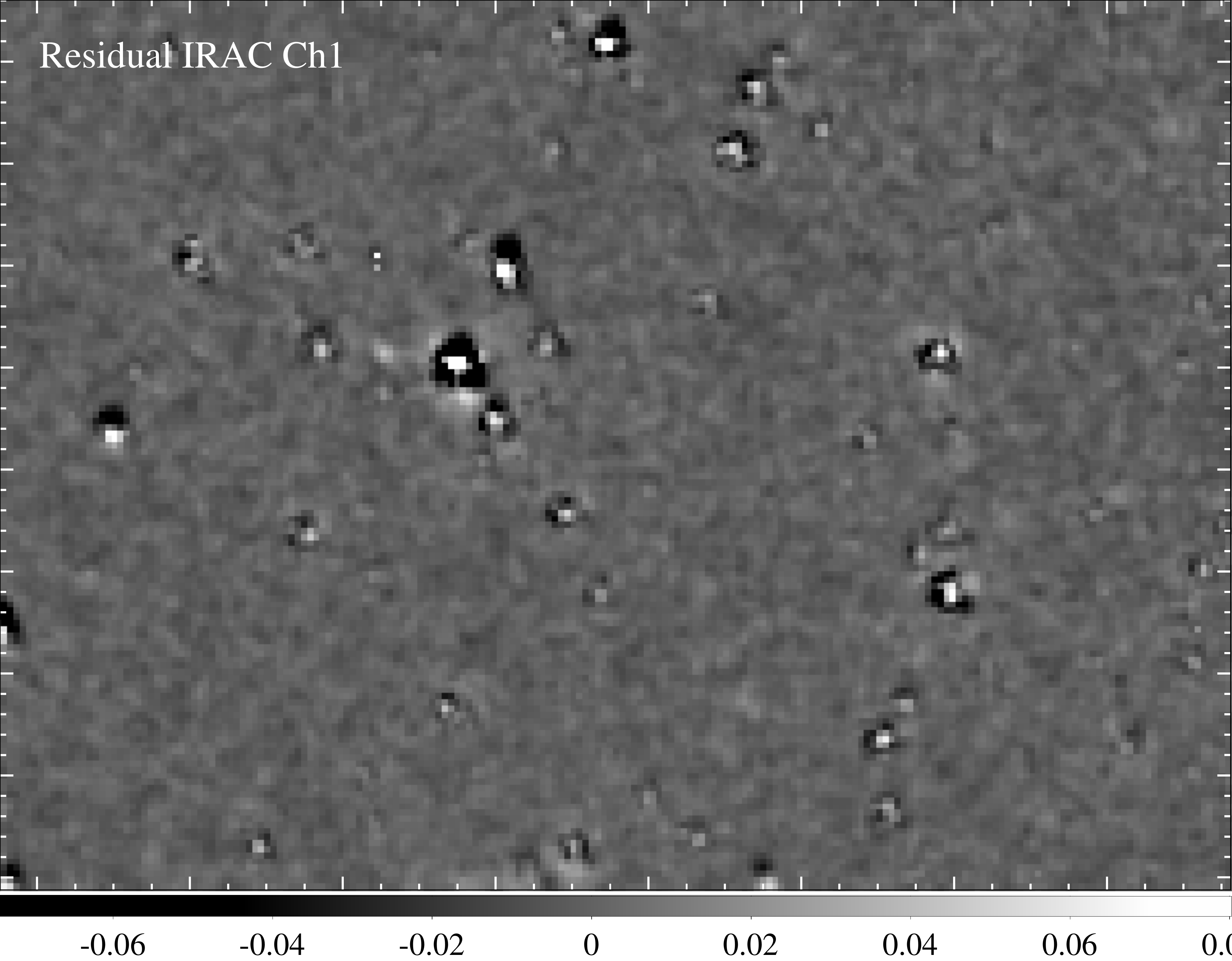}
\caption{The WFC3 F160W source selection band (top left) along with TFIT
  residual maps in the optical (top right), near infrared (bottom
  left) and IRAC infrared (bottom right). The maps all show the
    same area as the WFC3 F160W band, are in $\mu$Jy/pixel units and are all
    scaled linearly as shown by the respective color bars. The background noise for the Subaru $V$, UltraVISTA
  $Y$ and IRAC 3.6\,$\mu$m are at the levels of 0.001\,$\mu$Jy,
  0.004\,$\mu$Jy and 0.006\,$\mu$Jy respectively (images are
  background subtracted as discussed in Section 3). The arrow marked on
  the WFC3 maps shows a reference object at a flux density of 74\,$\mu$Jy.}
\end{figure*}

\begin{figure}
\centering
\includegraphics[trim=2cm 0cm 0cm 0cm, scale=0.5]{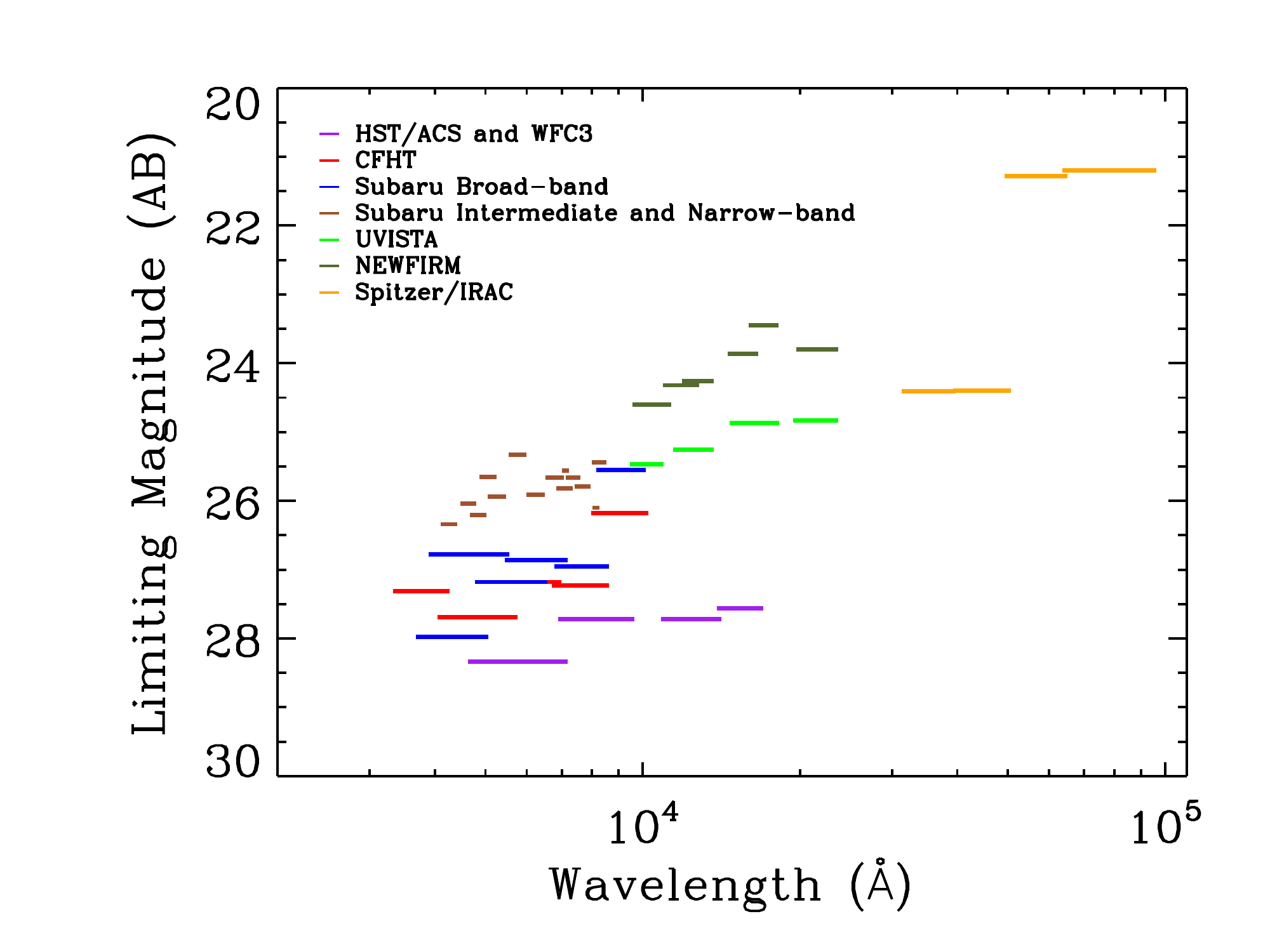}
\caption{The 5$\sigma$ limiting magnitude of the different
  observations in the CANDELS COSMOS (Tables 1 and 2) as a function of the
  wavelength. The symbol sizes are proportional to the filter response
function widths.}
\end{figure}

The COSMOS field was also observed by the
Subaru/Suprime-Cam in the $B$, $g^+$, $V$, $r^+$,
$i^+$ and $z^+$ broad-band filters. The Suprime-Cam has
a field of view of $34^{\prime} \times 27^{\prime}$ with a
pixel scale of 0.202\,arcsec\,pixel$^{-1}$. The data were processed
using the IMCAT package\footnote{\url{http://www.ifa.hawaii.edu/~kaiser/imcat/}}. The
individual frames were combined, flat-fielded and photometry and
astrometry calibrated. We refer the reader to \citet{Capak2007}
  and \citet{Taniguchi2007, Taniguchi2015} for further description of
  the observations and data processing. There is also further observations of the COSMOS field by
Subaru/Suprime-Cam in twelve intermediate bands along with narrow-band data across
two filters \citep{Capak2007, Taniguchi2007, Taniguchi2015} covering the
wavelength range of $\rm \sim 4000-8500\,\AA$. These observations were
processed similarly to the broad-band optical data
discussed above \citep{Taniguchi2015}. Although slightly
shallower than the optical broad-band observations, these filters have
higher resolving power than the former (with $R=\lambda/\Delta\lambda
\sim 23$; \citealp{Taniguchi2015}), equivalent to low
resolution spectroscopy in the optical. The resolving
power is even higher for the two narrow-band filters ($R \sim 50-100$;
\citealp{Taniguchi2015}) although with smaller wavelength coverage.
This provides a unique dataset for studying emission line galaxies and
high-redshift systems such as Lyman-$\alpha$ emitters
\citep{Shimasaku2006, Iwata2009, Koyama2014}. Table
2 summarizes the Subaru intermediate and narrow-band
observations. We used the Subaru V2 mosaics for broad-band and
  NB816 observations and V1 mosaics for intermediate and NB711 data available from the
IRSA\footnote{\url{http://irsa.ipac.caltech.edu/data/COSMOS/images/subaru/mosaics/}}
in this work.

The ground-based near infrared
observations are from the Visible and Infrared Survey Telescope for
Astronomy (VISTA; \citealp{Emerson2010}) 4.1\,m telescope VIRCAM
large-format array camera \citep{Dalton2006} in the $Y$, $J$, $H$ and $K_s$
bands with mean pixel scale of 0.34\,arcsec\,pixel$^{-1}$. The
complete contiguous $\rm \sim1.5\,deg^2$ of
UltraVISTA observations were done using a stripes pattern with
$\rm \sim0.7\,deg^2$ of the field observed with longer exposure (in
four stripes) separating the observations into deep and ultra-deep
regions \citep{McCracken2012} with the CANDELS {\it HST} observations
inside one of the ultra-deep stripes. The data were pre-processed at
CASU\footnote{\url{http://casu.ast.cam.ac.uk/surveys-projects/vista/technical/data-processing}},
which includes dark subtraction, flat fielding, gain normalization and
initial sky subtraction \citep{Irwin2004, McCracken2012}. The data were
further processed at TERAPIX using an iterative sky-background removal technique
and resampled to a pixel scale of 0.15\,arcsec\,pixel$^{-1}$. \citet{McCracken2012}
give more details on the data processing. We used the final DR1 mosaics of the
UltraVISTA data for the CANDELS multi-wavelength catalog\footnote{\url{http://irsa.ipac.caltech.edu/data/COSMOS/images/Ultra-Vista/}}.

The COSMOS field also has been observed by the NOAO Extremely
Wide-Field Infrared Imager (NEWFIRM) on the Mayall 4\,m telescope as
part of the NEWFIRM Medium Band
Survey\footnote{\url{http://www.astro.yale.edu/nmbs/Overview.html}}
(NMBS; \citealp {Whitaker2011}). The NEWFRIM observations in the
COSMOS cover an area of $27^{\prime}.6 \times 27^{\prime}.6$ encompassing the
CANDELS {\it HST} observations. The observations are over five medium-band
filters of $J_1$, $J_2$, $J_3$, $H_1$ and $H_2$ covering the
wavelength of $1-1.8\,\mu$m and the $K$ filter centered at
2.2\,$\mu$m. The three medium-band $J_1$, $J_2$ and $J_3$ filters
are a single broad-band $J$ filter split into three and the two
medium-band $H_1$ and $H_2$ filters combine into a single broad-band
$H$ \citep{vanDokkum2009, Whitaker2011}. The final mosaic has
a pixel scale of 0.3\,arcsec\,pixel$^{-1}$. \citet{Whitaker2011}
further discuss the data processing. Here we used the first data
release of the NMBS data (DR1) of the COSMOS field available from the
NOAO science archive\footnote{\url{http://r2.sdm.noao.edu/nsa/nsa_form.html}}.

\subsection{{\it Spitzer} InfraRed Observations}

The COSMOS field was observed by the {\it Spitzer Space Telescope}
IRAC instrument \citep{Fazio2004} at 3.6\,$\mu$m, 4.5\,$\mu$m,
5.8\,$\mu$m, 8.0\,$\mu$m as part of the S-COSMOS
\citep{Sanders2007}. The 3.6\,$\mu$m and 4.5\,$\mu$m bands
have much deeper data from the {\it Spitzer} Extended Deep
Survey\footnote{The S-CANDELS data \citep{Ashby2015} were not
    used here because they were not available at the time of catalog compilation.} (SEDS; \citealp{Ashby2013}). The
SEDS observations cover a strip of $10^\prime\times1^\circ$ oriented North-South coinciding
with the deep VISTA data mentioned above and
incorporate previous 3.6\,$\mu$m and 4.5\,$\mu$m data from the
S-COSMOS providing a uniform depth of 26\,mag
(3$\sigma$) for all observations \citep{Ashby2013}. The 5$\sigma$
limiting magnitude and FWHM size of the PSF in each IRAC band are
reported in Table 1. In this work we used the SEDS first data
  release (V1.2) \citep{Ashby2013} available from
  \url{https://www.cfa.harvard.edu/SEDS/data.html} for photometry
  measurements in the 3.6\,$\mu$m and 4.5\,$\mu$m bands and the S-COSMOS
  {\it Spitzer}/IRAC 5.8\,$\mu$m and 8.0\,$\mu$m GO2 (V2) data
  \citep{Sanders2007} available from \url{http://irsa.ipac.caltech.edu/data/SPITZER/S-COSMOS/}. Figure 2 shows the sky coverage of the different
ground based and space data in the COSMOS CANDELS.

\begin{table*}
\begin{center}
\caption{Summary of the CANDELS COSMOS Broad-Band Data.}
\begin{tabular}{llccccc}
\hline
\hline
Instrument & Filter$^{\star}$ & Effective & PSF & 5$\sigma$ limiting depth$^{\ddagger}$&Version & Reference \\
 & & Wavelength$^{\dagger}$ & FWHM & &  &\\
& & (\AA) & (arcsec)  & (AB Magnitude) & &\\
\hline
CFHT/MegaPrime & $u^*$ & 3817 & 0.93  & 27.31 &July 2009& \citet{Gwyn2012} \\
 & $g^*$ & 4860 & 1.08  & 27.69 & $-$ & $-$ \\
 & $r^*$ & 6220 & 0.84 & 27.18 & $-$ & $-$ \\
 & $i^*$ & 7606 & 0.85 & 27.23 & $-$& $-$ \\
 & $z^*$ & 8816 & 0.84 & 26.18 & $-$ & $-$ \\
Subaru/Suprime-Cam & $B$ & 4448 & 0.95 & 27.98 &V2& \citet{Taniguchi2007} \\ 
 & $g^+$ & 4761 & 1.58 & 26.78 & $-$ & $-$ \\
 & $V$ & 5470 & 1.33 & 26.86 & $-$ & $-$ \\
 & $r^+$ & 6276 & 1.05 & 27.18 & $-$  & $-$ \\
 & $i^+$ & 7671 & 0.95 & 26.95 & $-$ & $-$\\
 & $z^+$ & 9096 & 1.15 & 25.55 & $-$ & $-$\\
{\it HST}/ACS & F606W & 5919 & 0.10 & 28.34 &V0.5& \citet{Koekemoer2011} \\
 & F814W & 8060 & 0.10 & 27.72 & $-$  & $-$\\
{\it HST}/WFC3 & F125W & 12486 & 0.14  & 27.72 & $-$  & $-$\\
 & F160W & 15369 & 0.17 & 27.56 & $-$  & $-$\\
VISTA/VIRCAM & $Y$ & 10210 & 1.17 & 25.47 &DR1& \citet{McCracken2012}\\
 & $J$ & 12524 & 1.07 & 25.26 & $-$& $-$\\
 & $H$ & 16431 & 1.00 & 24.87 & $-$ & $-$\\
 & $K_s$ & 21521 & 0.98 & 24.83 & $-$& $-$\\
{\it Spitzer}/IRAC & 3.6\,$\mu$m & 35569 & 1.80 & 24.41  &V1.2& \citet{Ashby2013}\\
 & 4.5\,$\mu$m & 45020 & 1.86 & 24.40 & $-$ & $-$\\
 & 5.8\,$\mu$m & 57450 & 2.13 & 21.28 & V2& \citet{Sanders2007}\\
 & 8.0\,$\mu$m & 79158 & 2.29 & 21.20 & $-$ & $-$\\

\hline

\end{tabular}
{\footnotesize \newline Notes. $^{\star}$: Filters adopted from \url{http://cosmos.astro.caltech.edu/page/filterset}. $^{\dagger}$: Calculated as: $\lambda_{\rm
  eff}=\sqrt{(\int S(\lambda)\lambda d \lambda)/(\int
  S(\lambda)\lambda^{-1} d \lambda)}$ with $S(\lambda)$ the filter
  response function \citep{Tokunaga2005}.
$^{\ddagger}$: The 5$\sigma$ limiting magnitude calculated within a circular
aperture with a radius $\rm r_{ap}=FWHM$ of the PSF in each filter.}
\end{center}
\end{table*}

\begin{table*}[t]
\begin{center}
\caption{Summary of the CANDELS COSMOS Medium-Band and Narrow-Band Data.}
\begin{tabular}{llccccc}
\hline
\hline
Instrument & Filter & Effective & PSF & 5$\sigma$ limiting depth &Version & Reference \\
 & & Wavelength & FWHM &  &\\
& & (\AA) & (arcsec)  & (AB Magnitude) &\\
\hline
Subaru/Suprime-Cam & $IA484$ & 4849 & 1.14  & 26.34&V1 & \citet{Taniguchi2007,Taniguchi2015} \\
 & $IA527$ & 5261 & 1.60  & 26.04 & $-$  & $-$ \\
 & $IA624$ & 6232 & 1.05 & 26.21 & $-$ & $-$ \\
 & $IA679$ & 6780 & 1.58 & 25.65 & $-$ & $-$ \\
 & $IA738$ & 7361 & 1.08 & 25.94 & $-$ & $-$ \\
 & $IA767$ & 7684 & 1.65 & 25.33 & $-$ & $-$ \\
 & $IB427$ & 4263 & 1.64 & 25.91 & $-$ & $-$ \\
 & $IB464$ & 4635 & 1.89 & 25.66 & $-$ & $-$ \\
 & $IB505$ & 5062 & 1.44 & 25.82 & $-$ & $-$\\
 & $IB574$ & 5764 & 1.71 & 25.66 & $-$  & $-$\\
 & $IB709$ & 7073 & 1.58 & 25.79 & $-$  & $-$\\
 & $IB827$ & 8244 & 1.74 & 25.44 & $-$  & $-$\\
 & $NB711$ & 7120 & 0.79 & 25.56 & $-$ & $-$\\
 & $NB816$ & 8149 & 1.00 & 26.10 & V2  & $-$\\
Mayall/NEWFIRM & $J_1$ & 10460 & 1.19 & 24.60 & DR1 & \citet{Whitaker2011}\\
 & $J_2$ & 11946 & 1.17 & 24.32 & $-$ & $-$\\
 & $J_3$ & 12778 & 1.12 & 24.26 & $-$ & $-$\\
 & $H_1$ & 15601 & 1.03 & 23.86 & $-$ & $-$\\
 & $H_2$ & 17064 & 1.24 & 23.45 & $-$ & $-$\\
 & $K$ & 21700 & 1.08 & 23.80 & $-$ & $-$\\

\hline
\end{tabular}
\end{center}
\end{table*}

\section{Catalog Photometry}

In generating the multi-wavelength catalog, the high resolution data
({\it HST}/ACS and WFC3) were treated differently from the low resolution data
(ground-based and {\it Spitzer} IRAC).

\subsection{{\it HST} Photometry}

We performed photometry on the high resolution (ACS + WFC3) data using
\texttt{SExtractor} software version 2.8.6 \citep {Bertin1996} in dual
mode with the WFC3 F160W as the detection band consistently with the multi-wavelength
catalogs in the other four CANDELS fields \citep{Galametz2013,
  Guo2013}. \texttt{SExtractor} software was
modified in several ways to enhance the sky measurement, add a new
cleaning procedure and fix isophotal-corrected magnitude calculations as discussed by \citet
{Galametz2013}. In order to measure the photometry in the visible
bands we PSF-matched the ACS and WFC3 images and extracted the
photometry from the matched images in dual mode.

As shown in the
photometric studies of the CANDELS UDS and CANDELS GOODS-S fields
\citep {Galametz2013, Guo2013}, it is impossible to
identify all galaxies using a single set of parameters (signifying
the area, signal to noise, background, etc) for the extraction. The
challenge is to detect the brightest targets while avoiding blending
and also detect the faintest objects without
introducing spurious sources into the catalog. To this end we
used two sets of \texttt{SExtractor} input parameters. One set of
parameters is aimed at bright source detection with a focus
on deblending extended sources (cold mode), and a second set on faint
galaxies (hot mode). The two catalogs generated by the hot and
cold parameters were then combined following a routine adopted from
GALAPAGOS\footnote{\url{http://astro-staff.uibk.ac.at/~m.barden/galapagos/}}
\citep{Barden2012}. The combined catalog includes all the sources from the cold
mode catalog plus sources in the hot mode catalog that do not exist in
the cold mode as identified by the Kron ellipse of a cold mode
detected source as discussed by \citet {Galametz2013}. Figure 3 shows the 1$\sigma$ detection limit
of the combined catalog computed over a circular aperture with a
radius of 1\,arcsec along with the cumulative distribution of the
detection area and the exposure time distribution in the F160W band. Figure 4 shows the magnitude distribution of
the sources in the hot, cold and combined catalogs along with the
comparison of the F160W combined counts with the CANDELS UDS
\citep{Galametz2013} and 3D-HST \citep{Skelton2014}. Table 3 gives the
number counts in magnitude bins for the combined catalog along with the
associated Poissonian uncertainties.

At this stage we also assigned a photometry flag to every object in
the catalog. The flagging system
is the same as that adopted by the other CANDELS fields
\citep{Galametz2013, Guo2013} and discussed in detail by
\citet{Galametz2013}. We use a zero for a good photometry in the
flagging and assigned a value of one for bright stars and spikes
associated with those stars as photometry for objects contaminated by
this would be unreliable. A photometric flag of two is associated with
the edges of the image as measured from the F160W RMS maps.

\begin{table}
\begin{center}
\caption{{\it HST}/WFC3 F160W number counts from the combined hot+cold
  catalog.}
\begin{tabular}{*{3}{c}}
\hline
\hline
WFC3 F160W$^{\dagger}$ & N ($\text{deg}^{-2}\text{mag}^{-1}$) & Poisson Uncertainty \\
\hline
15.25 &       70 &       50 \\
      15.75 &       387 &       117 \\
      16.25 &       563 &       141 \\
      16.75 &       633 &       149 \\
      17.25 &       1196 &       205 \\
      17.75 &       1653 &       241 \\
      18.25 &       2638 &       305 \\
      18.75 &       3166 &       334 \\
      19.25 &       4960 &       418 \\
      19.75 &       6965 &       495 \\
      20.25 &       9990 &       593 \\
      20.75 &       13789 &       696 \\
      21.25 &       17377 &       782 \\
      21.75 &       27086 &       976 \\
      22.25 &       33735 &       1089 \\
      22.75 &       44956 &       1257 \\
      23.25 &       62720 &       1485 \\
      23.75 &       82841 &       1707 \\
      24.25 &       106200 &       1933 \\
      24.75 &       133426 &       2166 \\
      25.25 &       156045 &       2343 \\
      25.75 &       178769 &       2508 \\
      26.25 &       201810 &       2664 \\

\hline
\end{tabular}
\end{center}
\footnotesize $^{\dagger}$: Bin center magnitude.
\end{table}

\subsection{Ground-based and {\it Spitzer} Photometry}

In order to measure the photometry in the ground-based
and {\it Spitzer} bands, we used the Template FITting method (TFIT; \citealp {Laidler2007})
similarly to the other CANDELS multi-wavelength catalogs (\citealp
{Galametz2013, Guo2013}, Stefanon et al. 2016). TFIT is a robust
algorithm for measuring photometry in mixed
resolution data sets. Sources that are well separated in the
high resolution image ({\it HST}) could be blended in the low resolution
image (ground-based or {\it Spitzer}). TFIT uses position
and light profiles from the high resolution image to calculate
templates that are used to measure the photometry in the low
resolution image. It does that by smoothing the high resolution image
to match the PSF of the low resolution image using a convolution kernel
\citep {Galametz2013, Guo2013}. Fluxes in the low resolution image are then
measured using these templates while fitting the sources
simultaneously.

TFIT requires some pre-processing of low resolution
image in terms of orientation and pixel scale. The
individual steps taken are:

{\bf Background Subtraction}: The low resolution images must be
background-subtracted before running TFIT. We used a background
subtraction routine with several iterations that included a first
estimate through smoothing the image on large scales followed by
PSF smoothing and source masking that led to a noise map that was
interpolated to determine the background \citep{Galametz2013}.

{\bf Image Scale}: TFIT requires the low resolution image pixel
scale to be an integer multiple of the high resolution detection image
(0.06\,arcsec\,pixel$^{-1}$ for the F160W) and that both images have the
same orientation. Furthermore the
astrometry of the low resolution images must be consistent with the
high resolution observations. We used \texttt{SWARP}
to resample low resolution data sets to the next larger pixel
scale that is an integer multiple of the WFC3 mosaic and also used it
for astrometry and image alignment. 

{\bf Point Spread Function and Kernel}: The point spread function of
both the high resolution and low resolution images are needed for the
TFIT pre-processing. We constructed the PSF by stacking isolated and
unsaturated stars in each band using custom IDL routines. We also
constructed a kernel to convolve
the high resolution templates to the low resolution ones. The
  kernel was constructed using a Fourier
space analysis technique similar to
\citet{Galametz2013}, which takes the ratio of the Fourier transform
of each PSF. This gives the Fourier transform of
the kernel which is then transformed back into normal space generating
the kernel. As discussed by \citet{Galametz2013} a low passband filter
is applied in the Fourier space to cancel the high frequency
fluctuations and remove the effect of noise. For {\it Spitzer}/IRAC
, which has the largest difference in resolution from {\it HST}, one
could use the PSFs directly as the convolution kernel
\citep{Galametz2013}. We generated a model PSF by averaging a set of
oversampled PSFs that measure PSF variations across the detector. The
final PSF is a boxcar kernel smoothed and flux normalized model which
also incorporates all the PAs associated with different Astronomical
Observation Requests (AORs). We refer the reader to
\citet{Galametz2013} and \citet{Guo2013} for more detail.

{\bf Dilation Correction for High Resolution Images}: TFIT uses the
area of the galaxy identified from the high resolution {\it HST} bands to
measure the photometry. These are pixels defined by the high resolution
segmentation maps and are fed into TFIT in the form of the
isophotal areas of the high resolution image. However, as demonstrated
by \citet{Galametz2013}, \texttt{SExtractor} usually underestimates the isophotal area of faint or
small galaxies, and this leads to an underestimate of the flux for
such systems. \citet{Galametz2013} performed extensive tests to quantify and
correct for this effect, the so-called ``dilation correction'', by
refining and applying the public {\sc dilate} code. These simulations showed that the correction factor is
negligible for objects with large isophotal areas and that it is
largest for objects with area $<$ 60\,pixels. The original isophotal area size
from \texttt{SExtractor} hence defines the dilation factor applied. We used
the same criteria to correct our high resolution \texttt{SExtractor}
segmentation maps as outlined by \citet{Galametz2013} and refer the
reader to this work for further details. Even
  after this correction, the total flux measurement is always
  bound by uncertainties incorporated into the above assumptions and
  this constitutes one of the limitations of photometry estimation.

We ran TFIT in two stages. During the first step TFIT measured any remaining
mis-alignment in the form of distortion or mis-registration between the
high resolution and low resolution images in the form of shifted
kernels \citep{Laidler2007, Galametz2013, Guo2013}. In the second
step, TFIT used the kernels measured from the first step to correct
the misalignment and construct a difference residual map of the low
resolution bands. Figure 5
shows the TFIT residual maps in the low resolution
visible, near-infrared and infrared bands along with the high
resolution F160W detection band. The residual maps in the visible and
near-infrared  are close to zero and only show residuals in the center
of very bright objects. However, as argued by
\citet{Guo2013}, the residual maps are qualitative representations of
the TFIT photometry measurements, and we later
show in our data quality checks that the photometry is properly
measured for these bright objects. Tables 1 and 2 summarize the PSF
FWHMs used for the high and low resolution images. 

The final F160W \texttt{SExtractor} catalog has 38671
sources over the 216\,arcmin$^2$ area of the CANDELS COSMOS
field with WFC3 F160W observations. TFIT keeps the original F160W \texttt{SExtractor} ID and
coordinates for each object in the combined hot+cold catalog and
therefore we only need to combine corresponding entries from the
high resolution and low resolution catalogs. Figure 6 shows the 5$\sigma$
limiting depth for all the filters in the catalog as computed and
tabulated in Tables 1 and 2. We further derive and report a weight
for each target in the catalog calculated from the F160W RMS maps at
the \texttt{SExtractor} positions for each object as described in \citet{Guo2013}. 

\section{Data Quality Checks}

We checked the TFIT measured photometry by comparing it with
other independently measured photometry in the field. Additionally we
checked the colors of point sources in the catalog against model
predictions and color-color plots.

\subsection{Stars Color Checks}

The color of stars changes as a function of their
spectral type which in turn depends on the mass (and hence temperature) among other
parameters \citep{Kurucz1979, Vandenberg1985, Baraffe1998}. Using
this, we could compare the measured
color of the point-like objects in our catalog against predictions of
the colors of stars derived from stellar physics. The predicted
colors of stars were computed from the 
stellar library of BaSeL \citep{Lejeune1997, Westera2002} to measure the model
stellar colors. We present the comparisons on color-color plots using
several observed filters. To measure the predicted photometry
from the templates, we integrated the model stellar SED over the wavelength range of each
filter taking into account the filter response functions. This
provides the predicted colors of point sources. Figure 7
shows our TFIT measured colors for the point like objects compared to the
colors from the BaSeL stellar library. The color trend
of our point-like objects agrees with the general distribution of colors
predicted by the stellar models. This further confirms our measured
photometry, specifically for the brighter sources, and shows no
systematic bias in the photometry. The
scatter at the redder
color is mostly associated with the fainter sources in the catalog and
also due to the intrinsic scatter of colors inherent to the library
because of the degeneracies among the different populations of stars.

\begin{figure*}
\centering
\leavevmode
\includegraphics[scale=0.35]{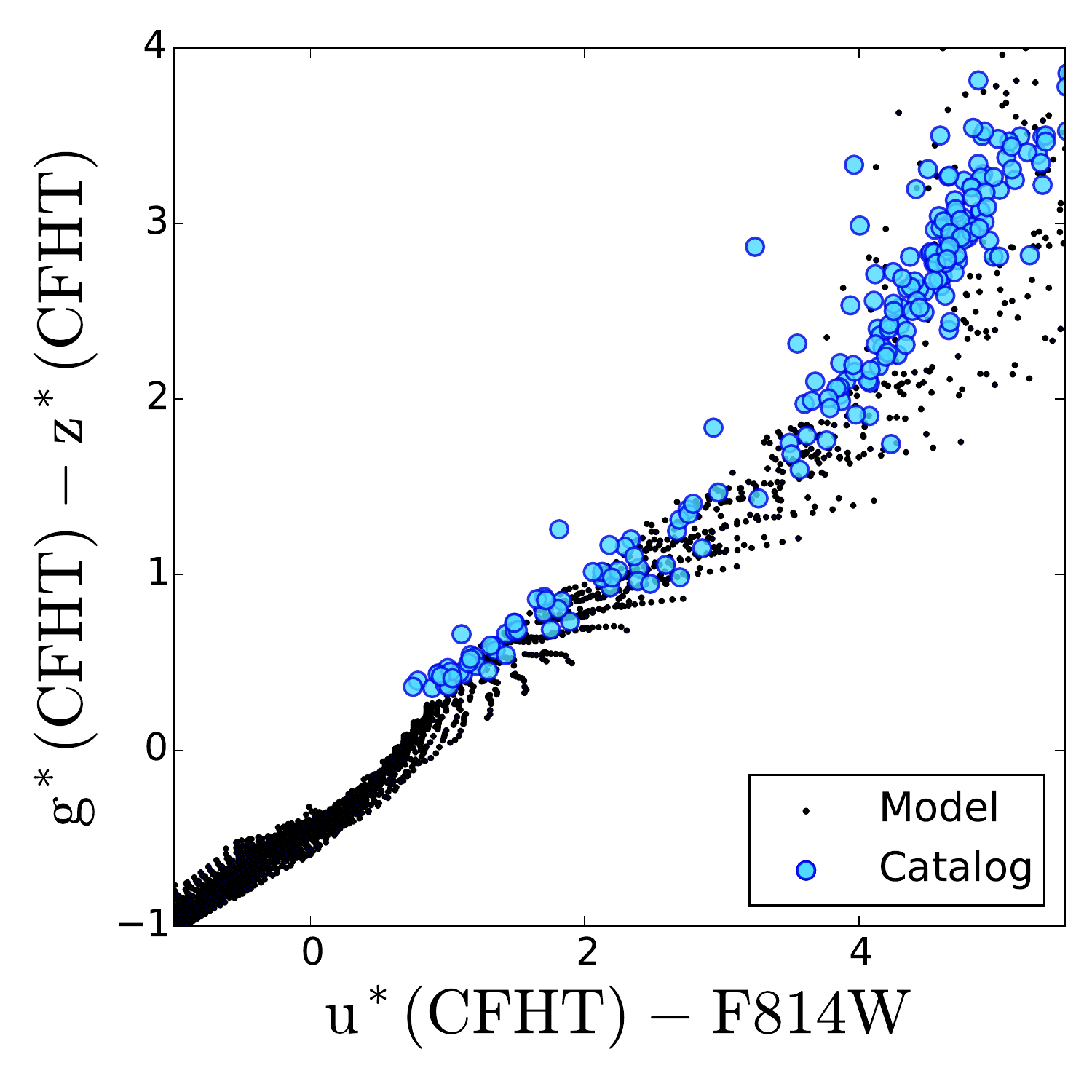}
\includegraphics[scale=0.35]{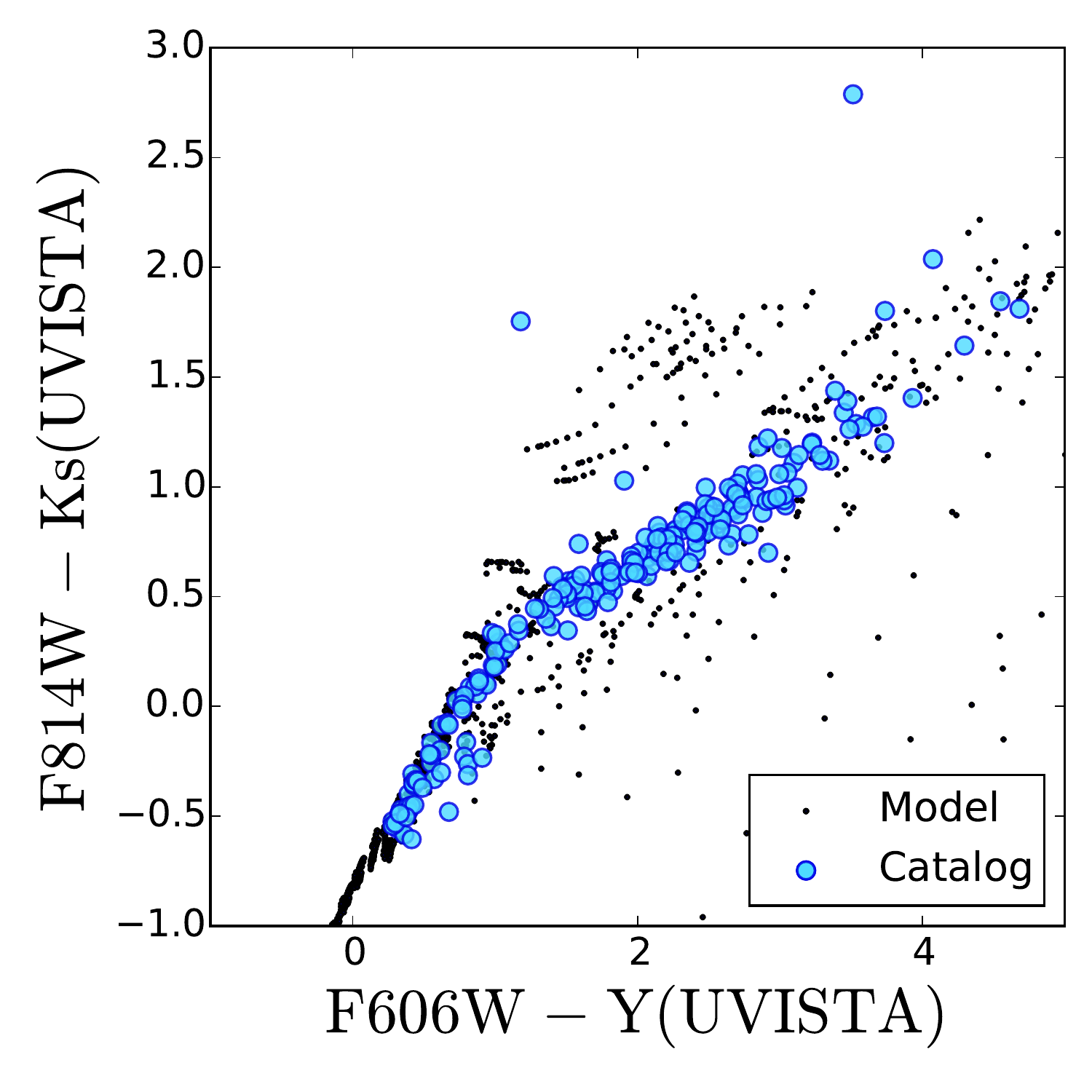}
\includegraphics[scale=0.35]{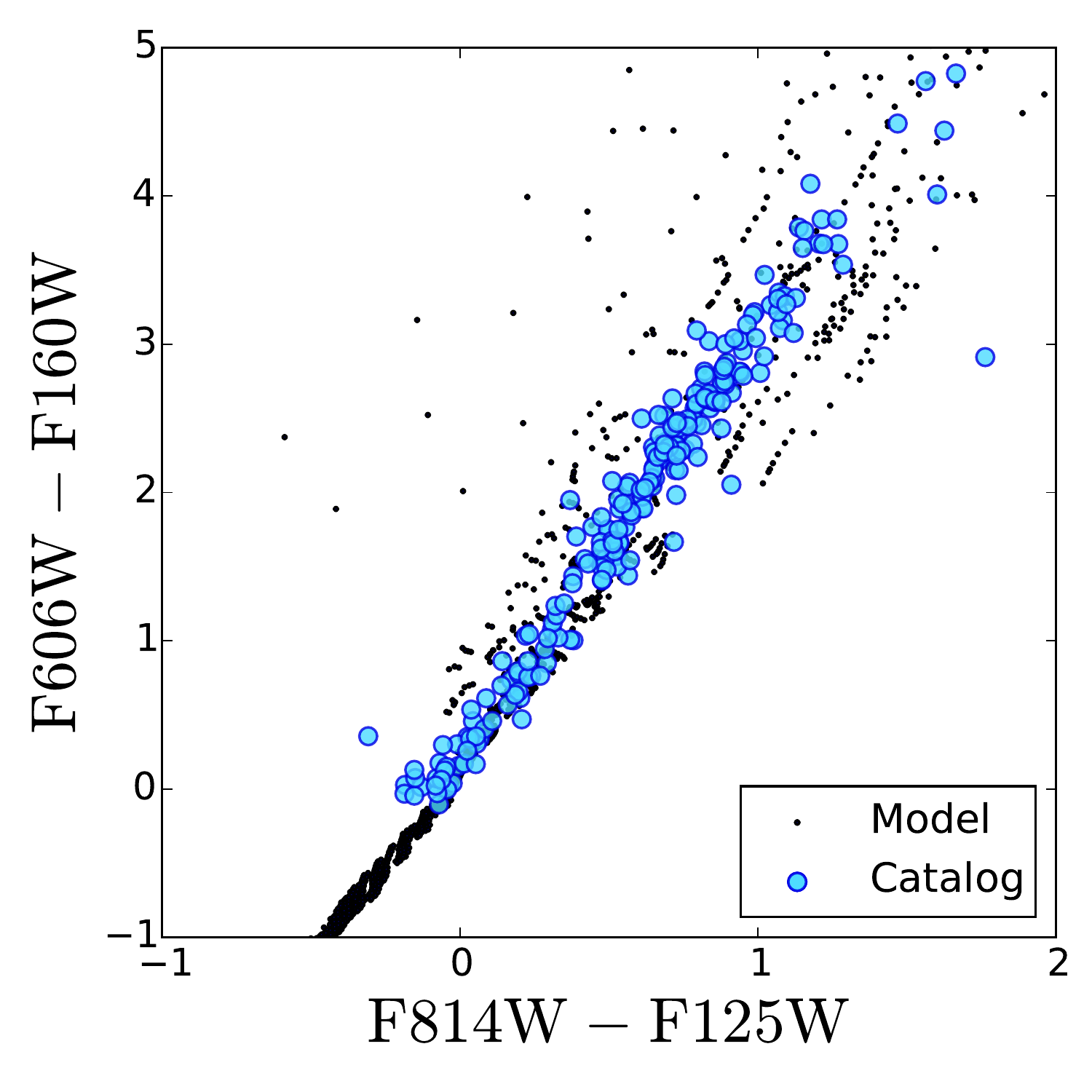} \\
\includegraphics[scale=0.35]{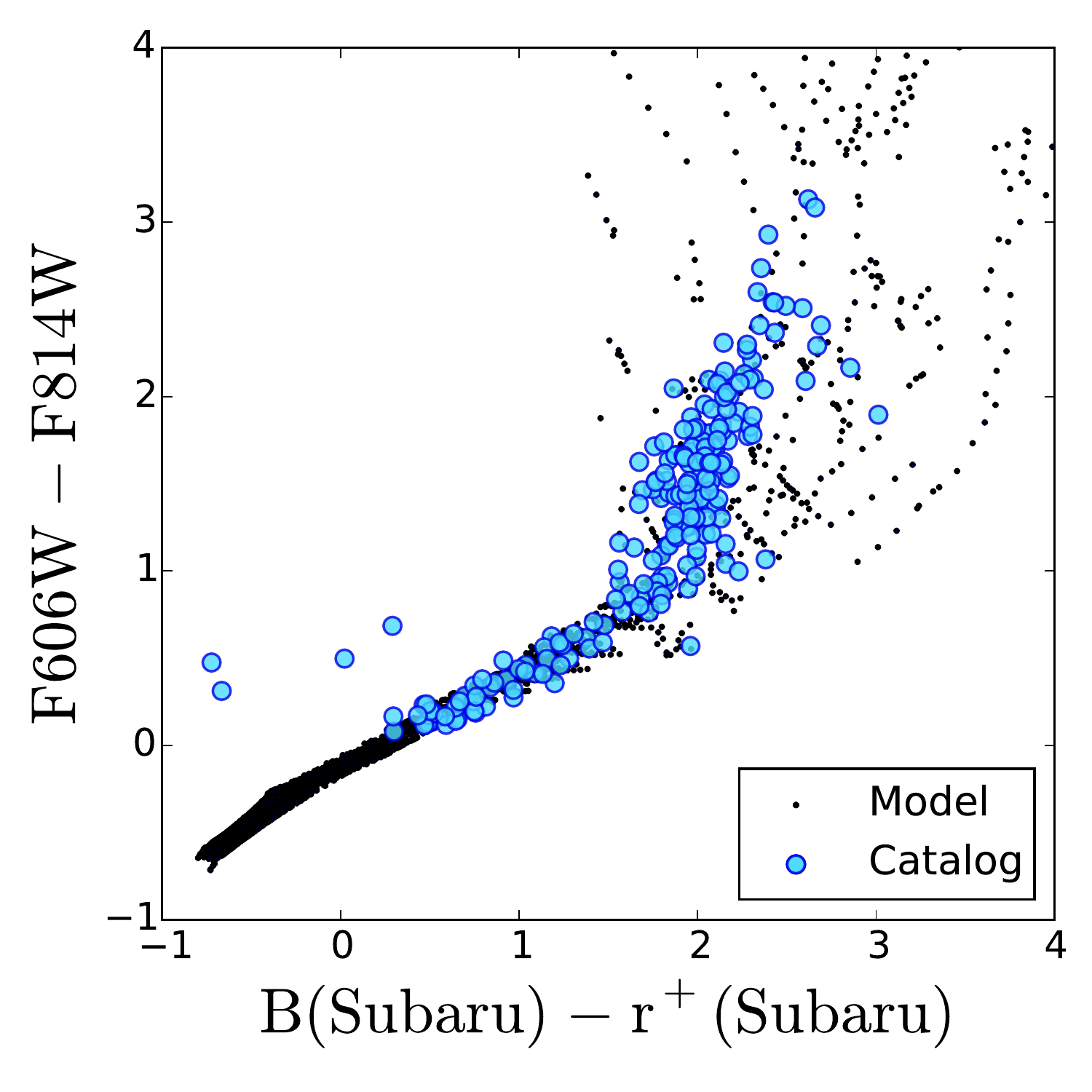}
\includegraphics[scale=0.35]{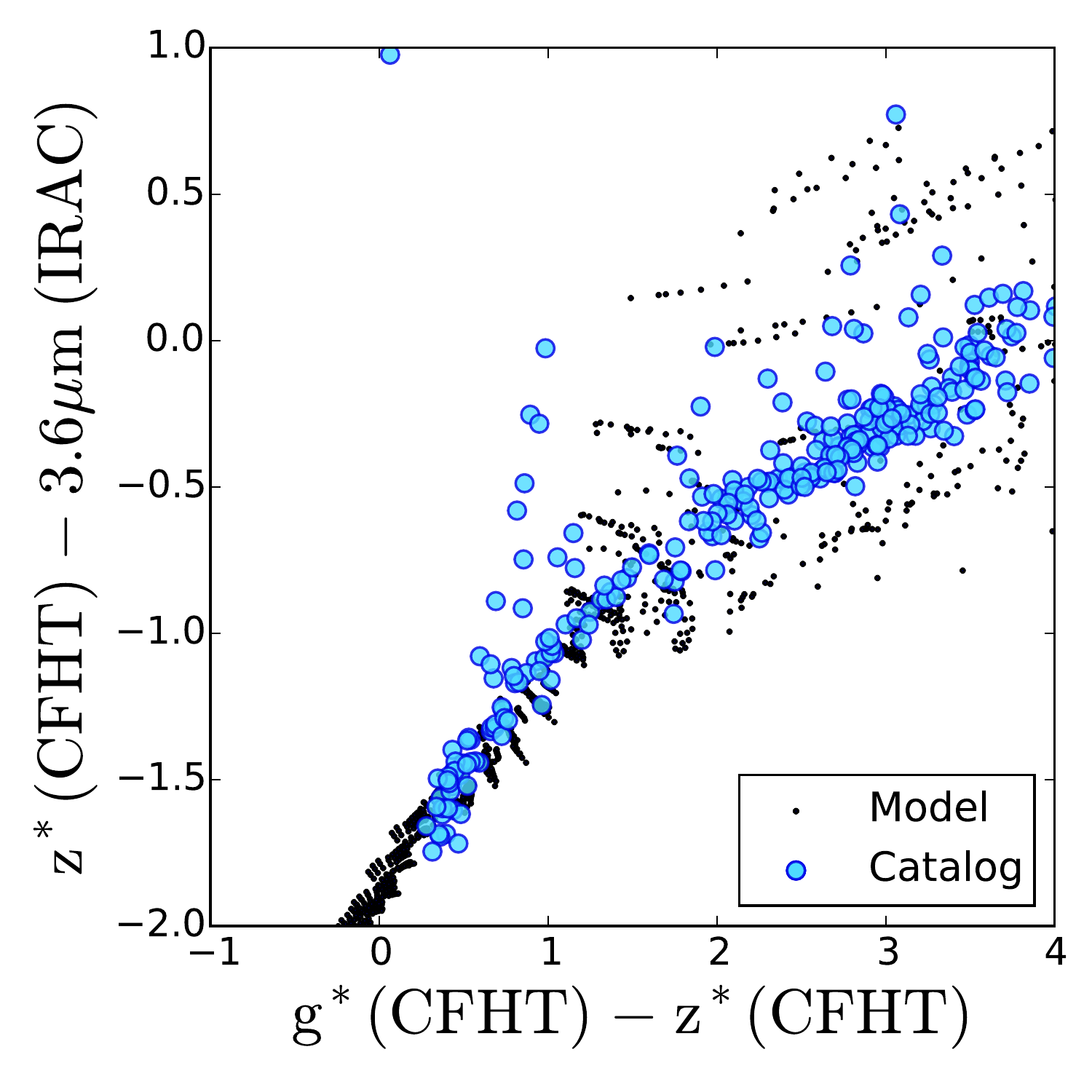}
\includegraphics[scale=0.35]{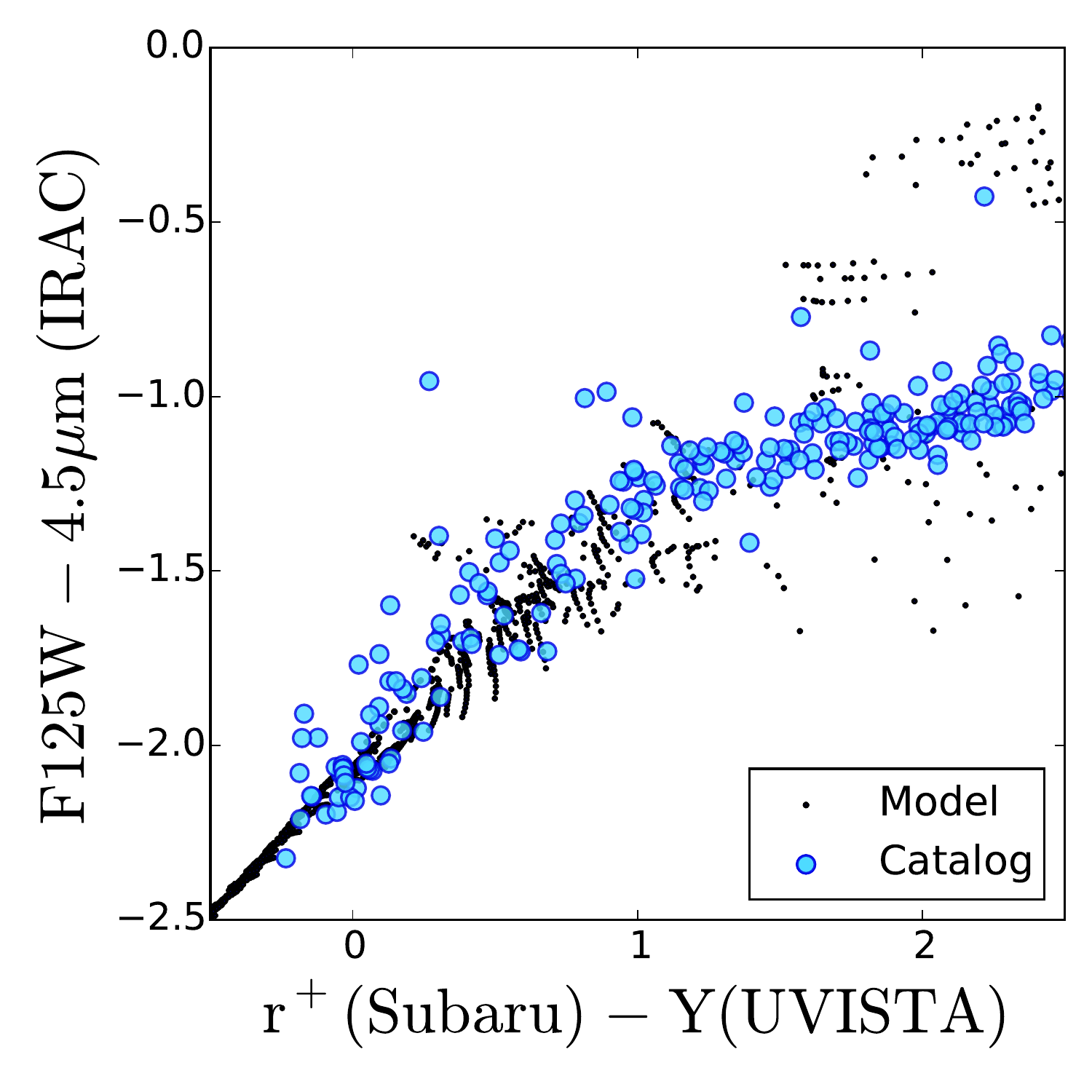}
\caption{Color-color diagrams showing the TFIT color of
  stars (determined from \texttt{SExtractor} as objects with
  $\rm CLASS\_STAR>0.9$ and $\rm H_{160}<22$ mag) in CANDELS COSMOS
  (blue) and model stars from the BaSeL
  library (black) \citep{Lejeune1997, Westera2002}. The model colors
  of stars were computed in each filter by integrating the model SED of
stars from the library over the filter transmission
curves while the observed colors of the stars are directly from
  the TFIT catalog with no SED inferred zero-point corrections (Table
  5) applied. The filters used are from
  \url{http://cosmos.astro.caltech.edu/page/filterset}.}
\end{figure*}

\begin{figure}
\centering
\includegraphics[trim=1.5cm 6cm 0cm 6cm, scale=0.45]{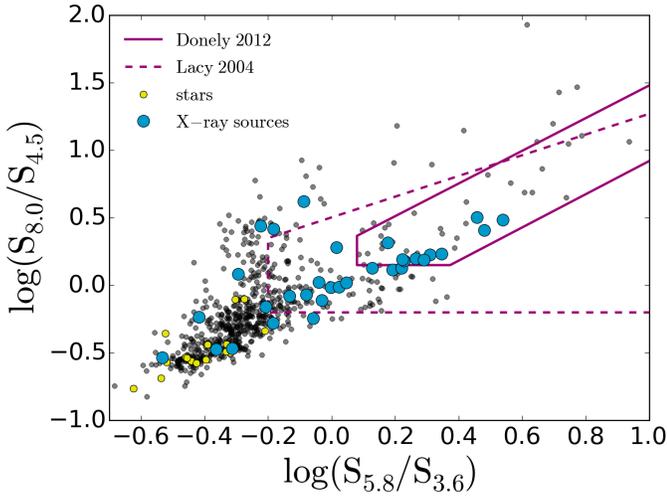}
\caption{IRAC color-color diagram and the AGN selection criteria from
  \citet{Lacy2004} (dashed magenta line) and \citet{Donley2012}
  (solid magenta line). Black circles are galaxies from the catalog
  with $\rm S/N>5$ in all four IRAC bands. Sources with $\rm
  CLASS\_STAR>0.95$ and $\rm H_{160}<21$ are shown by yellow
  circles. X-ray point-like detected sources from XMM-Newton
  wide-field observations of COSMOS \citep{Brusa2010} are shown by large blue circles.}
\end{figure}

\subsection{Infrared Color Validation Check}

A validation check of the infrared colors of objects in the
catalog was done by using the {\it Spitzer}/IRAC TFIT measured photometry
of galaxies to identify luminous Active Galactic Nuclei
(AGNs). Mid-infrared observations of galaxies have been used
extensively to identify and study AGN host
galaxies (e.g. \citealp{Laurent2000, Farrah2007, Petric2011, Yan2013,
  Lacy2015}). Several recent studies have used wide and deep
observations in the mid-infrared by {\it Spitzer}/IRAC to successfully identify large
samples of AGNs using flux ratios \citep{Lacy2004,
  Lacy2007, Stern2005, Stern2012, Donley2012, Messias2012}. These
color selections are based on the power-law behavior of the
mid-infrared continuum of luminous AGNs caused by heated dust
producing a thermal continuum \citep{Neugebauer1979, Ivezic2002,
  Donley2012, Messias2012}.

Figure 8 shows the TFIT measured IRAC color distributions of galaxies
in our catalog that have a $\rm S/N>5$ in all four
channels. According to these selections the red IRAC colors are
expected to be dominated by emission from AGN-heated dust, as
also predicted by previous studies of SDSS and radio-selected quasars
\citep{Lacy2004}, whereas any stellar components usually shifts the
$S_{5.8}/S_{3.6}$ ratio to bluer colors. The 
\citet{Donley2012} criteria are more conservative in selecting AGNs by
removing star-forming and
quiescent galaxies identified through other selection
methods from optical and near-infrared observations (such as the $BzK$
and LBG selections; \citealp{Madau1996, Giavalisco2002}). Figure 8
also shows the IRAC color distribution of the X-ray
detected sources in COSMOS \citep{Cappelluti2007} along with
the IRAC color distribution of point sources. The majority
of the X-ray detected AGNs have IRAC color distributions consistent
with the selections by \cite{Lacy2004} and \citet{Donley2012}. As
  discussed by \citet{Donley2012}, not all the X-ray luminous and IRAC
detected sources are identified by AGN color criteria. In fact
\citet{Donley2012} argue that the X-ray detected QSOs that fall
outside the color criteria seem to be more heavily obscured with lower
luminosity AGNs such that the host galaxy contributes more to the
optical-near IR flux. The
stellar sources and the general population however follow different
IRAC color distributions as demonstrated by \cite{Lacy2004}
and \cite{Donley2012}. We also present a check of stellar colors using
IRAC 3.6\,$\mu$m and 4.5\,$\mu$m bands in Section 6.2 and Figure 22,
where our photometry is found to be consistent with synthetic colors.

\subsection {Validation Checks with Public Photometry}

We also checked the TFIT photometry by comparing it with
the public photometry available from the 3D-HST survey
\footnote{\url{http://3dhst.research.yale.edu/Home.html}}
\citep{Skelton2014}. The 3D-HST photometry was measured by \texttt{SExtractor} on
PSF-matched combined {\it HST}/WFC3 images in three bands (F125W,
F140W and F160W) as the detection \citep{Skelton2014}. The {\it HST}
images were reduced similarly to CANDELS using the same pixel scale
and tangent point \citep{Skelton2014}. Photometry on
the low resolution bands was performed using the MOPHONGO code
\citep{Labbe2005, Labbe2006, Wuyts2007, Labbe2013} which takes into
account the variations in the PSF size across different filters and in
particular the source confusion problem in the low resolution images
by using a combined PSF-matched WFC3 images as a high resolution image
prior for photometry estimation \citep{Labbe2005, Skelton2014}. The
3D-HST adjusted the AUTO fluxes by an aperture
correction derived from growth curves and furthermore performed
Galactic extinction corrections. The Galactic extinction correction was measured at the
center of each filter and was based on \citet{Finkbeiner1999}. These
corrections are relatively small ($\lsim 0.07$; Table 5 in
\citealp{Skelton2014}). We took this into account when comparing our fluxes with that of the 3D-HST. All
fluxes in the 3D-HST catalogs were converted to AB magnitudes using a zero point of 25
\citep{Skelton2014}. 

Figure 9 shows the comparison between TFIT measured
photometry and the public photometry from the 3D-HST. For the source matching we used
a radius equal to the FWHM size of the PSF in the F160W ($\rm \sim
0.17 arcsec$). We find good agreement between the measured fluxes in our catalog and that of the
3D-HST. The offset is generally $\lsim 0.1$\,mag. There is
  however a magnitude dependent trend when comparing the CANDELS and
  3D-HST photometry. This is related to the difference in the
  photometry extraction between the CANDELS (TFIT) and the 3D-HST
  (aperture photometry with fixed apertures for each band; Tables 4-8
  of \citealp{Skelton2014}). Figure 10 shows the
comparison between CANDELS COSMOS TFIT measured $\rm (Band-F160W)$
color and the corresponding colors of sources
measured from the 3D-HST catalog \citep{Skelton2014} as a function of
the F160W magnitude. For
both catalogs the F160W band is taken as the reference band for
measuring the color. While comparing the $\rm {\it Spitzer}-F160W$ color for
the different fields as a function of photometric redshift, we noticed a deviation of
$\sim 0.6$ mag between the colors in the COSMOS and GOODS-S fields for
the red objects in the {\it Spitzer} 8.0\,$\mu$m band. There is an offset (though smaller) in similar color between the
3D-HST COSMOS and GOODS-S fields. Looking at the variation of this
color difference between the CANDELS COSMOS and 3D-HST as a function of the
F160W, we notice that most of the difference is associated with
objected fainter than $\sim 22$ mag which is similar to our 5$\sigma$
detection limit. 

\begin{figure*}
\centering
\includegraphics[scale=0.5]{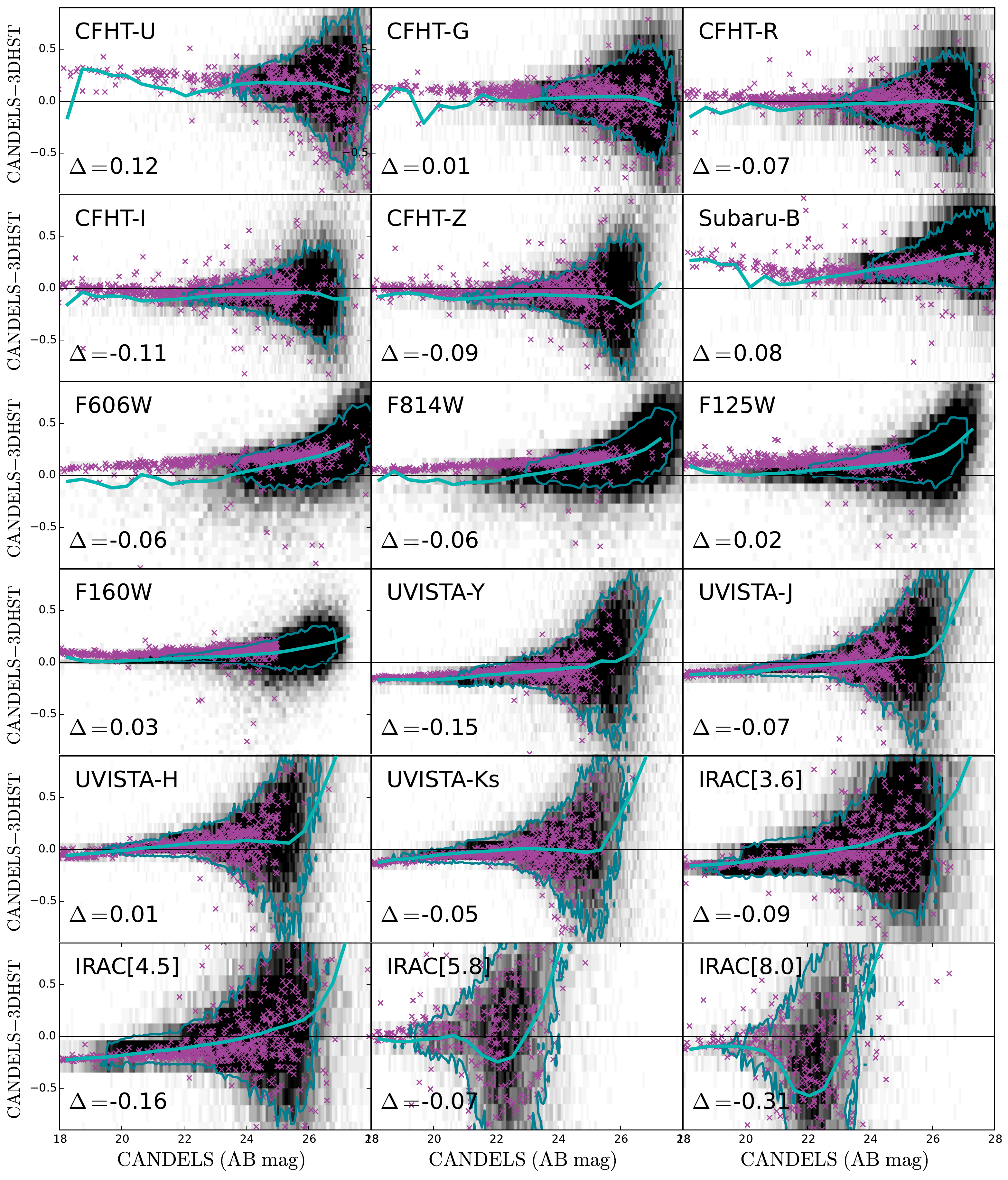}
\caption{Photometry comparison between CANDELS and 3D-HST \citep{Skelton2014}. The grey
  scale density map shows all sources and the
magenta shows point sources (identified with \texttt{SExtractor} $\rm
CLASS\_STAR>0.95$ and $\rm H_{160}<25$ mag). The thick and thin
cyan lines show the median of the distribution and the corresponding 1$\sigma$
confidence intervals. The number reported in each panel represents the
median of the offset for the bright-end of the distribution between the
CANDELS and the 3D-HST photometry (arbitrarily chosen to be $\rm H_{160}<22$ mag).}
\end{figure*}

\begin{figure*}
\centering
\includegraphics[scale=0.5]{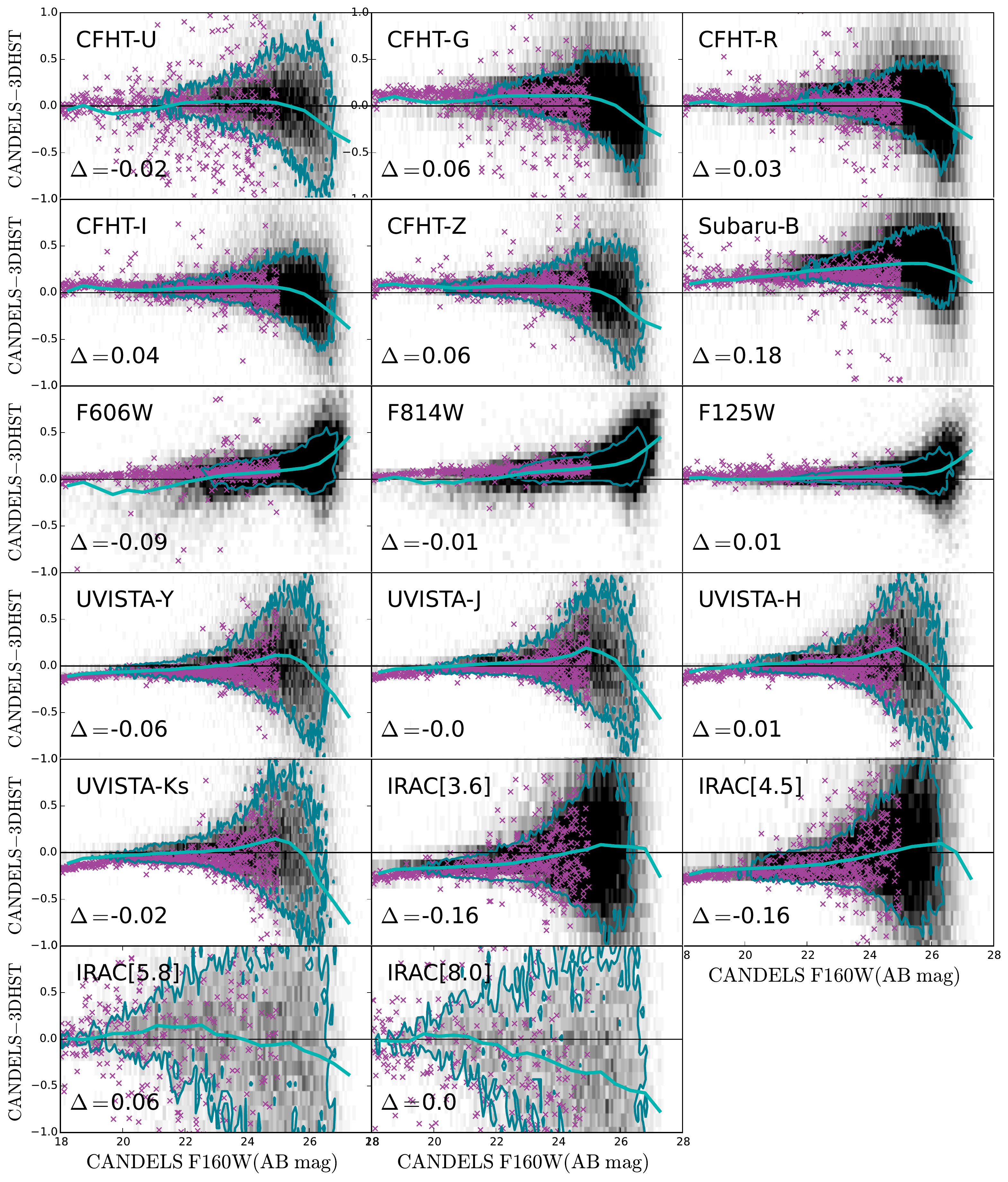}
\caption {Color comparison of the CANDELS COSMOS and
  the 3D-HST \citep{Skelton2014} as a function of the F160W magnitude. Each
  panel shows the difference of the $\rm (Band-F160W)$ color
  between CANDELS and 3D-HST (with F160W being the reference band for
  calculating the colors in both). The cyan lines show the median and
  1$\sigma$ variations in the color difference. Point sources in the plot are
  shown as magenta data points and the median of the offset is
  reported at the bottom left of each panel.}
\end{figure*}

\section {Photometric Redshift and Stellar Mass Estimates}

The methods we used to measure the photometric
redshifts and stellar masses are presented in \citet {Dahlen2013}
and \citet {Mobasher2015} respectively. Measurements of these parameters 
are carried out for other CANDELS fields, including CANDELS UDS,
CANDELS GOODS-S \citep {Santini2015}, CANDELS EGS (Stefanon et al.
2016) and CANDELS GOODS-N (Barro et al. 2016, in prep.).

\subsection {Photometric Redshifts}

We provided the CANDELS COSMOS photometric catalog to individual
teams in the collaboration. The teams were asked to estimate
photometric redshifts to galaxies using a calibrating sample containing
spectroscopic redshifts. The spectroscopic redshifts were taken from the zCOSMOS
compilation \citep{Lilly2007}. Only redshifts for sources in the CANDELS
COSMOS area with clear emission line features were used with uncertain
redshifts left out. The spectroscopic redshifts used here are all in
public domain. However, for training purposes, a larger sample of
unpublished spectroscopic redshifts were used from zCOSMOS (Mara
Salvato - private communication). Measurements from
different teams are in good agreement and also agreed with an independent
sample of 448 high quality spectroscopic
redshifts not used to calibrate the photometric redshift
methods. A total of six individuals participated. Methods included various fitting codes using
minimum $\chi^2$ and MCMC along with varying Star Formation Histories
(SFH). The details of these methods are outlined in Table 7 in
  the Appendix. Using simulations, we showed
that one could determine redshifts to an accuracy of 0.025 in
$\Delta z = (z_{\rm phot} - z_{\rm spec})/(1 + z_{\rm spec})$. For
each galaxy, we derived the median redshift from different methods and
consider that as the redshift estimate for that galaxy. 
The confidence intervals were measured by combining the intervals from
different methods, following the procedure used by
\citet{Dahlen2013}. Taking the median of the photometric
  redshifts does not necessarily produce a {\it better} measurement as
this depends heavily on the codes and templates used in calculating
the redshift and the corresponding scatter when compared to the
spectroscopic redshifts. Comparing with an independent sample of
spectroscopic redshifts not used for photometric redshift training, we
find that while taking the median reduces the outlier fractions
compared to some of the individual codes, marginally lower outlier
fractions are obtained using only a subset of the codes (those of
Wuyts, Gruetzbach and Salvato for this field). Given the relatively
small number of spectroscopic redshifts available for this test (262),
we do not consider the difference significant and choose to report the
median of all the codes as our recommended best photometric redshift.
Figure 11 shows a direct comparison between the photometric and spectroscopic
redshifts for galaxies in the CANDELS COSMOS field, only using high quality
spectroscopic redshifts. The training was done using a larger sample of
spectroscopic redshifts in CANDELS COSMOS area from the zCOSMOS
survey. These do not overlap with the comparison sample here, which is
a smaller sub-sample that is in public domain. Figure 12 shows
  the spectroscopic redshift distributions of the training and
  comparison samples. The high quality
  spectroscopic redshifts used for the training of
  the photometric redshifts are available over
the range $0<z<1$ and hence photometric redshifts reported are most
reliable for that redshift range. The values of the
redshift difference, as defined in \citet{Dahlen2013}, along with
the outlier fractions are listed in Table 4. As mentioned above, reliable spectroscopic
redshifts are available out to $z\sim1$ and the numbers
listed in Table 4 are only representative for galaxies in this range.

\begin{table*}[t]
\begin{center}
\caption{Spectroscopic redshift comparison Table.}
\begin{tabular}{*{6}{c}}
\hline
\hline
Survey & OLF$^a$ & $\sigma_{F}$$^b$ & $\sigma_{\rm NMAD}$$^c$ & $\sigma_{O}$$^d$&  Number of Galaxies$^{e}$ \\
\hline
CANDELS COSMOS & 0.008 & 0.035 & 0.011 & 0.016 & 506 \\
COSMOS Team & 0.05 & 0.071& 0.008 & 0.017 & 504 \\
3D-HST & 0.012 & 0.045 & 0.008 & 0.015 & 499 \\
\hline
\end{tabular}
\end{center}
\footnotesize
$^{a}$: Defined as fraction of objects with $|\Delta
z|/(1+z_{\rm spec})>0.15$ where $\Delta z = (z_{\rm phot} - z_{\rm
  spec})/(1 + z_{\rm spec})$ \citep{Dahlen2013}, $^{b}$: $\sigma_{F} \equiv {\rm
  rms}(\Delta z/(1+z_{\rm CANDELS}))$, $^{c}$: $\sigma_{\rm NMAD}
\equiv 1.48\times{\rm median}(|\Delta z|/(1+z_{\rm CANDELS}))$, $^{d}$:
$\sigma_{O} \equiv {\rm rms}(\Delta z/(1+z_{\rm CANDELS}))$ after
removing the outliers, $^{e}$: with reliable spectroscopic
redshift used for the comparison. This is for objects with
  $0<z<1$ and hence the reported numbers are valid for this redshift
  range where high quality spectroscopic redshifts are available.
\end{table*}

\begin{figure}
\centering
\includegraphics[trim=2cm 0cm 0cm 0cm,scale=0.45]{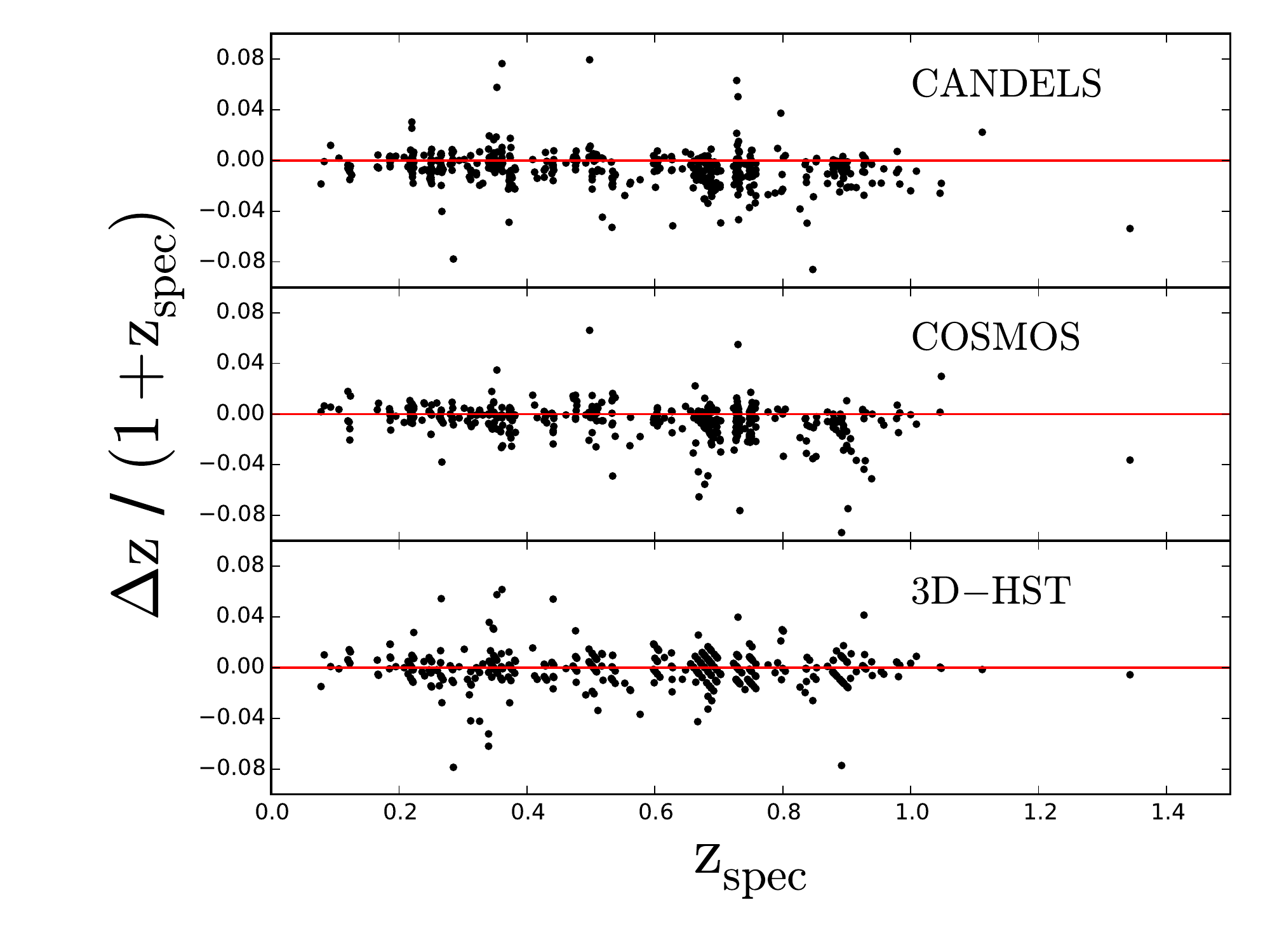}
\caption{Comparison between photometric redshift and spectroscopic
  redshift for the CANDELS, COSMOS \citep{Ilbert2013} and the 3D-HST
  \citep{Skelton2014}. The CANDELS photometric redshift correspond to the median value measured
by the different methods.}
\end{figure}

\begin{figure}
\centering
\includegraphics[trim=1cm 0cm 0cm 0cm,scale=0.45]{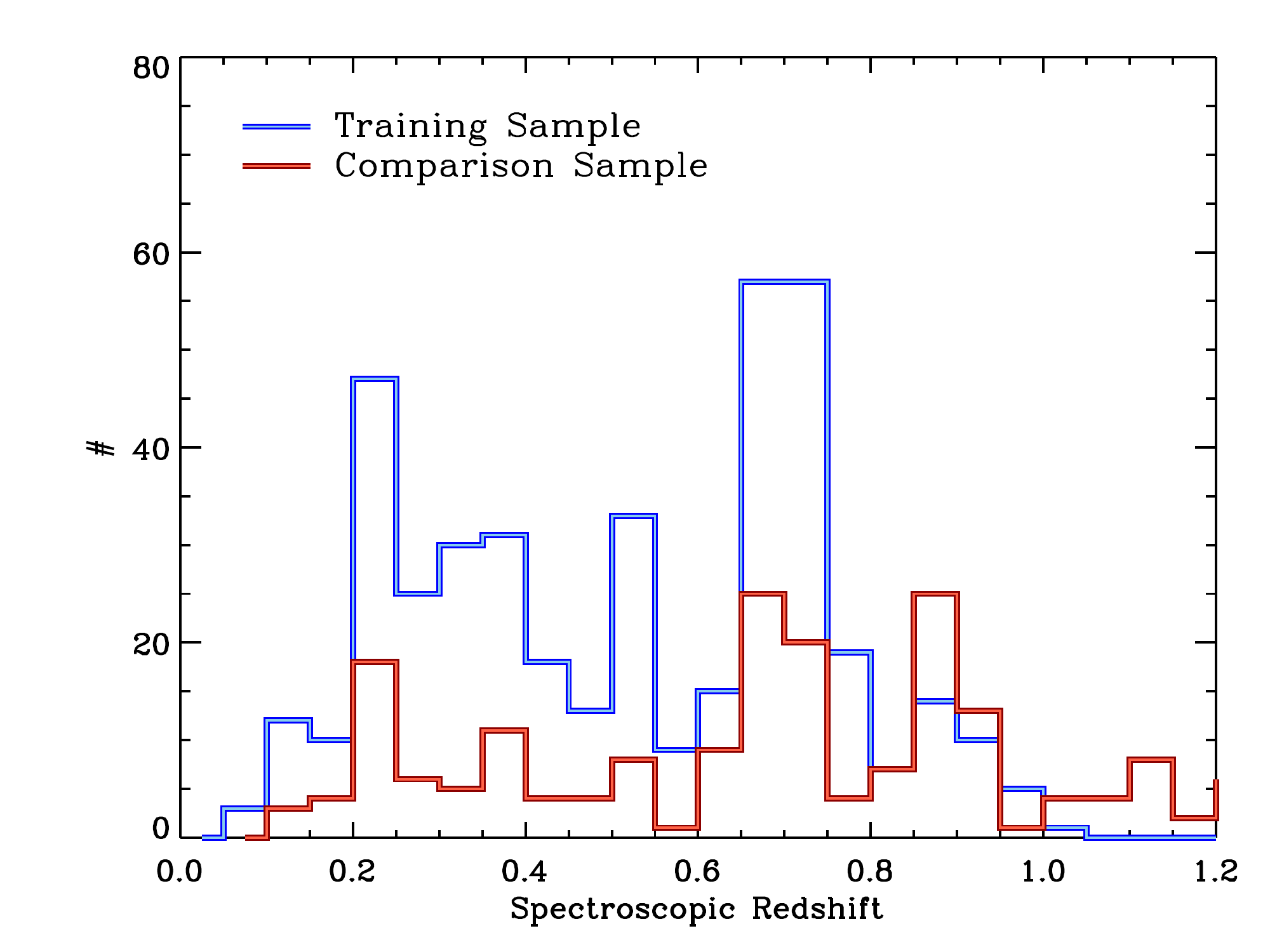}
\caption{Spectroscopic redshift distributions of the training set used
to calibrate the photometric redshifts (in blue) and the independent
comparison set (in red).}
\end{figure}

\begin{figure*}
\centering
\includegraphics[trim=1cm 0cm 0cm 0cm, scale=0.5]{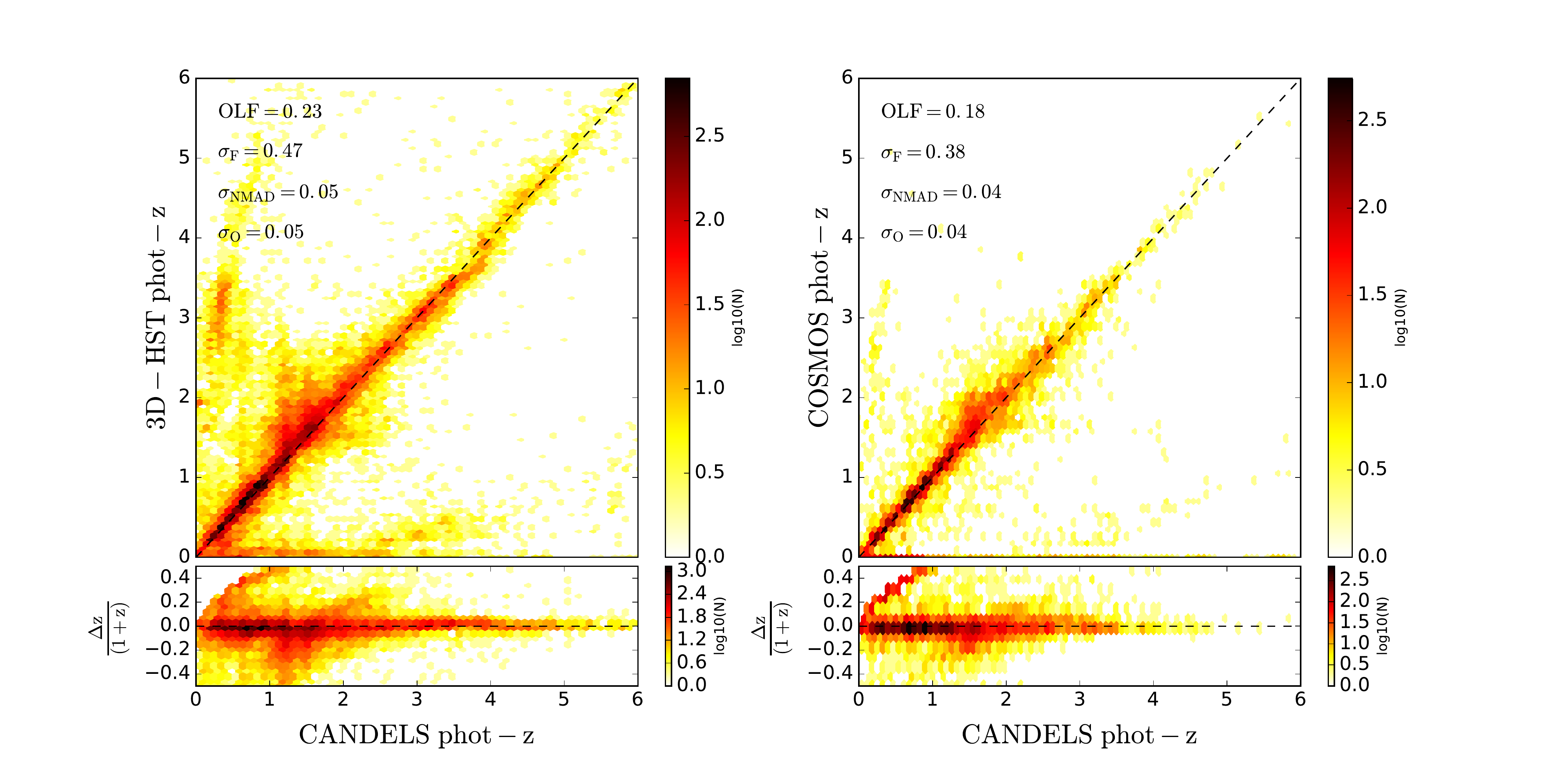}
\caption{Photometric redshift comparison plots of CANDELS COSMOS with 3D-HST (left) and
  COSMOS (right) catalogs. The comparison is shown in the form of
  2D-histogram with logarithmic bins to help see the small outlier
  fraction. The sub-panels in both plots show $\Delta
  z/(1+z_{\rm CANDELS})$ as a function of CANDELS photometric redshifts,
  where $\Delta z = (z_{\rm CANDELS} -z_{\rm other})$ for photometric redshift measurements.} 
\end{figure*}

\begin{table}
\begin{center}
\caption{SED fitting measured photometric offsets.}
\begin{tabular}{lc}
\hline
\hline
Filter & Median Offset$^{\dagger}$ \\
 & (AB Mag) \\
\hline
CFHT-$u^*$ & 0.041 \\
CFHT-$g^*$ & -0.044 \\
CFHT-$r^*$ & -0.005 \\
CFHT-$i^*$ & -0.025 \\
CFHT-$z^*$ & 0.000 \\
Subaru-$B$ & -0.004 \\
Subaru-$r^+$ & -0.071 \\
Subaru-$i^+$ & -0.074 \\
Subaru-$z^+$ & -0.150 \\
ACS-F606W & 0.072 \\
ACS-F814W & 0.019 \\
WFC3-F125W & 0.104 \\
WFC3-F160W & 0.091 \\
UVISTA-$Y$ & -0.027 \\
UVISTA-$J$ & 0.046 \\
UVISTA-$H$ & 0.067 \\
UVISTA-$K_s$ & -0.031 \\
IRAC-3.6\,$\mu$m & -0.061 \\
IRAC-4.5\,$\mu$m & -0.176 \\
IRAC-5.8\,$\mu$m & -0.069 \\
IRAC-8.0\,$\mu$m & -0.697 \\
NEWFIRM-$J_1$ & -0.061 \\
NEWFIRM-$J_2$ & -0.027 \\
NEWFIRM-$J_3$ & 0.000 \\
NEWFIRM-$H_1$ & -0.025 \\
NEWFIRM-$H_2$ & -0.060 \\
NEWFIRM-$K$ & -0.073 \\
\hline
\end{tabular}
\end{center}
\footnotesize
$^{\dagger}$: positive offset: measured flux fainter than expected from template
\end{table}

We cross-compared the CANDELS photometric redshift catalog with those
from COSMOS \citep{Ilbert2013} and 3D-HST \citep{Skelton2014}. Figure
11 shows the comparison between the photometric
and spectroscopic redshifts in the COSMOS and 3D-HST catalogs. The rms values
are listed in Table 4 and show similar trends as CANDELS COSMOS. We
compare photometric redshifts between the
CANDELS COSMOS and those
from the COSMOS \citep{Ilbert2013} and 3D-HST \citep{Skelton2014}
catalogs, in Figure 13. The two photometric redshift
comparison plots present the full scatter ($\sigma_{F} \equiv {\rm
  rms}(\Delta z/(1+z_{\rm CANDELS}))$), the
normalized median absolute deviation ($\sigma_{\rm NMAD} \equiv
1.48\times{\rm median}(|\Delta z|/(1+z_{\rm CANDELS}))$) and the scatter after
excluding outliers ($\sigma_{O} \equiv {\rm rms}(\Delta z/(1+z_{\rm CANDELS}))$) as well as the
outlier fraction (defined as fraction of objects with $|\Delta
z|/(1+z_{\rm CANDELS}) > 0.15$). There is consistency in the
photometric redshift measurements for the majority of galaxies. The
spread seen in Figure 13, is due to log-binning of the histograms and
does not indicate any large inconsistency. Table 5 shows the photometric
offsets measured from SED fitting for different bands. The offsets are
measured through the SED fitting, simultaneously with photometric
redshifts. Different independent SED fitting methods estimated the
offsets and they agreed fairly well. The magnitude offsets were then
used in the photometry to estimate the final photometric
redshifts. The photometric redshifts have the correction for
the photometric offsets but the photometry presented here is not
corrected for the offset.

One method for estimating the photometric redshift uncertainties that
is specially useful for fainter flux limits where not enough
spectroscopic redshift information is available is the pair
statistics estimates \citep{Quadri2010, Huang2013, Dahlen2013, Hsu2014}. The method, as
outlined in \citet{Dahlen2013}, relies on the fact that spatially close pairs (defined as objects with
separation less than 15\,arcsec) have a high probability of being
associated with each other and therefore being at similar
redshifts \citep{Dahlen2013}. This close-pair association would show
up as excess power at small separations when compared to the
distribution based on random galaxies \citep{Dahlen2013}. Figure
14 shows the random and close pair photometric redshift difference
distribution along with the excess distribution of the close pairs in photometric
redshift after subtracting the photometric redshift distribution of
the random galaxies. By fitting a Gaussian function to this excess
distribution we measured an uncertainty of 0.017 ($\rm \sqrt{2}\times
\sigma_{Gaussian}$) in the photometric redshift distribution.

\begin{figure}
  \centering
  \includegraphics[trim=0cm 0cm 0cm 0cm, scale=0.5]{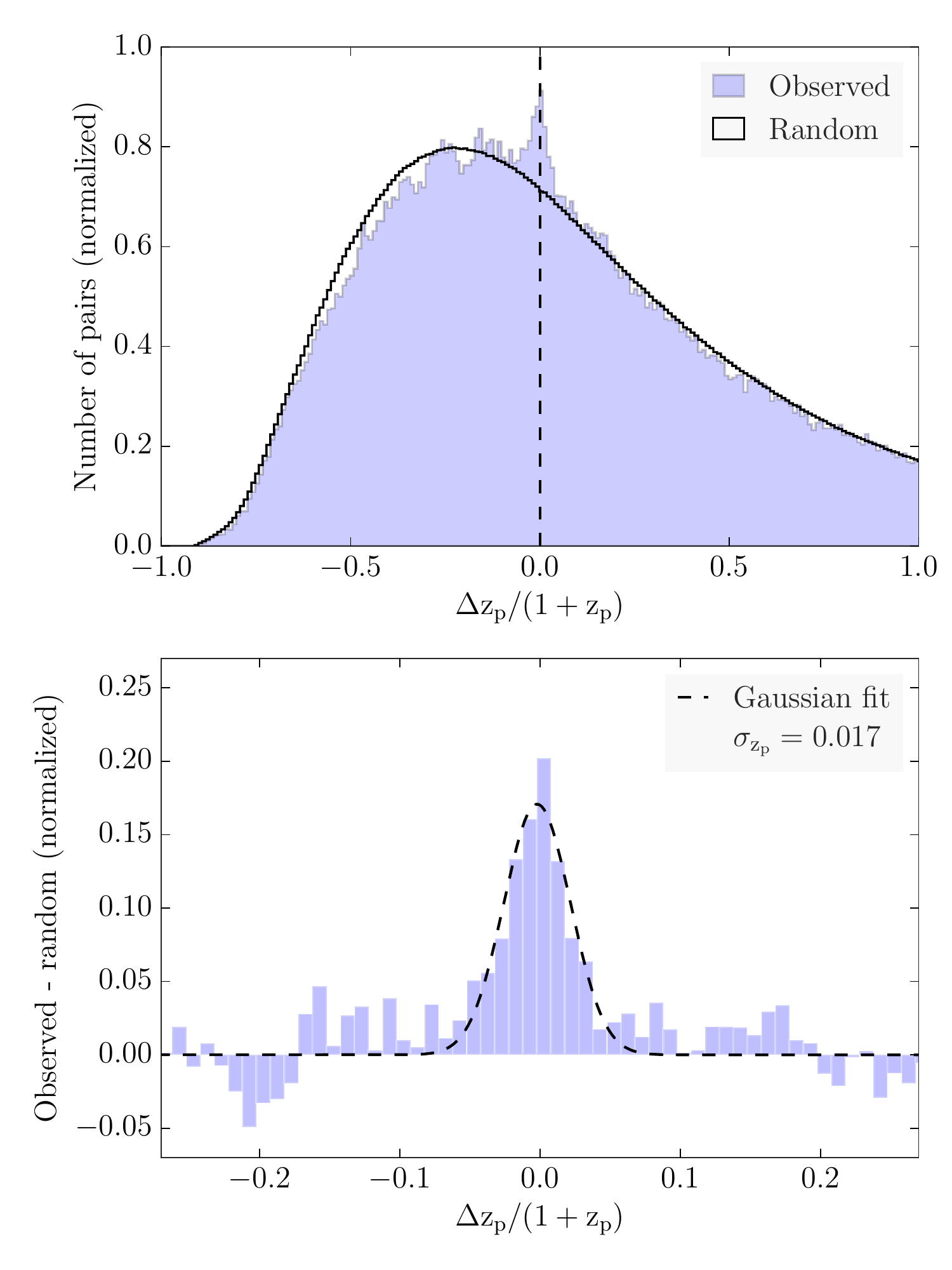}
  \caption{Top: Distribution of the photometric redshift difference
    for the random pair (black) and the close pairs, defined as
    objects with separation less than
    15\,arcsec (blue). Bottom: Distribution of the overdensity of the
    photometric redshift difference for close pairs. The distribution is observed in excess of the
    photometric redshift difference distribution of random galaxies
    that are subtracted (top panel) \citep{Dahlen2013}. A Gaussian
    fit to the distribution gives the uncertainties associated with
    the redshifts. This test only characterizes the core of the
    photometric redshift errors and not the outlier rate.} 
\end{figure}

\subsection {Stellar Masses}

\citet {Mobasher2015} studied stellar mass measurement using CANDELS
data. With extensive simulations, they explored
different sources of uncertainty in stellar mass measurements and
estimated the error budget associated with them. The stellar masses were
measured from different methods independently and were compared with their
expected (input) mass. All methods produced stellar masses in good
agreement. We use the median mass (among all the methods) as the
reported CANDELS estimate. As we discussed earlier, taking the median does
not necessarily produce a more robust mass estimate because the results of
individual codes used are not instances of the same stochastic
process.

The TFIT multi-waveband photometric catalog for the CANDELS COSMOS
field was provided to the CANDELS teams outlined in Table 8 and they
measured the stellar masses through separate SED fitting techniques. For all the
independent measurements, redshifts were fixed to their median values
(as described in the previous section). Given that the
  photometric redshifts are calibrated with a spectroscopic sample
  with $0<z<1$, the stellar mass estimates are also well calibrated
  and most robust within this redshift range, as discussed in the
  previous section, although we report stellar mass estimates out to
  $z\sim5$ in this catalog. Some of the methods included nebular emission when
fitting the SEDs as outlined in Table 8. A total of eight entries were received from different
teams. We measured the median stellar mass between different
methods. We used the Hodges-Lehmann\footnote{\url{http://www.jstor.org/stable/2238406}}
method to estimate the median
stellar mass which accounts for the
small number of entries when measuring the median value.

\begin{figure*}
\centering
\includegraphics[trim=2cm 0cm 0cm 0cm, scale=0.35]{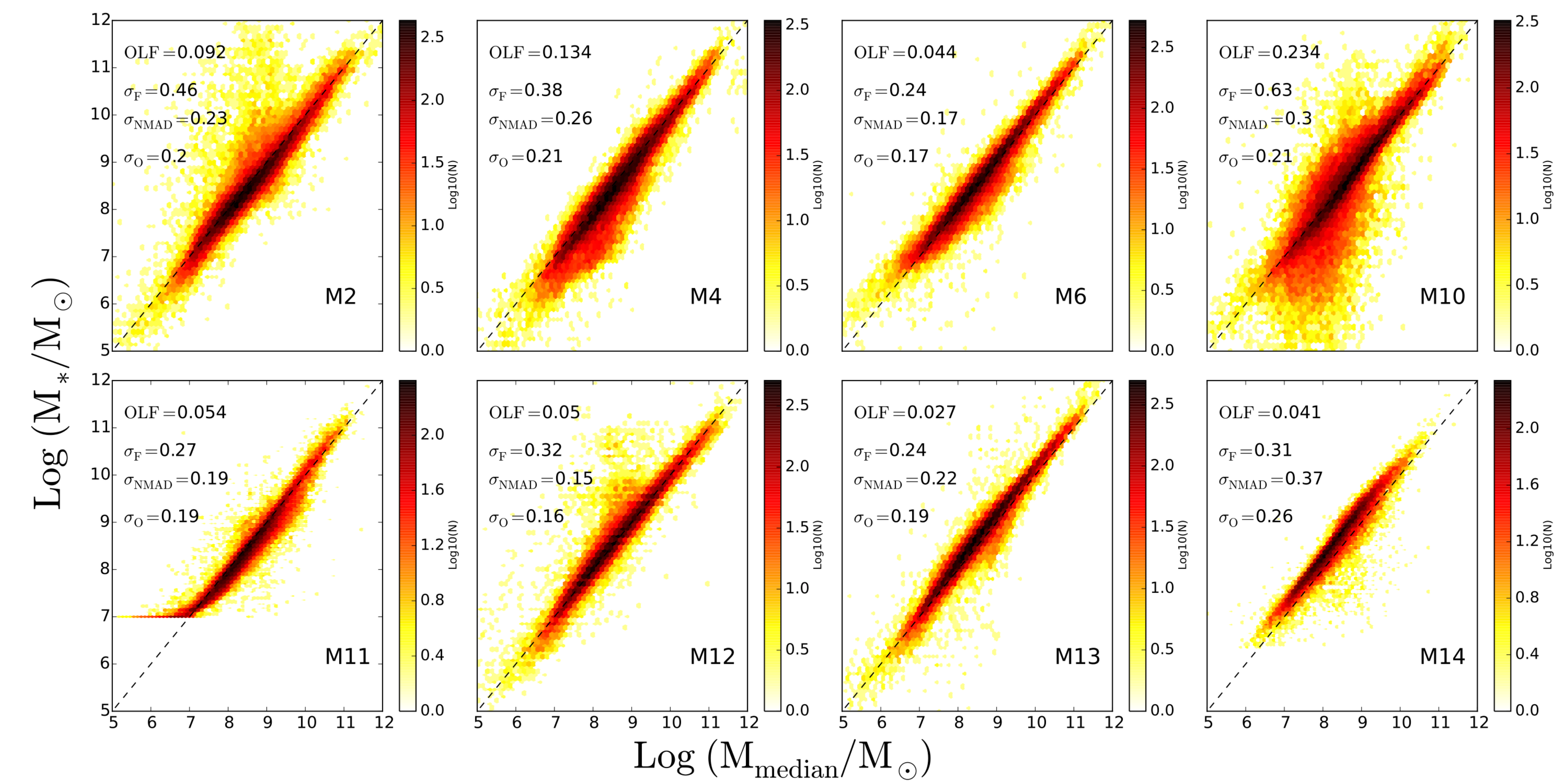}
\caption{Comparisons of the reported stellar masses using eight different
  methods outlined in Table 8 versus the median stellar mass of all the other
  methods. The plots are shown as 2D histograms with logarithmic bins. We report
  the variations in the mass difference and the outliers as defined in
Section 5.1 in each panel.}
\end{figure*}

Figure 15 presents a direct comparison between different methods used to
measure stellar masses. Here, we plot estimates from each method against the
median from all the rest of the methods. There is good internal consistency
between different methods and the tails seen in
Figure 15 help identify where the
outliers are in each method compared to the rest. These outliers will
not bias our measurements as we report the median of all methods for
our final stellar mass measurements as demonstrated in \citet{Mobasher2015}.

The median stellar mass estimates from the CANDELS are compared
to those from the COSMOS \citep{Ilbert2013} and 3D-HST catalogs \citep{Skelton2014}
in Figure 16. In these comparisons, we measure the scatter similarly to
the photometric redshift where $\sigma_F \equiv 
  rms({\rm log}(M_{\rm CANDELS})- {\rm log}(M_{\rm other}))$ and $\sigma_O \equiv 
  rms({\rm log}(M_{\rm CANDELS})- {\rm log}(M_{\rm other}))$ after removing the outliers
where outlier fraction is defined as the fraction of objects with $
\Delta {\rm log}(M) \equiv ({\rm log}(M_{\rm CANDELS})- {\rm
  log}(M_{\rm other}) > 0.5$ \citep{Mobasher2015}, as shown
on Figure 16. The CANDELS stellar mass measurements are
consistent with both COSMOS and 3D-HST stellar masses. The stellar
mass offsets between CANDELS
median measurement and 3D-HST and COSMOS are also plotted as a
function of F160W magnitude in Figure 17. As expected, the
larger discrepancies between the stellar mass estimates occur at
fainter magnitudes. The rms values of the stellar mass estimate
offsets as defined above are presented in Table 6 as a function of $H$-band magnitude,
confirming a good agreement at brighter magnitudes and an overall
consistency for the majority of galaxies.

\begin{figure*}
\centering
\includegraphics[trim=1.5cm 0cm 0cm 0cm, scale=0.5]{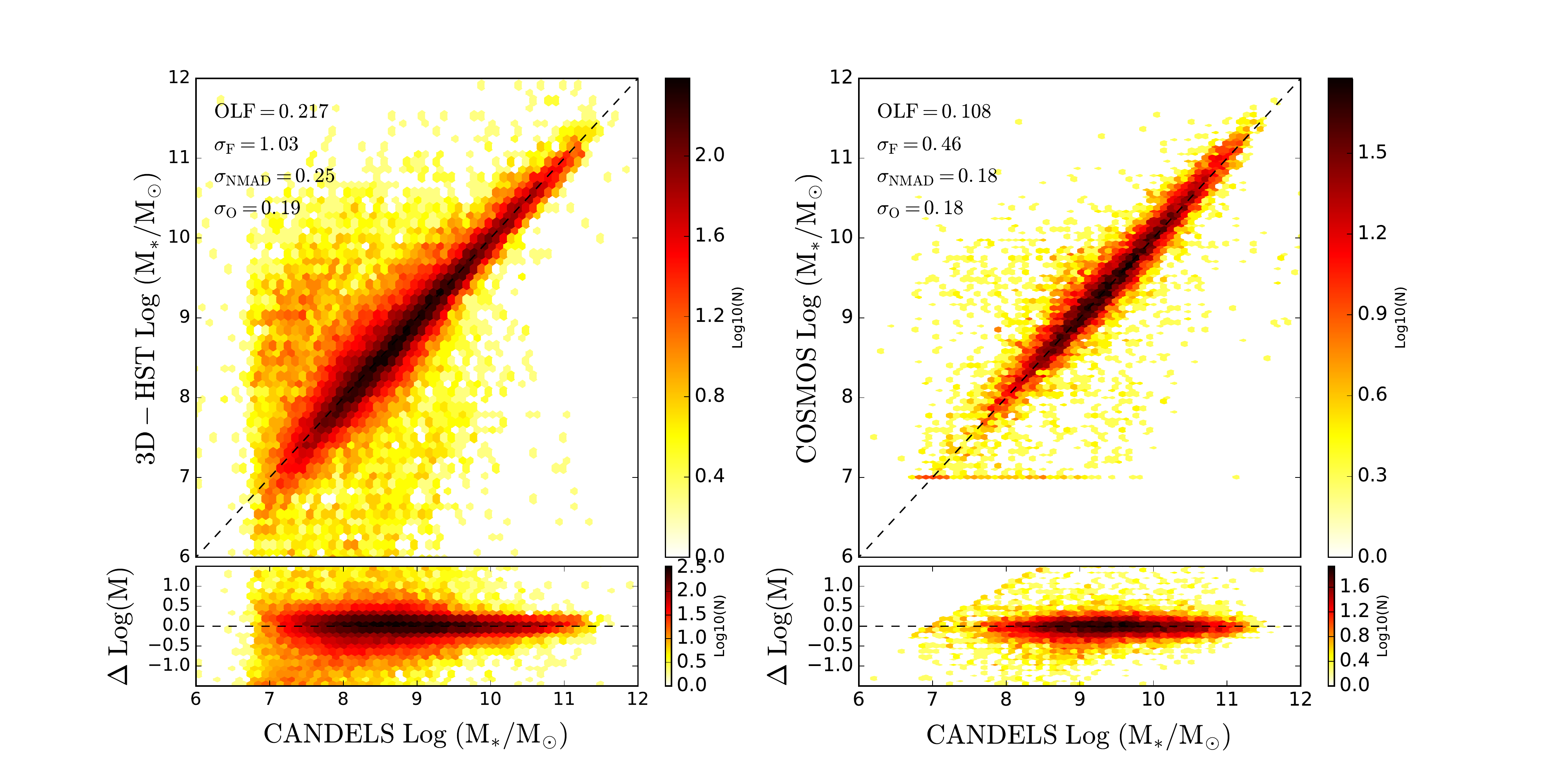}
\caption{Stellar mass comparison plots of CANDELS COSMOS with 3D-HST (left) and
  COSMOS (right) measurements. The comparison is shown in the form of
  2D-histogram with logarithmic bins to help see the small outlier
  fraction. In each panel we report the variations in the mass
  difference and the outlier fractions as defined in Section 5.1. The
  sub-panels show $\Delta {\rm log}(M)$ as a
  function of CANDELS ${\rm log}(M_{*}/M{\odot})$ where $\Delta {\rm log}(M) =
  {\rm log}(M_{\rm CANDELS})-{\rm log}(M_{\rm other})$ for stellar mass measurement.}
\end{figure*}

\begin{figure*}
\centering
\includegraphics[trim=0cm 0cm 0cm 0.5cm, scale=0.45]{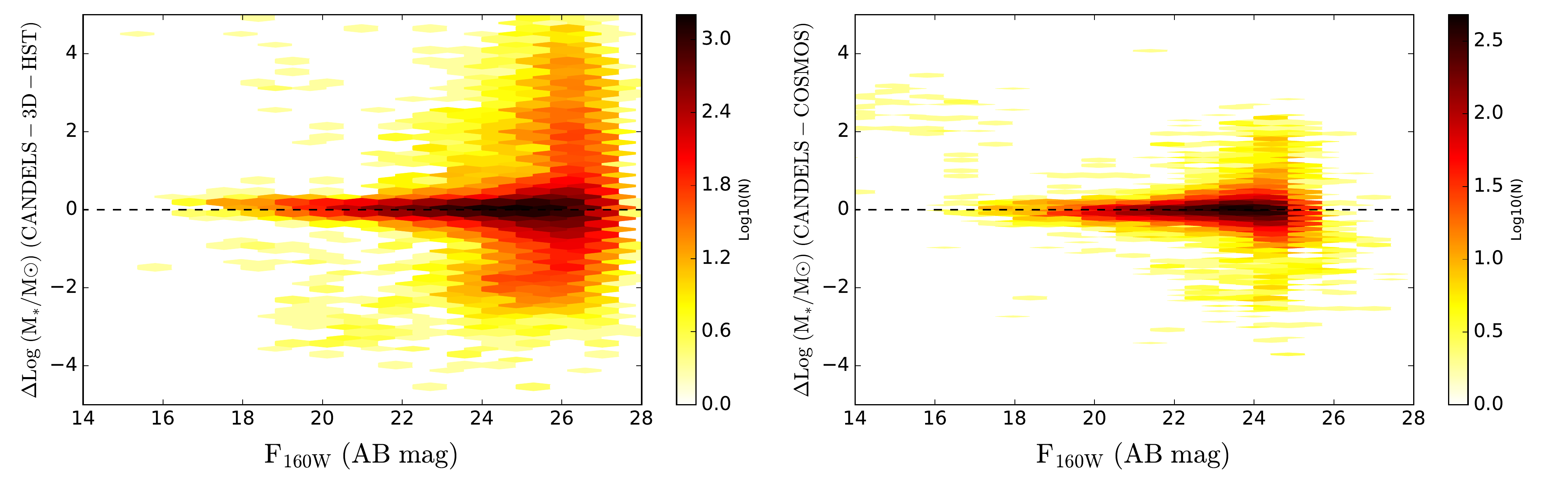}
\caption{Stellar mass offsets between CANDELS and 3D-HST (left) and
  COSMOS (right) as a function of the F160W magnitude. Both plots are
  in forms of 2D histograms with logarithmic bins, showing the
  fraction of inconsistent measurements increase at fainter magnitudes.}
\end{figure*}

\begin{figure*}
\centering
\leavevmode
\includegraphics[trim=1cm 0cm 0cm 1cm, scale=0.7]{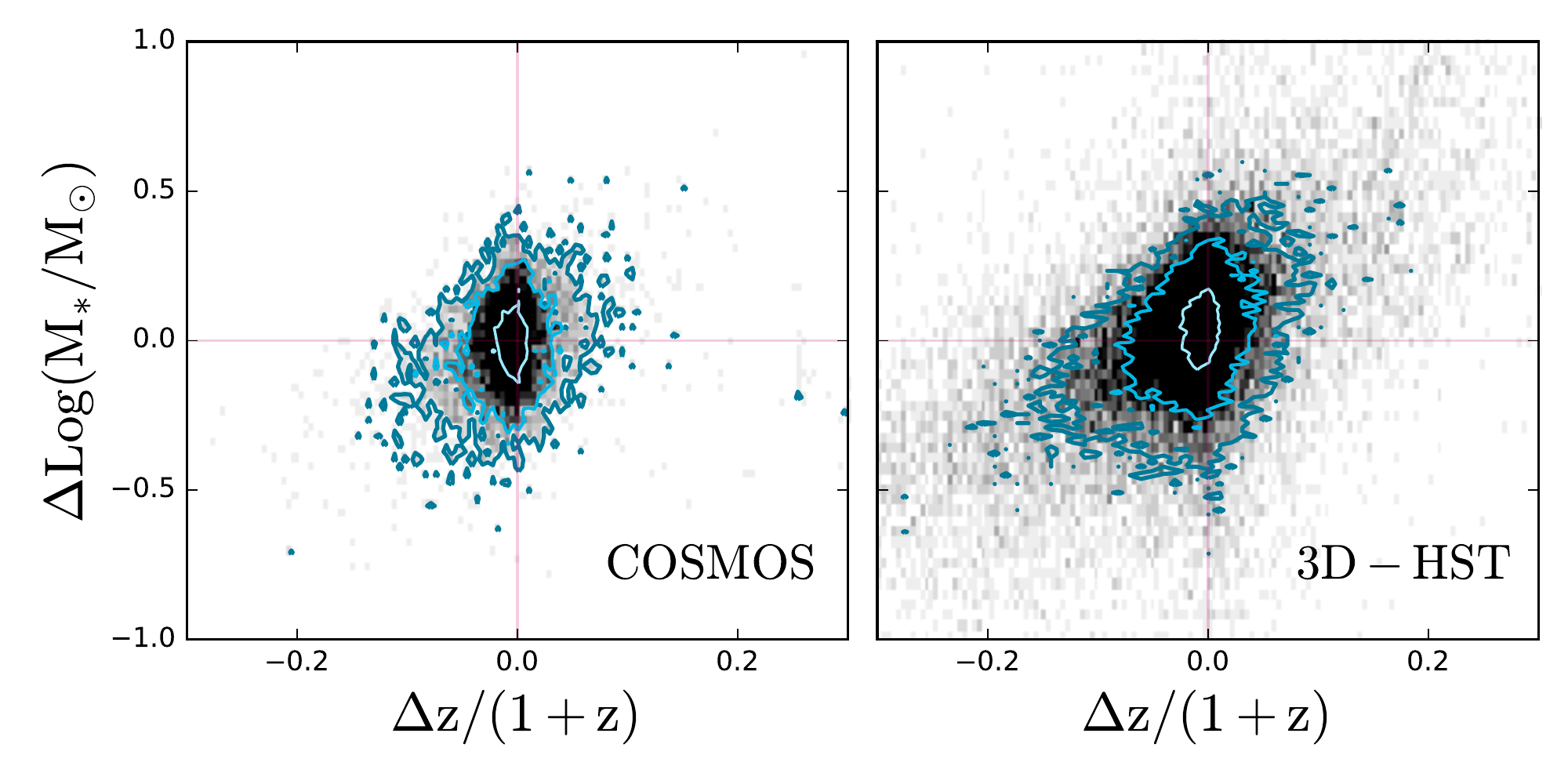}
\caption{Stellar mass offsets versus photometric redshifts offsets
  between CANDELS measurements and those from COSMOS (left) and
  3D-HST team (right). The solid red lines show the 1:1 relations.}
\end{figure*}

\begin{figure}
\centering

\includegraphics[trim=1cm 0cm 0cm 0cm, scale=0.45]{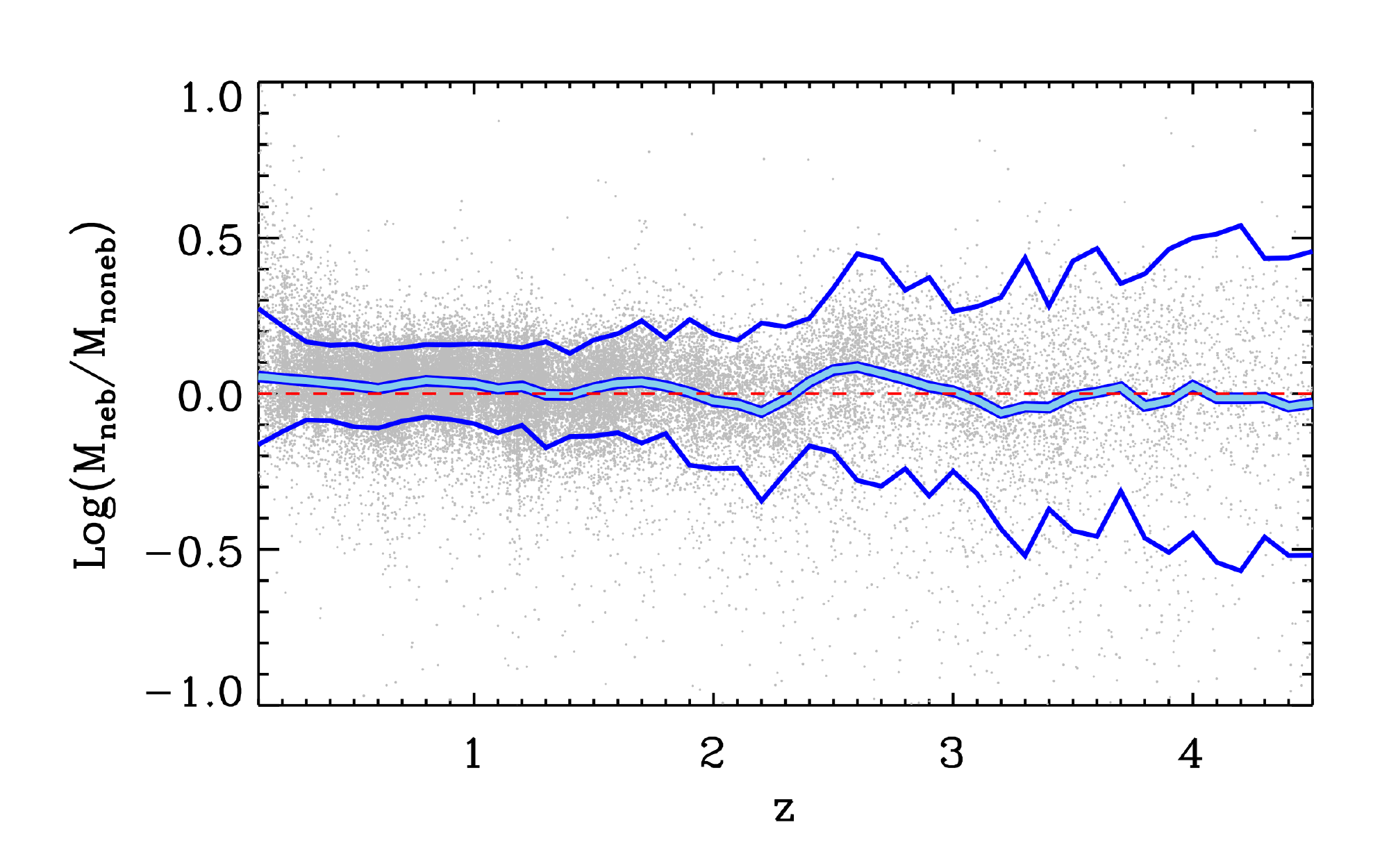}
\caption{The difference of the median of the stellar mass measured
  with and without nebular emission as a function of the redshift. The dashed
  red line shows the 1:1 relation. The median and 1$\sigma$ variations
are shown with the light and dark blue respectively. The median is
consistent with no evolution as a function of redshift for the two
mass estimates. }
\end{figure}

\begin{figure}
\centering
\includegraphics[trim=0.5cm 0cm 0cm 0cm, scale=0.75]{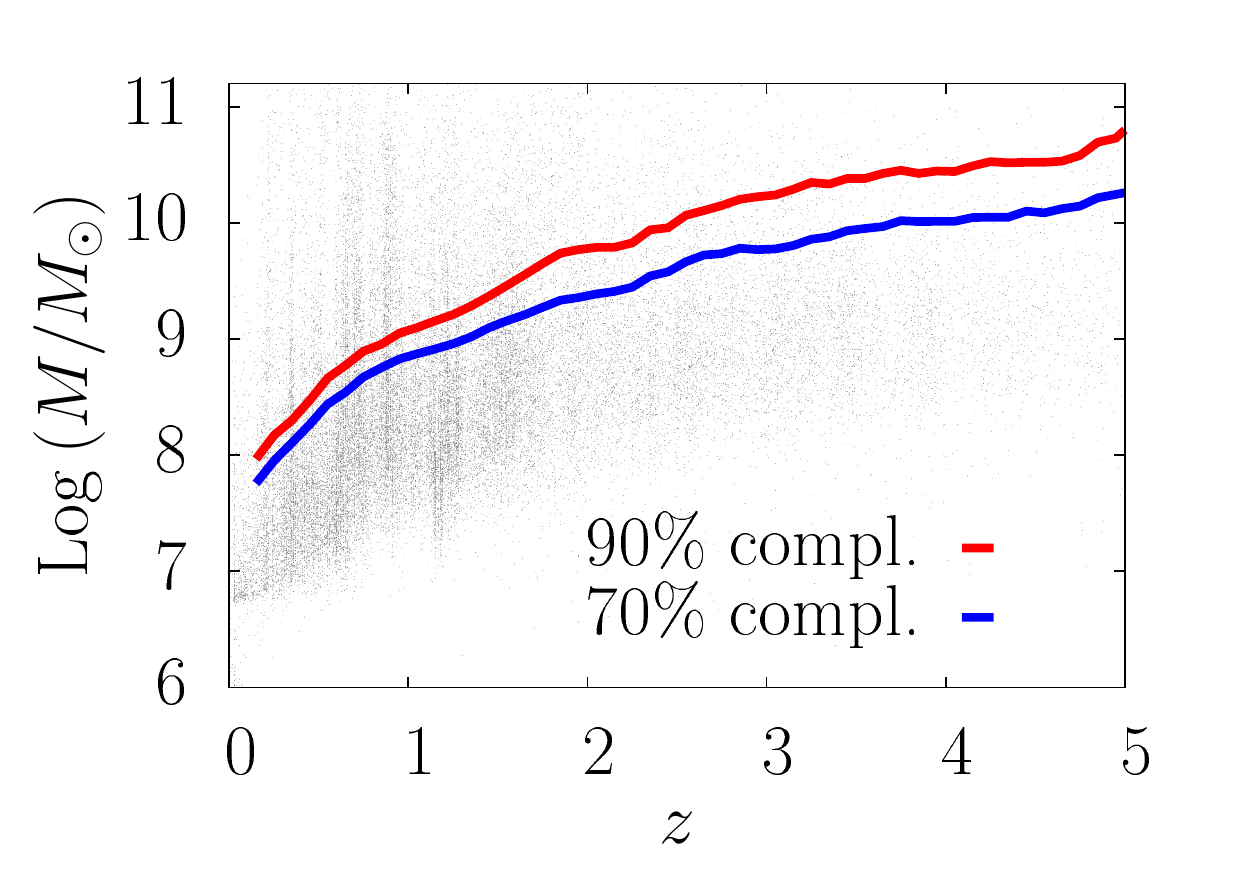}
\caption{Stellar mass as a function of redshift. Points show measured
  stellar masses for the CANDELS COSMOS sample. Red and blue lines
  show the 90\% and 70\% stellar mass completeness limits for the
  general population of galaxies. The 90\% (70\%) completeness limit is defined in a 
sense that only $<$ 10\% (30\%) of the galaxies could be missed in the
low-mass end of the galaxy distribution.}
\end{figure}

\begin{table*}
\begin{center}
\caption{Variations and outlier fractions in mass measurements.}
\begin{tabular}{*{6}{c}}
\hline
\hline
 & Magnitude Cut (AB) & OLF$^{\dagger}$ & $\sigma_{\rm F}$ & $\sigma_{\rm NMAD}$ & $\sigma_{\rm O}$ \\
\hline			
COSMOS & & & & & \\
 & $21<H < 22$	& 0.032 &	0.307	&		0.124	&	0.149 \\
 & $22<H < 23$	& 0.048 &	0.309	&		0.138	&	0.153 \\
 & $23<H < 24$	& 0.077 &	0.384	&		0.179	&	0.175 \\
 & $24<H < 25$	& 0.172 &	0.549	&		0.251	&	0.204 \\
3D-HST & & & & & \\
 & $21<H < 22$	& 0.049 &	0.597	&		0.133	&	0.131 \\
 & $22<H < 23$	& 0.054 &	0.463	&		0.138	&	0.144 \\
 & $23<H < 24$	& 0.089 &	0.563	&		0.148	&	0.148 \\
 & $24<H < 25$	& 0.138 &	0.687	&		0.201	&	0.179 \\
\hline
\end{tabular}
\end{center}
\footnotesize
$^{\dagger}$: Defined as $|\Delta {\rm log}(M)|>0.5$ \citep{Mobasher2015}.
\end{table*}

When measuring the stellar mass, we fit the SEDs by fixing redshifts
to the median of the photometric redshifts from the CANDELS COSMOS
team, as discussed in Section 5.1. Therefore, if the redshifts for the
same galaxies are different in those from 3D-HST and COSMOS teams,
the effect would propagate to the estimated stellar masses. As a
result, the observed offsets between the stellar mass values in
Figures 16 and 17 could partly be explained by the discrepancy
between the redshifts. To explore this, we studied the residual diagrams
between redshifts and stellar masses for the three measurements. For
this we look at the ratio of the stellar masses as measured by
different methods (${\rm log}(M_{\rm CANDELS})-{\rm log}(M_{\rm other})$) as a
function of the redshift difference $(z_{\rm CANDELS}-z_{\rm other})/(1+z_{\rm CANDELS})$ in Figure 18
showing a 0.25\,dex scatter in stellar mass for galaxies with similar
redshifts. This is in agreement with results of \citet{Mobasher2015},
who measured the combined error budget in stellar mass values
due to different parameters. Therefore, the distribution in residual
mass here is consistent with the expected uncertainties in the stellar mass
measurements (i.e the vertical scatter). The
galaxies with deviant redshifts also have deviant stellar mass
estimates, partly explaining the observed scatter between the
CANDELS and 3D-HST and CANDELS and COSMOS stellar masses.

Figure 19 compares the difference between stellar mass measurements with
and without correction for nebular emission lines. This shows a small scatter (0.25\,dex) in the stellar
mass, consistent with \citet{Mobasher2015}, but no significant offset
over the whole population over the large redshift range.

\begin{figure*}
\centering
\leavevmode
\includegraphics[scale=0.44]{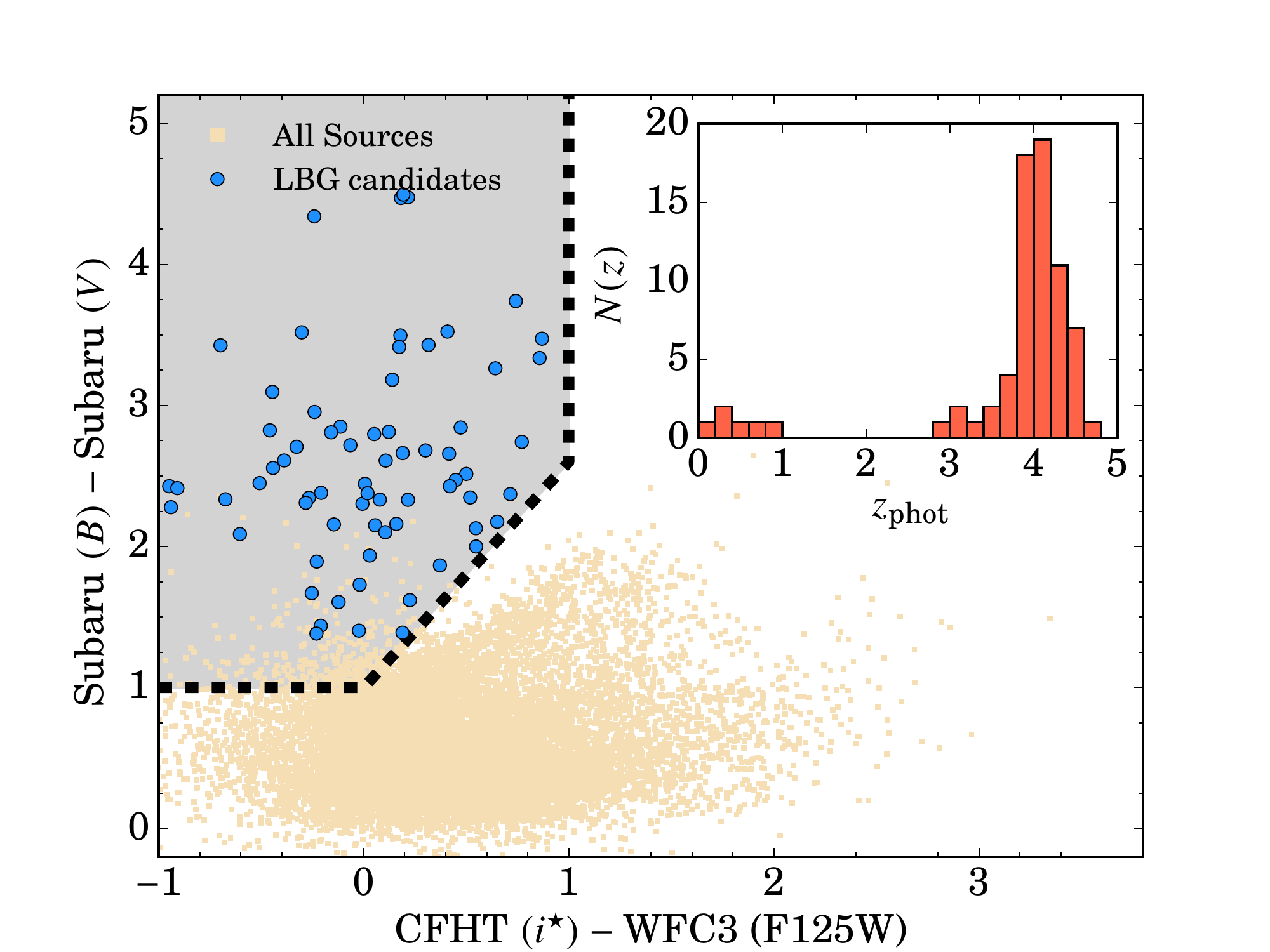}
\includegraphics[scale=0.44]{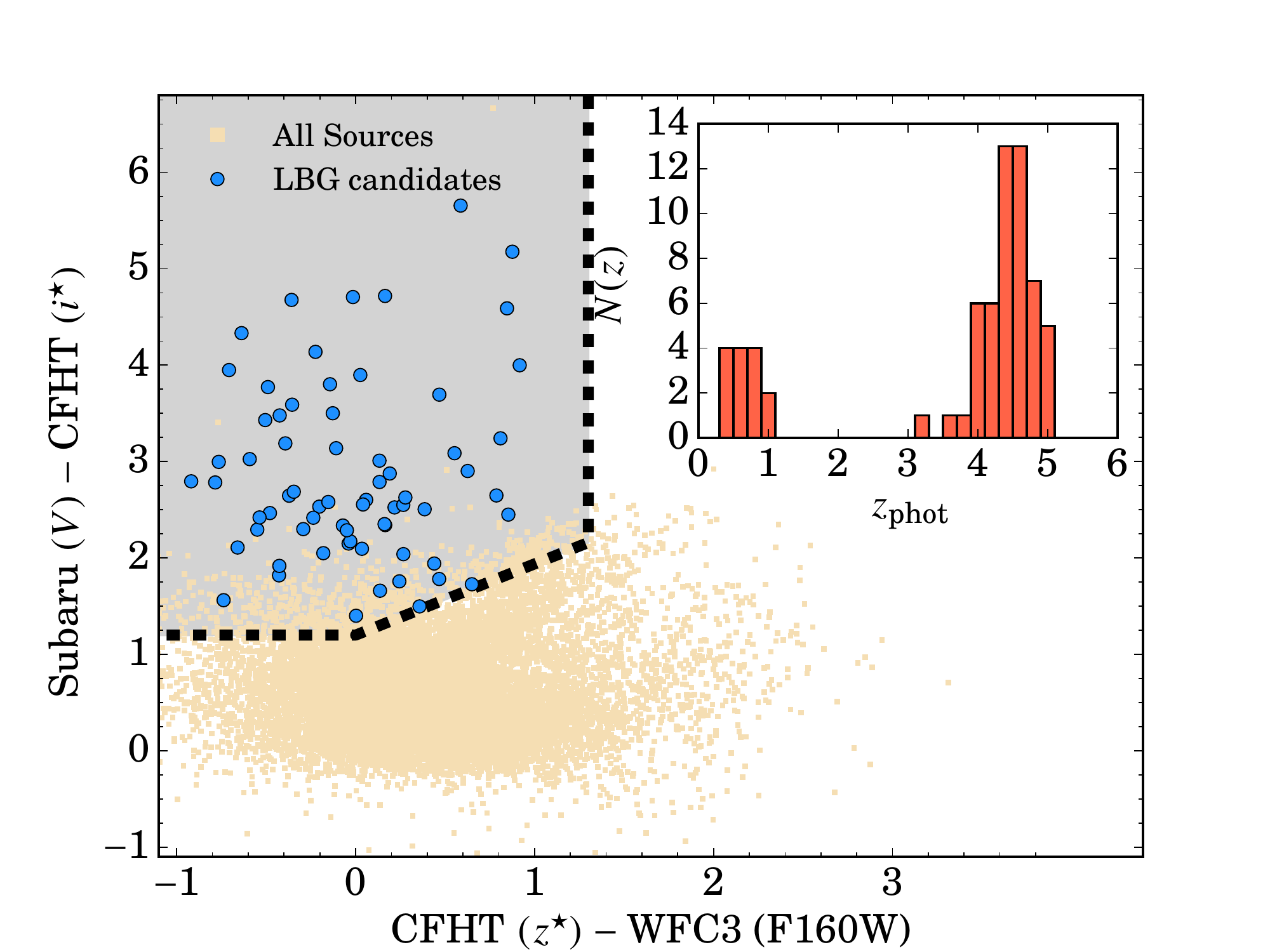} 
\caption{The LBGs selected from CANDELS COSMOS catalog using the
  selection criteria from \citet{Bouwens2015} as outlined in the text. $B$-dropout and
  $V$-dropout galaxies are shown by blue points in the shaded
  selection areas with other sources shown by orange points. Sources that fall within the color selection but are not
detected do not satisfy the non-detection S/N limit on the blue
bands and mostly sit at the boundary of the selection. The subplots in each panel shows the
  distribution of photometric redshifts of selected galaxies.}
\end{figure*}

\begin{figure*}
\centering
\includegraphics[trim=2cm 0cm 0cm 0cm, scale=0.33]{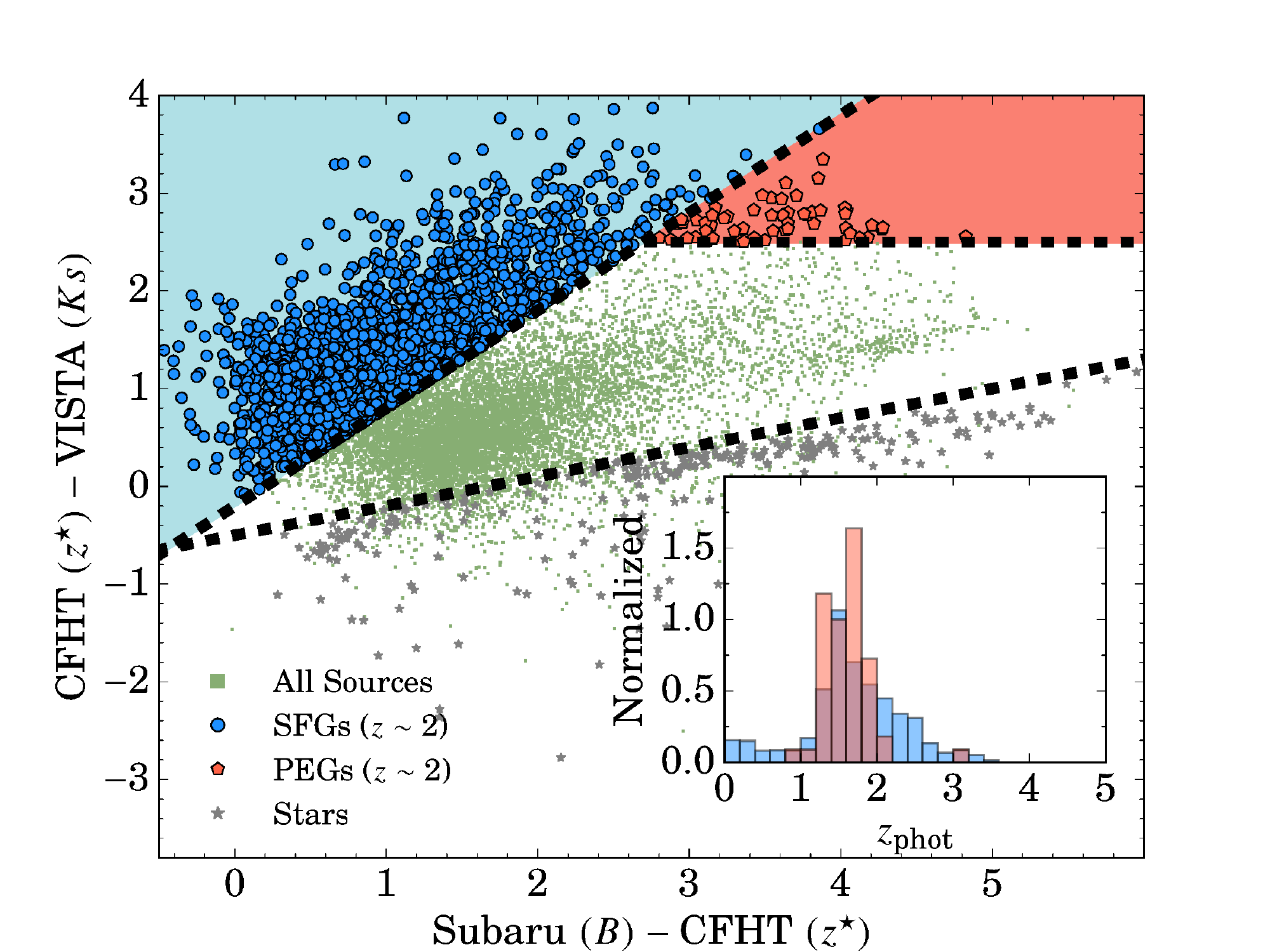}
\includegraphics[trim=2cm 0cm 0cm 0cm, scale=0.33]{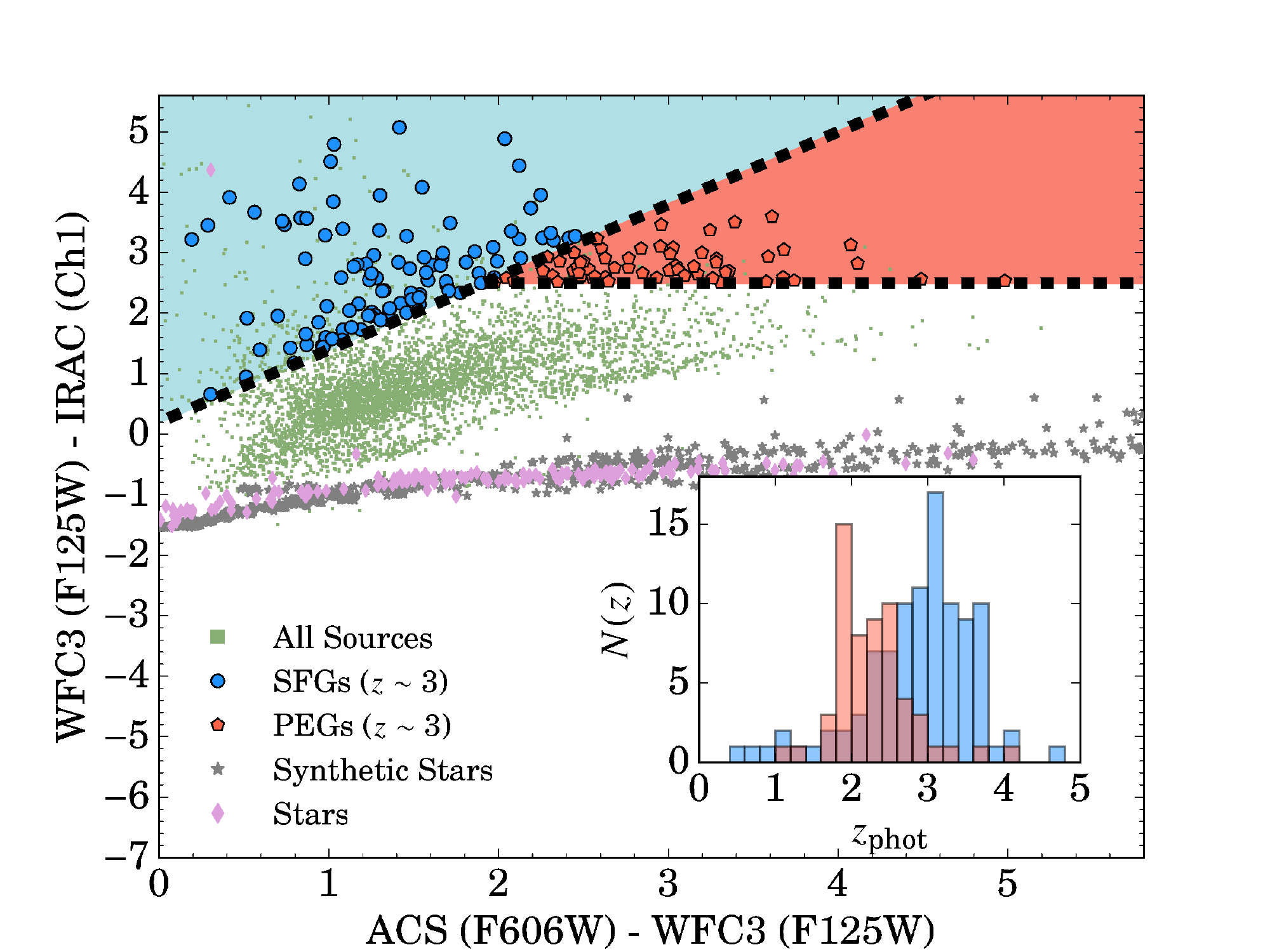}
\includegraphics[trim=2cm 0cm 2cm 0cm, scale=0.33]{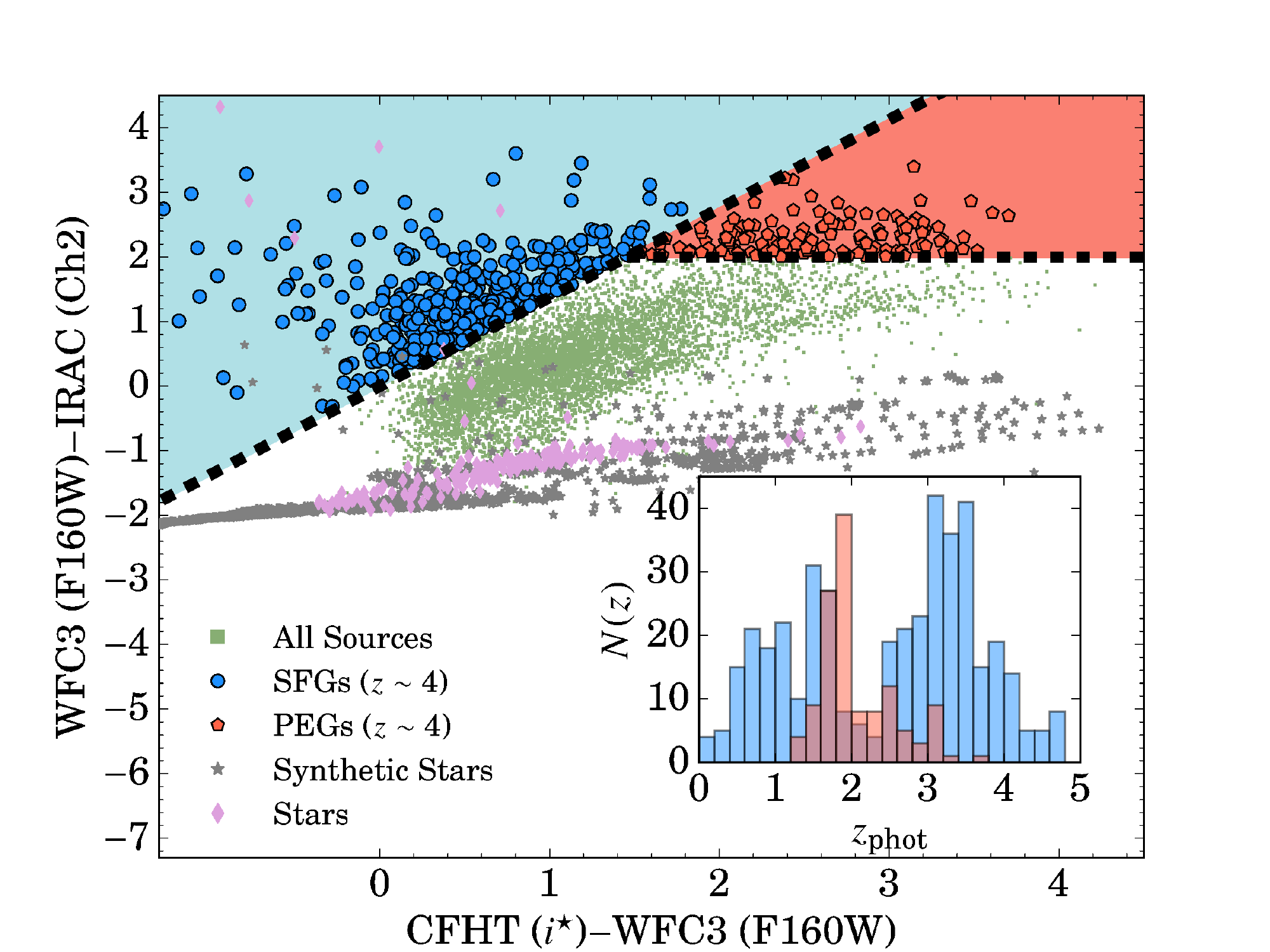}
\caption{The $BzK$ (left), $VJL$ (middle) and $iHM$ (left) color-color plots
  \citep{Daddi2004, Guo2012, Guo2013} showing the star-forming and passive galaxies
  positions at $z \sim 1.5$, $z \sim 2.5$ and $z \sim 3.5$ respectively. The star
  forming and quiescent population are plotted in blue and red in both
  diagrams. The corresponding colors of stars in our catalog
  (identified from \texttt{SExtractor} CLASS\_STAR parameter) are shown
  in each plot with grey symbols. $BzK$ plot shows the predicted
  colors of stars as reported by \citet{Daddi2004}. We further show
  the colors of model stars from the BaSeL library \citep{Lejeune1997,
    Westera2002} as magenta
  diamonds in the $VJL$ and $iHM$ plots. The inset in the plots show the
photometric redshift distribution (Section 5) of the star forming and
quiescent populations identified by the corresponding color
selections. The redshift distributions of the identified sources are
consistent with expectations from the well calibrated color selections
further verifying our photometry.}
\end{figure*}

\subsection{Mass Completeness}

We use the method introduced by \citet{Pozzetti2010} to estimate 
the stellar mass completeness limit of the general population of
galaxies (see also \citealp{Ilbert2013} and \citealp{Darvish2015b}). 
Given the magnitude limit of the sample ($H_{\rm lim}=27.56$;
$5\sigma$ limiting magnitude in the F160W detection band), we assigned a 
limiting stellar mass ($M_{\rm lim}$) to each galaxy. $M_{\rm lim}$ is the stellar mass
that a galaxy would have at its estimated redshift, if its apparent
magnitude was the same as the magnitude limit of our sample ($H_{\rm lim}=27.56$). This was 
evaluated by ${\rm log}(M_{\rm lim}/M_{\odot})={\rm log}(M/M_{\odot})+0.4(H-H_{\rm lim})$,
where M is the estimated stellar mass of the galaxy with its apparent magnitude H.
This results in a distribution of $M_{\rm lim}$ values at any given redshift. The 90\% (70\%)
stellar mass completeness limit at each redshift is therefore equivalent to the mass with 
90\% (70\%) of the galaxies having their $M_{\rm lim}$ value below the stellar mass completeness 
limit. In general, the stellar mass completeness limit depends on the $M/L$ ratio 
and is higher for quiescent and dusty galaxies. The stellar mass
  estimate is also sensitive to the presence of the Balmer break in
  the SED of galaxies. At $z\gsim3$ the WFC3 $H$-band will no longer be
  probing the Balmer break and a redder filter (such as $K_s$-band or 3.6\,$\mu$m) is
  more suited at estimating the stellar mass at these redshifts
  \citep{Ilbert2013}. Using the UltraVISTA $K_s$ band limiting magnitude
  reported in Table 1, we found that while the completeness estimates
  change at $z\gsim3$, the deviations are at the level of 0.1-0.2 dex
  which is within the stellar mass uncertainties. Figure 20 shows the
distribution of stellar mass as a function of redshift, along with the 
estimated 90\% and 70\% completeness limits.

\section{Application to $2<z<5$ Galaxies}

\subsection{Star-forming Galaxies at $z \sim 4-5$}

One of the main methods of identifying high redshift galaxies is
by targeting the pronounced Lyman break at rest-frame 912\,\AA\ that
exist in the SED of these galaxies (e.g. \citealp{Pettini2002, Shapley2003,
  Bolton2013, Faucher2015, Williams2015}). The break is caused by
the absorption of the UV light from hot and young stars by the neutral
Hydrogen \citep{Madau1995, Madau1996}. Because of this break in the SED, these galaxies would be
un-detected in the bluer bands and appear in the
redder filters (the so-called drop-out technique or LBG selection;
\citealp{Madau1996, Steidel1999, Giavalisco2002, Stark2009}). This
technique has been used extensively over the past few years in
conjunction with deep multi-waveband data and spectroscopic observations to
identify and study star-forming galaxies all the way to the cosmic dawn and
epoch of reionization (e.g. \citealp{Yan2004, Stark2011, Capak2011,
  Steidel2011, Oesch2013, Treu2013, Bouwens2014, Oesch2015}).

Here we use the following LBG color selection from \citet{Bouwens2015}
to identify candidates at $\langle z \rangle \sim 4$ and $\langle z
\rangle \sim 5$ respectively:

\begin{subequations}\label{grp}
\begin{align}
    &({\rm Subaru}(B)-{\rm Subaru}(V)) >1,\label{first}\\
    &({\rm CFHT}(i^*)-{\rm WFC3(F125W)}) <1,\label{second}\\
    &\notag ({\rm Subaru}(B)-{\rm Subaru}(V)) >1.6\\
    &\times({\rm CFHT}(i^*)-{\rm WFC3(F125W)})+1. \label{third}
\end{align}
\end{subequations}

for the $B$-dropout and: 

\begin{subequations}\label{grp}
\begin{align}
    &({\rm Subaru}(V)-{\rm CFHT}(i^*)) >1.2,\label{first}\\
    &({\rm CFHT}(z^*)-{\rm WFC3(F160W)}) <1.3,\label{second}\\
    &\notag ({\rm Subaru}(V)-{\rm CFHT}(i^*)) >0.8\\
    &\times({\rm CFHT}(z^*)-{\rm WFC3(F160W)})+1.2 \label{third}
\end{align}
\end{subequations}

for the $V$-dropout. Figure 21 shows the color-color diagrams for the $B$-dropout
and $V$-dropout selections and the corresponding photometry used
along with the photometric redshift distribution
of the selected candidates, as measured in the previous
section. The LBG selected sources have photometric redshifts consistent with the
selections. There are five spectroscopically
  confirmed sources (three in the $B$-dropout and two in the $V$-dropout)
  among the selected candidates with colors consistent with the
  corresponding criteria.

\subsection{Passively Evolving Galaxies at $z>1$}

The presence of passively evolving galaxies at $z \sim 2$ has been
established for some time now \citep{Daddi2005, Papovich2006,
  Kriek2006, Kriek2008, Kriek2009, Ilbert2013, Muzzin2013b,
  Papovich2015}. These galaxies are argued to
be the progenitors of the most massive systems that form the red
sequence at low redshifts \citep{Cassata2013, Williams2014,
  Barro2014b}. In the past few years there have
been several studies that predict the presence of these systems at $z
\sim 3$ \citep{Nayyeri2014, Wang2016} and some predictions as high as
$z \sim 5-6$ \citep{Mobasher2005, Wiklind2008}. Identification of
these objects requires deep observations in infrared wavelengths where
they have the peak of their emission due to colder and redder stars
\citep{Wuyts2007, Nayyeri2014}. The CANDELS COSMOS deep {\it HST}/WFC3 near infra-red
observations and the {\it Spitzer}/IRAC
TFIT photometry measurements are crucial in separating red galaxies
from blends of fainter sources and lower redshift dusty star-forming
galaxies. The multi-band data in the CANDELS COSMOS, especially the
wealth of data available on both sides of the Balmer/4000\,\AA\ break
help constrain the SEDs of old systems and separate them from
the red dusty star-forming galaxies. 

One well-known color-color diagram is the $BzK$ color plot that is used to
identify star-forming and quiescent galaxies at $z \sim 1-2$ \citep
{Daddi2004}. The $BzK$ method uses $(z-K)$ vs. $(B-z)$ colors of
galaxies to separate the two populations at high redshift
\citep{Daddi2004, Daddi2007}. This
is mostly based on the presence of the Balmer/4000\,\AA\ break in the
SED of older galaxies that is observed in the $(z-K)$ color at $z \sim
1-2$. The color selection diagram is such that objects with a
$BzK \equiv (z-K)-(B-z)>-0.2$ are identified as actively star-forming
galaxies at $z \sim 1-2$ while objects with a $BzK<-0.2$ and
$(z-K)>2.5$ are identified as passively evolving systems at similar
redshifts. Variations of the $BzK$ diagram can be used to extend this method to higher
redshifts \citep{Daddi2004, Guo2012}. The so-called $VJL$ and $iHM$ diagrams use
the $(J-3.6\,\mu\rm m)$ vs. $(V-J)$ and $(H-4.5\,\mu\rm m)$ vs. $(i-H)$ colors to identify star-forming
and quiescent galaxies at $z\sim2.5$ and 3.5 respectively. It is based on the same
principle as the $BzK$ diagram with the $(J-3.6\,\mu\rm m)$ and
$(H-4.5\,\mu\rm m)$ colors probing the 4000\,\AA\ break at higher redshifts.

We use the $BzK$, $VJL$ and $iHM$ selections \citep{Daddi2004,
  Guo2012} to identify passively evolving
systems in the CANDELS COSMOS at $z>1$. Figure 22 shows the $BzK$
color-color plot along with the higher redshift
$VJL$ and $iHM$ plots. The star-forming and quiescent galaxies in each plot are identified in
the blue and red regions respectively. The inset in each plot shows
the photometric redshift distribution.  As
expected, these sources are very red in both the $(B-z)$ and $(z-K)$
colors and the corresponding ones in the $VJL$ and $iHM$ diagrams. This is
indicative of the pronounced Balmer/4000\,\AA\ break in the SED of
these galaxies and the general old stellar population. Figure 22 inset
shows the redshift distribution of the quiescent galaxies in red. The
$BzK$ identified passive systems have a mean photometric redshift of
$z = 1.61$ and a distribution that is consistent with the population
being at $z>1$. Furthermore this population has a mean stellar mass of
$\rm 10^{10.97}\,M_{\odot}$. The photometric
redshift and stellar mass distributions and mean values along with the
colors of the passive $BzK$ galaxies are consistent with the quiescent
galaxy selection at high redshift. One of the passive galaxies
identified in the CANDELS COSMOS area has been spectroscopically
confirmed at $z=1.265$. This galaxy has a stellar mass of
$\rm 10^{10.98}\,M_{\odot}$ and photometric redshift of
$z_{phot}=1.29$ consistent with the spectroscopic redshift and with
the galaxy being massive and old at $z>1$. The $VJL$ selected passive
systems have mean photometric redshift of $z=2.33$ and mean stellar
mass of $\rm 10^{11.03}\,M_{\odot}$. The photometric redshift and mass distributions
are consistent with passive old galaxy selection at $z \sim 2-3$. The
$iHM$ redshift distribution shows a bi-modality with a large fraction
of lower redshift galaxies in the selection. \citet{Guo2012} discussed
the 50\% contaminant fraction in the $iHM$ color selection. We have 16
spectroscopically confirmed star-forming galaxies in the $BzK$ plot at
$z\sim1.5$ and two at $z\sim2.5$ in the $VJL$. All these sources have
colors consistent with the expected values from the selection
criteria. There is only one source in the spectroscopic sample with
$z_{\rm spec}=1.564$ with inconsistent $(B-z)$ and $(z-K)$ colors
compared to what is expected for the $sBzK$ sample. This source has
$z_{\rm phot}=0.682$ and is one of the outliers in Figure 11. 

The $BzK$ color diagram is well calibrated using large
and highly complete spectroscopic samples of galaxies
\citep{Daddi2004} and furthermore it predicts the
location of stars on the $(z-K)$ vs. $(B-z)$ plane (through
the criterion $(z-K)<0.3(B-z)-0.5$;
\citealp{Daddi2004}). \citet{Guo2012} recently extended this analysis
to higher redshifts using redder filters and in particular the {\it Spitzer}
3.6\,$\mu$m and 4.5\,$\mu$m observations. This allows us to examine
the colors of the stars measured in our catalog against predictions
made by the $BzK$, $VJL$ and $iHM$ color-color diagrams. Figure 22
shows the $BzK$ colors of point sources from our catalog, identified from
the \texttt{SExtractor} CLASS\_STAR parameter, along with the
predicted locus from \citet{Daddi2004}. We further show the
$(J-3.6\,\mu\rm m)$ vs. $(V-J)$ and $(H-4.5\,\mu\rm m)$ vs. $(i-H)$
colors of stars in our catalog along with the corresponding synthetic
colors of stars from BaSeL stellar library \citep{Lejeune1997, Westera2002} similar to
\citet{Guo2012}. We see from Figure 22 that the measured near infrared
and specifically infrared {\it Spitzer} colors of the stars in our catalog
are consistent with the predictions from stellar models.

%The passive galaxies identified in the CANDELS COSMOS area at $z \sim
%1.5$ from the $BzK$ selections have number density of ...

%more on the number density and mass density and comparisons with other works
%more on the SED inferred physical parameters (SFR and age)

\section{Summary}

We used CANDELS {\it HST}/WFC3 and ACS observations over a 216\,arcmin$^2$ area of the
COSMOS field to construct a multi-wavelength catalog of galaxies that
is selected in the WFC3 F160W band. The catalog contains photometry
for 38671 sources from $\sim 0.3-8\,\mu$m along with
physical properties.

\begin{itemize}

\item We used \texttt{SExtractor} to measure the photometry of objects in the
  high resolution ({\it HST}/WFC3 and ACS) bands. This also becomes the
  reference catalog to measure the photometry in the low resolution
  (ground-based and {\it Spitzer}/IRAC) bands. \texttt{SExtractor} was run using
  two sets of parameters (the so-called hot and cold modes) that were
  adjusted to detect both faint/small and bright/blended objects with
  the combined catalog getting contributions from both. 

\item We used TFIT to measure the photometry in the low resolution
  bands. This involves using prior information from the high
  resolution \texttt{SExtractor} runs (the position and light distribution
  profiles) to construct templates that were fit to objects in the
  low resolution image and from which the photometry was measured.

\item Our combined final catalog contains photometry in all four
  {\it HST} bands observed by CANDELS (two ACS and two WFC3), the optical
  broad-band and narrow-band observations by Subaru, optical data by
  CFHT, near infrared medium-band and broad-band observations by
  NEWFIRM and UVISTA respectively and infrared observations by
  {\it Spitzer}/IRAC in all four channels.

\item We measured the photometric redshift and physical properties of
  all the objects in our catalog through SED fitting using various
  codes and reporting the median values with corresponding uncertainties. 

\item We selected star-forming galaxies from LBG selections using our
  measured photometry at $3<z<5$ and quiescent systems from $BzK$ and
  $VJL$ selections at $z>1$. The photometric redshift distribution of
  the candidates along with spectroscopic confirmations further
  verifies the measured photometry.

\end{itemize}

\section*{Acknowledgement}

We wish to thank the anonymous referee for carefully reading the
original manuscript and providing very useful suggestions. This work is
based on observations taken by the CANDELS Multi-Cycle Treasury
Program with the NASA/ESA HST, which is operated by the
Association of Universities for Research in Astronomy, Inc., under
NASA contract NAS5-26555. This work is based in part on observations made with the {\it Spitzer} Space
Telescope, which is operated by the Jet Propulsion Laboratory,
California Institute of Technology under a contract with
NASA. Financial support for this work was provided by NSF through
AST-1313319 for HN and AC. HN further acknowledges support from NASA
(grant No. NNX16AF39G). This work is based in part on data
products from observations made with ESO Telescopes at
the La Silla Paranal Observatories under ESO programme
ID 179.A-2005 and on data products produced by TER-
APIX and the Cambridge Astronomy survey Unit on be-
half of the UltraVISTA consortium. This study was based in
part on observations obtained with MegaPrime/MegaCam,
a joint project of CFHT and CEA/DAPNIA, at the Canada-
France-Hawaii Telescope (CFHT) which is operated by the
National Research Council (NRC) of Canada, the Institut
National des Science de l'Univers of the Centre National de
la Recherche Scienti que (CNRS) of France, and the University of
Hawaii. This work is based in part on data products produced at
TERAPIX and the Canadian Astronomy
Data Centre as part of the Canada-France-Hawaii Telescope
Legacy Survey, a collaborative project of NRC and CNRS.

\bibliographystyle{apj}
\bibliography{cosmos}

\newpage

\section*{Appendix A: The Hot/Cold mode SExtractor Parameters}

The cold mode \texttt{SExtractor} parameters were optimized to detect the bright
galaxies while avoiding deblended sources. For this reason we use a
Tophat filter with a small deblending threshold and minimum count. The
hot mode was adjusted to detect low surface
brightness and small galaxies. Therefore we chose a Gaussian filter
which is more suitable for smaller targets and larger deblending
parameters. The main \texttt{SExtractor} parameters are listed below with the
differences in the cold/hot mode marked in bold.
\\

\noindent \#-------- Catalog -------- \\

\noindent CATALOG\_TYPE	ASCII\_HEAD	\\

\noindent\#-------- Extraction -------- \\

\noindent DETECT\_TYPE	CCD	\\
FLAG\_TYPE       OR \\
{\bf DETECT\_MINAREA	5.0/10.0} \\		
{\bf DETECT\_THRESH	0.75/0.70}	 \\
{\bf ANALYSIS\_THRESH	5.0/0.70}	 \\
FILTER		Y \\
{\bf FILTER\_NAME	tophat\_9.0\_9x9.conv/gauss\_4.0\_7x7.conv} \\
{\bf DEBLEND\_NTHRESH	16/64} \\
{\bf DEBLEND\_MINCONT	0.0001/0.001} \\
CLEAN		Y \\
CLEAN\_PARAM	1.0 \\
MASK\_TYPE	CORRECT	 \\

\noindent\#-------- Photometry -------- \\

\noindent PHOT\_FLUXFRAC   0.2, 0.5, 0.8    \\
PHOT\_APERTURES 1.47,2.08,2.94,4.17,5.88,8.34, \\
11.79,16.66,23.57,33.34,47.13 \\
{\bf SATUR\_LEVEL     120.0/3900.0}          \\
PIXEL\_SCALE     0.060           \\
MAG\_GAMMA	4.0		 \\

\noindent\#-------- Star/Galaxy Separation -------- \\

\noindent {\bf SEEING\_FWHM	0.18/0.19}	 \\
STARNNW\_NAME	default.nnw \\	

\noindent\#-------- Background -------- \\

\noindent {\bf BACK\_SIZE	256/128}	 \\
{\bf BACK\_FILTERSIZE	9/5} \\
BACKPHOTO\_TYPE  LOCAL \\
{\bf BACKPHOTO\_THICK	100/48}	 \\	

\noindent\#-------- Check Image -------- \\

\noindent CHECKIMAGE\_TYPE  SEGMENTATION \\

\noindent\#-------- Memory -------- \\

\noindent MEMORY\_OBJSTACK 4000             \\
MEMORY\_PIXSTACK 400000           \\
MEMORY\_BUFSIZE  5000            \\
 
\noindent\#--------  Miscellaneous --------  \\

\noindent VERBOSE\_TYPE	NORMAL		 \\

\noindent\#--------  New Stuff --------  \\

\noindent WEIGHT\_TYPE     MAP\_RMS,MAP\_RMS \\
WEIGHT\_THRESH   10000.0,10000.0 \\
PHOT\_AUTOPARAMS 2.5, 3.5 \\
GAIN 3070.790 \\
MAG\_ZEROPOINT 25.960 \\

\section*{Appendix B: Summary of the Assumptions used in Measuring the
  Photometric Redshift and Stellar Mass}

\begin{table*}[h]
\begin{center}
\caption{Summary of the SED fitting codes used in estimating the photometric redshift.}
\begin{tabular}{llclcl}
\hline
\hline
PI & Code & Fitting Method & Template Set & Emission Line & Reference \\
\hline
Finkelstein & EAZY & min $\chi^2$ & EAZY+BX418 & Yes &
                                                                \citet{Brammer2008},
  \citet{Erb2010}\\
Gruetzbauch & EAZY & min $\chi^2$ & EAZY & Yes &
                                                          \citet{Brammer2008}
  \\
Pforr & HyperZ & min $\chi^2$ & Maraston05 & No &
                                                           \citet{Bolzonella2000},
                                                           \citet{Maraston2005}
  \\
Salvato & LePhare & min $\chi^2$ & BC03+Polletta & Yes &
                                                                  \citet{Arnouts2011},
                                                                  \citet{Bruzual2003},
                                                                  \\
 & & & & & \citet{Polletta2007} \\
Wiklind & WikZ & min $\chi^2$ & BC03 & No &
                                                     \citet{Wiklind2008}
  \\

Wuyts & EAZY & min $\chi^2$ & EAZY & Yes &
                                                          \citet{Brammer2008}
  \\

\hline
\end{tabular}
\end{center}
{\footnotesize Notes. See \citet{Dahlen2013}
  for more detail.}
\end{table*}

\begin{table*}[h]
\begin{center}
\caption{Summary of the SED fitting codes used in estimating the
  stellar mass.}
\begin{tabular}{lllccccl}
\hline
\hline
ID & PI & Code & Fitting Method & Template Set & Emission Line & IMF & Reference \\
\hline
M2 & Barro & FAST & min $\chi^2$ & BC03 & No & Chabrier &
                                                                \citet{Kriek2009}, \citet{Barro2013}
  \\
M4 & Finkelstein & own code & min $\chi^2$ & CB07 & Yes & Salpeter &
                                                          \citet{Bruzual2007}, \citet{Finkelstein2012b}
  \\
M6 & Fontana & zphot & min $\chi^2$ & BC03 & Yes$^{\dagger}$ & Chabrier &
                                                          \citet{Giallongo1998},
                                                                     \citet{Fontana2000},                                                             
  \\
 & & & & & & & \citet{Fontana2006} \\
M14 & Lee & SpeedyMC & MCMC & BC03 & Yes & Chabrier &
                                                           \citet{Acquaviva2012}
  \\
M10 & Pforr & HyperZ & min $\chi^2$ & M05 & No & Chabrier &
                                                                  \citet{Bolzonella2000},
                                                      \citet{Maraston2005},
                                                                  \\
 & & & & & & & \citet{Maraston2006},
                                                      \citet{Daddi2005},\\
& & & & & & & \citet{Pforr2012}, \citet{Pforr2013} \\
M11 & Salvato & LePhare & median of PDFs & BC03 & Yes & Chabrier &
                                                                  \citet{Arnouts2011}
                                                                  \\
M12 & Wiklind & WikZ & min $\chi^2$ & BC03 & No & Chabrier &
                                                     \citet{Wiklind2008}
  \\
M13 & Wuyts & FAST & min $\chi^2$ & BC03 & No & Chabrier &
                                                          \citet{Kriek2009}, \citet{Wuyts2011}
  \\

\hline
\end{tabular}
\end{center}
{\footnotesize Notes. See \citet{Mobasher2015} and \citet{Santini2015}
  for more detail. $^{\dagger}$: also without nebular emission
  included.}
\end{table*}

\section*{Appendix C: Catalog Entries}

%\vspace{1cm}
\begin{longtable*}{ll}
\caption{Photometry Catalog Entries.}\\
\hline
Column Number & Column Name \\ 
\hline
\# 1  ID \\
\# 2 & IAU\_designation \\
\# 3 & RA \\
\# 4 & Dec \\
\# 5 & APCOR \\
\# 6  & CFHT\_uS\_FLUX  \\
\# 7 & CFHT\_uS\_FLUXERR  \\
\# 8 & CFHT\_gS\_FLUX  \\
\# 9 & CFHT\_gS\_FLUXERR  \\
\# 10 & CFHT\_rS\_FLUX  \\
\# 11 & CFHT\_rS\_FLUXERR  \\
\# 12 & CFHT\_iS\_FLUX  \\
\# 13 & CFHT\_iS\_FLUXERR  \\
\# 14 & CFHT\_zS\_FLUX  \\
\# 15 &  CFHT\_zS\_FLUXERR  \\
\# 16 & Subaru\_B\_FLUX  \\
\# 17 & Subaru\_B\_FLUXERR  \\
\# 18 & Subaru\_gp\_FLUX  \\
\# 19 & Subaru\_gp\_FLUXERR  \\
\# 20 & Subaru\_V\_FLUX  \\
\# 21 & Subaru\_V\_FLUXERR  \\
\# 22 & Subaru\_rp\_FLUX  \\
\# 23 & Subaru\_rp\_FLUXERR  \\
\# 24 & Subaru\_ip\_FLUX  \\
\# 25 & Subaru\_ip\_FLUXERR  \\
\# 26 & Subaru\_zp\_FLUX  \\
\# 27 & Subaru\_zp\_FLUXERR  \\
\# 28 & ACS\_F606W\_FLUX  \\
\# 29 & ACS\_F606W\_FLUXERR  \\
\# 30 & ACS\_F814W\_FLUX  \\
\# 31 & ACS\_F814W\_FLUXERR  \\
\# 32 & WFC3\_F125W\_FLUX  \\
\# 33 & WFC3\_F125W\_FLUXERR  \\
\# 34 & WFC3\_F160W\_FLUX  \\
\# 35 & WFC3\_F160W\_FLUXERR  \\
\# 36 & Ultravista\_Y\_FLUX  \\
\# 37 & UltraVISTA\_Y\_FLUXERR  \\
\# 38 & UltraVISTA\_J\_FLUX  \\
\# 39 & UltraVISTA\_J\_FLUXERR  \\
\# 40 & UltraVISTA\_H\_FLUX  \\
\# 41 & UltraVISTA\_H\_FLUXERR  \\
\# 42 & UltraVISTA\_Ks\_FLUX  \\
\# 43 & UltraVISTA\_Ks\_FLUXERR  \\
\# 44  & IRAC\_Ch1\_FLUX  \\
\# 45 & IRAC\_Ch1\_FLUXERR  \\
\# 46 & IRAC\_Ch2\_FLUX  \\
\# 47 &  IRAC\_Ch2\_FLUXERR  \\
\# 48 & IRAC\_Ch3\_FLUX  \\
\# 49 & IRAC\_Ch3\_FLUXERR  \\
\# 50 & IRAC\_Ch4\_FLUX  \\
\# 51 & IRAC\_Ch4\_FLUXERR  \\
\# 52 & NEWFIRM\_J1\_FLUX  \\
\# 53 & NEWFIRM\_J1\_FLUXERR  \\
\# 54 & NEWFIRM\_J2\_FLUX  \\
\# 55 & NEWFIRM\_J2\_FLUXERR  \\
\# 56 & NEWFIRM\_J3\_FLUX  \\
\# 57 & NEWFIRM\_J3\_FLUXERR  \\
\# 58 & NEWFIRM\_H1\_FLUX  \\
\# 59 & NEWFIRM\_H1\_FLUXERR  \\
\# 60 & NEWFIRM\_H2\_FLUX  \\
\# 61 & NEWFIRM\_H2\_FLUXERR  \\
\# 62 & NEWFIRM\_K\_FLUX  \\
\# 63 & NEWFIRM\_K\_FLUXERR  \\
\# 64 & Subaru\_IB\_427\_FLUX  \\
\# 65 & Subaru\_IB\_427\_FLUXERR  \\
\# 66 & Subaru\_IB\_464\_FLUX  \\ 
\# 67 & Subaru\_IB\_464\_FLUXERR  \\
\# 68 & Subaru\_IB\_484\_FLUX  \\
\# 69 & Subaru\_IB\_484\_FLUXERR  \\
\# 70 & Subaru\_IB\_505\_FLUX  \\
\# 71 & Subaru\_IB\_505\_FLUXERR  \\
\# 72 & Subaru\_IA\_527\_FLUX  \\
\# 73 & Subaru\_IA\_527\_FLUXERR  \\
\# 74 & Subaru\_IB\_574\_FLUX  \\
\# 75 & Subaru\_IB\_574\_FLUXERR  \\
\# 76 & Subaru\_IA\_624\_FLUX  \\
\# 77 & Subaru\_IA\_624\_FLUXERR  \\
\# 78 & Subaru\_IA\_679\_FLUX  \\
\# 79 & Subaru\_IA\_679\_FLUXERR  \\
\# 80 & Subaru\_IB\_709\_FLUX  \\
\# 81 & Subaru\_IB\_709\_FLUXERR  \\
\# 82 & Subaru\_NB\_711\_FLUX  \\
\# 83 & Subaru\_NB\_711\_FLUXERR  \\
\# 84 & Subaru\_IA\_738\_FLUX  \\
\# 85 & Subaru\_IA\_738\_FLUXERR  \\
\# 86 & Subaru\_IA\_767\_FLUX  \\
\# 87 & Subaru\_IA\_767\_FLUXERR  \\
\# 88 & Subaru\_NB\_816\_FLUX  \\
\# 89 & Subaru\_NB\_816\_FLUXERR  \\
\# 90 & Subaru\_IB\_827\_FLUX  \\
\# 91 & Subaru\_IB\_827\_FLUXERR  \\
\# 92 & FWHM\_IMAGE \\
\# 93 & FLAGS \\
\# 94 & CLASS\_STAR \\
\hline
\end{longtable*}
{\footnotesize\noindent Notes: 
Col. (1): F160W SExtractor ID. \\
Col. (2): IAU designation. \\
Col. (3) \& (4): Target coordinates (in degrees). \\
Col. (5): F160W FLUX\_AUTO/FLUX\_ISO, applied to ACS and WFC3 bands.\\
Col. (6) - (15): CFHT fluxes and errors from TFIT (microJansky).\\
Col. (16) - (27): Subaru broad-band fluxes and errors from TFIT (microJansky). \\
Col. (28) - (35): {\it HST}/ACS and WFC3 fluxes and errors from TFIT (microJansky). \\
Col. (36) - (43): UltraVISTA near infrared fluxes and errors from TFIT (microJansky). \\
Col. (44) - (51): {\it Spitzer} IRAC infrared fluxes and errors from TFIT (microJansky). \\
Col. (52) - (63): NEWFIRM medium band fluxes and errors from TFIT (microJansky). \\
Col. (64) - (91): Subaru medium and narrow-band fluxes and errors from TFIT (microJansky). \\
Col. (92): SExtractor F160W FWHM (pixel). \\ 
Col. (93): Photometry flags. \\
Col. (94): SExtractor stellar classification (1=Star). \\}

\begin{longtable}{ll}
\caption{Photometric Redshift Catalog Entries.}\\
\hline
Column Number & Column Name \\ 
\hline
\# 1 & ID \\
\# 2 & Spec\_z \\
\# 3 & Spec\_z\_dq \\
\# 4 & Photo\_z\_Wuyts \\
\# 5 & zinf68\_Wuyts \\
\# 6 & zsup68\_Wuyts \\
\# 7 & zinf95\_Wuyts \\
\# 8 & zsup95\_Wuyts \\
\# 9 & Photo\_z\_Pforr \\
\# 10 & zinf68\_Pforr \\
\# 11 & zsup68\_Pforr \\
\# 12 & zinf95\_Pforr \\
\# 13 & zsup95\_Pforr \\
\# 14 & Photo\_z\_Wiklind \\
\# 15 & zinf68\_Wiklind \\
\# 16 & zsup68\_Wiklind \\
\# 17 & zinf95\_Wiklind \\
\# 18 & zsup95\_Wiklind \\
\# 19 & Photo\_z\_Finkelstein \\
\# 20 & zinf68\_Finkelstein \\
\# 21 & zsup68\_Finkelstein \\
\# 22 & zinf95\_Finkelstein \\
\# 23 & zsup95\_Finkelstein \\
\# 24 & Photo\_z\_Gruetzbauch \\
\# 25 & zinf68\_Gruetzbauch \\
\# 26 & zsup68\_Gruetzbauch \\
\# 27 & zinf95\_Gruetzbauch \\
\# 28 & zsup95\_Gruetzbauch \\
\# 29 & Photo\_z\_Salvato \\
\# 30 & zinf68\_Salvato \\
\# 31 & zsup68\_Salvato \\
\# 32 & zinf95\_Salvato \\
\# 33 & zsup95\_Salvato \\
\hline
\end{longtable}
{\footnotesize\noindent Notes: \\
Col. (1): F160W SExtractor ID. \\
Col. (2): Spectroscopic redshift.  \\
Col. (3): Spectroscopic redshift data quality: 1=secure, 2=intermediate, 3=uncertain.  \\
Col. (4) - (8): Wuyts photometric redshift estimates and uncertainties.  \\ 
Col. (9) - (13): Pforr photometric redshift estimates and uncertainties.  \\
Col. (14) - (18): Wiklind photometric redshift estimates and uncertainties.  \\
Col. (19) - (23): Finkelstein photometric redshift estimates and uncertainties.  \\ 
Col. (24) - (28): Gruetzbauch photometric redshift estimates and uncertainties.  \\
Col. (29) - (33): Salvato photometric redshift estimates and uncertainties.  \\}

\begin{longtable}{ll}
\caption{Stellar Mass Catalog Entries.}\\
\hline
Column Number & Column Name \\ 
\hline
\# 1  &   ID \\
\# 2   &  Hmag          \\ 
\# 3  &   PhotFlag          \\   
\# 4  &   CLASS\_STAR        \\
\# 5  &   AGNFlag         \\ 
\# 6  &   zbest        \\
\# 7  &   zspec            \\  
\# 8  &  q-zspec       \\
\# 9  &  zphot          \\
\# 10  &  zphot\_l68      \\                
\# 11  &  zphot\_u68         \\        
\# 12  &  zphot\_l95              \\
\# 13  &  zphot\_u95                 \\    
\# 14  &  zphot\_Ilbert     \\
\# 15  &  M\_med      \\
\# 16  &  s\_med           \\   
\# 17  &  M\_neb\_med                   \\
\# 18  &  s\_neb\_med               \\
\# 19  &  M\_14\_cons       \\        
\# 20  &  M\_11\_tau               \\     
\# 21  &   M\_6\_tau\_NEB           \\
\# 22  &   M\_13\_tau           \\
\# 23  &   M\_12               \\
\# 24  &   M\_6\_tau                \\
\# 25  &   M\_2\_tau                      \\
\# 26  &   M\_6\_deltau                \\    
\# 27  &   M\_6\_invtau            \\
\# 28  &   M\_10                        \\
\# 29  &   M\_4b        \\
\# 30  &   M\_14\_lin    \\            
\# 31  &   M\_14\_deltau  \\  
\# 32  &   M\_14\_tau        \\
\# 33  &   M\_14\_inctau    \\
\# 34  &   M\_14       \\
\# 35  &   M\_neb\_med\_lin \\    
\# 36  &   s\_neb\_med\_lin    \\
\# 37  &   M\_med\_lin      \\
\# 38  &   s\_med\_lin     \\
\hline
\end{longtable}
{\footnotesize\noindent Notes: \\
Col. (1): F160W SExtractor ID. \\
Col. (2): F160W SExtractor MAG\_AUTO for convenience (mag). \\
Col. (3): 0 = good nonzero = use with caution or bad. \\
Col. (4): SExtractor stellar classification (1=Star). \\
Col. (5): AGN. \\
Col. (6): Best of photo-z or spec-z. This is the redshift used for SED-fitting.\\
Col. (7): Spectroscopic redshift. \\
Col. (8): Quality of spectroscopic redshift (good=1). \\
Col. (9): photometric redshift. \\
Col. (10) - (13): 68\% and 95\% confidence intervals on the photo-z.\\
Col. (14): Photometric redshift from COSMOS catalog. \\
Col. (15): CANDELS reference median stellar mass (dex(solMass)).\\
Col. (16): Standard deviation on M\_med (dex(solMass)).\\
Col. (17): Median stellar mass including nebular component (dex(solMass)).\\
Col. (18): Standard deviation on M\_neb\_med (dex(solMass)).\\
Col. (19): Stellar mass from Method 14\_cons (dex(solMass)). \\
Col. (20): Stellar mass from Method 11\_tau (dex(solMass)). \\
Col. (21): Stellar mass from Method 6\_tau\_NEB (dex(solMass)). \\
Col. (22): Stellar mass from Method 13\_tau (dex(solMass)). \\
Col. (23): Stellar mass from Method 12 (dex(solMass)). \\
Col. (24): Stellar mass from Method 6\_tau (dex(solMass)). \\
Col. (25): Stellar mass from Method 2\_tau (dex(solMass)). \\
Col. (26): Stellar mass from Method 6\_deltau (dex(solMass)). \\
Col. (27): Stellar mass from Method 6\_invtau (dex(solMass)). \\
Col. (28): Stellar mass from Method 10 (dex(solMass)). \\
Col. (29): Stellar mass from Method 4 (dex(solMass)). \\
Col. (30): Stellar mass from Method 14\_lin (dex(solMass)). \\
Col. (31): Stellar mass from Method 14\_deltau (dex(solMass)). \\
Col. (32): Stellar mass from Method 14\_tau (dex(solMass)). \\
Col. (33): Stellar mass from Method 14\_inctau (dex(solMass)). \\
Col. (34): Stellar mass best fitted from Method 14 (dex(solMass)). \\
Col. (35) - (36): Median stellar mass including nebular component
calculated by the Hodges-Lehmann estimator in the linear space and
standard deviation (dex(solMass)). \\
Col. (37) - (38): Median stellar mass with no nebular component
calculated by the Hodges-Lehmann estimator in the linear space and
standard deviation (dex(solMass)). \\}

\end{document}